# Relativistic De Broglie matter wave contains SU(n) symmetry

Huaiyang Cui
Department of Physics, Beihang University, Beijing, 100191, China
Email: hycui@buaa.edu.cn

**Abstract:** We show that the relativistic matter waves of two particle system contains the SU(2) symmetry; the relativistic matter waves of three particle system contains the SU(3) symmetry. Therefore, the relativistic matter wave is suitable for describing many-particle system such as quarks or gluons etc that are engaged with implicated symmetry problems. The weak interaction and strong interaction are investigated by using the relativistic de Broglie matter wave with SU(n) symmetry. Our results agree well with experiments. This work is to re-discover and re-explore de Broglie matter wave with the SU(n) symmetry and its applications to particle physics. This work was divided into two parts: Chapter 1, Relativistic de Broglie matter wave; Chapter 2, Factory of prosthesis.

**Keywords:** matter wave, SU(n) symmetry, weak interaction, strong interaction,

## Chapter 1 Relativistic de Broglie matter wave

We show that the relativistic matter waves of two particle system contains the SU(2) symmetry; the relativistic matter waves of three particle system contains the SU(3) symmetry. Therefore, the relativistic matter wave is suitable for describing many-particle system such as quarks or gluons etc that are engaged with implicated symmetry problems.

1. **Riding-wave momentum**

In the author's early paper [19], we have derived out the relativistic matter wave from the Newton second law as follows

$$\phi = \exp\left(\frac{i}{\hbar}\int_{x_0}^{x}(mu_\mu + qA_\mu)dx_\mu\right) \quad (1)$$

where the integral takes from the initial point $x_0$ to the final point $x$ by an arbitrary mathematical path in the uncertainty zone in which the particle has uncertain positions duo to the quantum hidden variable. In bound states, due to periodic boundary condition, the relativistic matter wave is quantized for a closed orbit $L$ by

$$\frac{1}{\hbar}\oint_L (m\mathbf{u} + q\mathbf{A})\cdot d\mathbf{x} = 2\pi n \quad n = 1,2,3... \quad . \quad (2)$$

For convenience, we define the **riding-wave momentum** $R_\mu$ by

$$R_\mu = mu_\mu + qA_\mu .\qquad(3)$$

Then the relativistic matter wave and its quantization condition are written as

$$\phi = \exp\left(\frac{i}{\hbar}\int_{x_0}^{x} R_\mu dx_\mu\right)$$
$$\frac{1}{\hbar}\oint_L \mathbf{R}\cdot d\mathbf{x} = 2\pi n \quad n=1,2,3...\qquad(4)$$

In a closed $N$ particle system, the $j$-th particle is in an electromagnetic field $A$ which is the sum of the contributions over other $N-1$ particles, as shown in Fig.1.

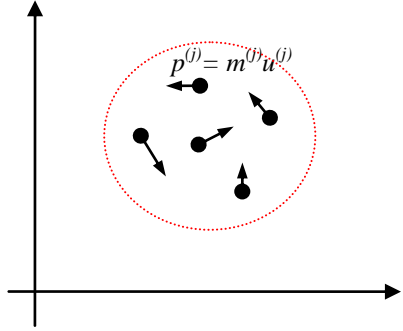

Fig.1 $N$ particle system.

$$A_\mu(x^{(j)}) = \sum_{k=1,k\neq j}^{N} A_\mu^{(k)} = \sum_{k=1,k\neq j}^{N} \frac{1}{4\pi\varepsilon_0}\frac{u_\mu^{(k)}}{|x^{(k)}-x^{(j)}|} .\qquad(5)$$

Where the superscript index ($j$) denotes the $j$-th particle under consideration, the superscript ($k$) denotes the others, we have chosen the Coulomb gauge for the 4-vector potentials. For our purposes, we rewrite the 4-vector potential in terms of momentum as follows

$$A_\mu(x^{(j)}) = \sum_{k=1,k\neq j}^{N} A_\mu^{(k)} = \sum_{k=1,k\neq j}^{N} \frac{1}{4\pi\varepsilon_0 m^{(k)}}\frac{p_\mu^{(k)}}{|x^{(k)}-x^{(j)}|} .\qquad(6)$$

In this closed $N$ particle system, we can safely say that the $j$-th particle has the riding-wave momentum in the general form as

$$R_\mu^{(j)} = p_\mu^{(j)} + \sum_{k=1}^{N} S^{(jk)} p_\mu^{(k)} .\qquad(7)$$

Where $S^{(jk)}$ is the coefficients of the expansion for the $j$-th particle over the $k$-th particle. In matrix form, that is

$$R = (1+S)p$$
or
$$\begin{bmatrix} R^{(1)} \\ R^{(2)} \\ ... \\ R^{(N)} \end{bmatrix} = (1+\begin{bmatrix} S^{(11)} & S^{(12)} & ... & S^{(1N)} \\ S^{(21)} & S^{(22)} & ... & S^{(2N)} \\ ... & ... & ... & ... \\ S^{(N1)} & S^{(N2)} & ... & S^{(NN)} \end{bmatrix})\begin{bmatrix} p^{(1)} \\ p^{(2)} \\ ... \\ p^{(N)} \end{bmatrix}\qquad(8)$$

Where 1 denotes unitary matrix, now, the matrix $S$ represents the interactions between each other, actually its entries are regarded as smaller quantities if the interactions are weak. In most cases for larger $N$ system, the coefficient matrix $S$ is an array hard to calculate like the 4-vector potential, so in practice $S$ is regarded as unknown quantity determined by experiments.

**Theorem 1:** The weak interaction matrix $S$ in the riding-wave momentum formula is a Hermitian matrix: $S^+ = S$.

**Proof:** The wave function of the $j$-th particle is given by

$$\phi^{(j)} = \exp\left(\frac{i}{\hbar}\int_{0(L)}^{x} R_\mu^{(j)} dx_\mu\right) = \exp\left(\frac{i}{\hbar}\int_{0(L)}^{x}[p_\mu^{(j)}(\mathbf{x},t) + S^{(jk)} p_\mu^{(k)}] dx_\mu\right), \qquad (9)$$

we define

$$\psi^{(j)} = \exp\left(\frac{i}{\hbar}\int_{0(L)}^{x} p_\mu^{(j)} dx_\mu\right),$$

$$|\psi^{(j)}|^2 = \psi^{(j)}\psi^{(j)*} = 1 \qquad (10)$$

regarding $S$ as smaller quantities for the weak interactions, then we have

$$\begin{aligned}
\phi^{(j)} &= \exp\left(\frac{i}{\hbar}\int_{0(L)}^{x}[p_\mu^{(j)}(\mathbf{x},t) + S^{(jk)} p_\mu^{(k)}] dx_\mu\right) \\
&= \psi^{(j)} \exp\left(\frac{S^{(jk)} i}{\hbar}\int_{0(L)}^{x}[\sum_{j=1}^{N} p_\mu^{(k)}] dx_\mu\right) = \psi^{(j)} \prod_{k=1}^{N} (\psi^{(k)})^{S^{(jk)}} \\
&= \psi^{(j)} \prod_{k=1}^{N}\left(1 + S^{(jk)} \ln\psi^{(k)}\right) + O(S^2) \\
&= \psi^{(j)} \left(1 + \sum_{k=1}^{N} S^{(jk)} \ln\psi^{(k)}\right) + O(S^2)
\end{aligned} \qquad (11)$$

Typically, we demand the matter wave function to meet the unity requirement:

$$|\phi^{(j)}|^2 = \phi^{(j)}\phi^{(j)*} = 1 . \qquad (12)$$

Using (here, do not sum over $j$)

$$\phi^{(j)} = \psi^{(j)}\left(1 + \sum_{k=1}^{N} S^{(jk)} \ln\psi^{(k)}\right)$$

$$\phi^{(j)+} = \left(1 + \sum_{k=1}^{N} \ln\psi^{(k)+} S^{(kj)*}\right)\psi^{(j)+}, \qquad (13)$$

$$\psi^{(j)+}\psi^{(j)} = 1$$

then we have (here, do not sum over $j$)

$$\phi^{(j)+}\phi^{(j)} = \left(1 + \sum_{k=1}^{N} \ln\psi^{(k)+} S^{(kj)*}\right) \psi^{(j)+}\psi^{(j)} \left(1 + \sum_{k=1}^{N} S^{(jk)} \ln\psi^{(k)}\right)$$

$$= \left(1 + \sum_{k=1}^{N} \ln\psi^{(k)+} S^{(kj)*}\right)\left(1 + \sum_{k=1}^{N} S^{(jk)} \ln\psi^{(k)}\right) \quad (14)$$

$$= 1 + \sum_{k=1}^{N} \ln\psi^{(k)+} S^{(kj)*} + \sum_{k=1}^{N} S^{(jk)} \ln\psi^{(k)} + O(S^2)$$

We know

$$\ln\psi^{(k)+} = \ln\psi^{(k)*} = -\ln\psi^{(k)} \quad . \quad (15)$$

thus

$$\phi^{(j)+}\phi^{(j)} = 1 + \sum_{k=1}^{N} (-S^{(kj)*} + S^{(jk)}) \ln\psi^{(k)} \quad . \quad (16)$$

Therefore the unity requirement leads to the conclusion

$$\begin{aligned} -S^{(kj)*} + S^{(jk)} &= 0 \\ S^+ &= S \end{aligned} \quad . \quad (17)$$

Proof finished.

**Theorem 2:** Since $S$ is a Hermitian matrix, according to the theory of group, for $N=3$, the Hermitian $S$ is a linear combination of the Gell-mann matrix set in terms of SU(3) symmetry:

$$S_{3X3} = \begin{bmatrix} S_{11} & S_{12} & S_{13} \\ S_{21} & S_{22} & S_{23} \\ S_{31} & S_{32} & S_{33} \end{bmatrix} = \sum_{a=1}^{8} \frac{1}{2} \alpha_a \lambda_a \quad . \quad (18)$$

Where $\alpha_a$ is eight real small coefficients of the expansion, $\lambda_a$ are the eight independent Gell-Mann matrices called as the eight generators of SU(3) group, the Gell-mann matrices are

$$\lambda_1 = \begin{bmatrix} 0 & 1 & 0 \\ 1 & 0 & 0 \\ 0 & 0 & 0 \end{bmatrix} \quad \lambda_2 = \begin{bmatrix} 0 & -i & 0 \\ i & 0 & 0 \\ 0 & 0 & 0 \end{bmatrix} \quad \lambda_3 = \begin{bmatrix} 1 & 0 & 0 \\ 0 & -1 & 0 \\ 0 & 0 & 0 \end{bmatrix}, \quad (19)$$

$$\lambda_4 = \begin{bmatrix} 0 & 0 & 1 \\ 0 & 0 & 0 \\ 1 & 0 & 0 \end{bmatrix} \quad \lambda_5 = \begin{bmatrix} 0 & 0 & -i \\ 0 & 0 & 0 \\ i & 0 & 0 \end{bmatrix} \quad \lambda_6 = \begin{bmatrix} 1 & 0 & 0 \\ 0 & 0 & 1 \\ 0 & 1 & 0 \end{bmatrix}, \quad (20)$$

$$\lambda_7 = \begin{bmatrix} 0 & 0 & 0 \\ 0 & 0 & -i \\ 0 & i & 0 \end{bmatrix} \quad \lambda_8 = \frac{1}{\sqrt{3}} \begin{bmatrix} 1 & 0 & 0 \\ 0 & 1 & 0 \\ 0 & 0 & -2 \end{bmatrix} \quad . \quad (21)$$

Proof: This is a mathematical theorem, was proved in the theory of group. See [1]. Proof finished.

**Theorem 3:** Since $S$ is a Hermitian matrix, according to the theory of group, for $N=2$, the Hermitian $S$ is a linear combination of the Pauli matrix set in terms of SU(2) symmetry:

$$\begin{bmatrix} R^{(1)} \\ R^{(2)} \end{bmatrix} = (1+S) \begin{bmatrix} p^{(1)} \\ p^{(2)} \end{bmatrix} = (1+\alpha_1\sigma_1+\alpha_2\sigma_2+\alpha_3\sigma_3) \begin{bmatrix} p^{(1)} \\ p^{(2)} \end{bmatrix}. \tag{22}$$

Where, the Pauli matrices (SU(2) group) are given by (corresponding to $\lambda_1, \lambda_2, \lambda_3$ in SU(3) group)

$$\sigma_1 = \begin{bmatrix} 0 & 1 \\ 1 & 0 \end{bmatrix}, \quad \sigma_2 = \begin{bmatrix} 0 & -i \\ i & 0 \end{bmatrix}, \quad \sigma_3 = \begin{bmatrix} 1 & 0 \\ 0 & -1 \end{bmatrix}. \tag{23}$$

and $\alpha_1$, $\alpha_2$, $\alpha_3$ are three independent real first order small parameters.
Proof: This is a mathematical theorem, was proved in the theory of group. See [1]. Proof finished.

**Theorem 4:** The riding-wave momentum **R** is recognized as the canonical momentum in the analytical mechanics.
Proof: by definition, we know

$$\begin{aligned} \mathbf{R}^{(j)} &= \mathbf{p}^{(j)} + q^{(j)}\mathbf{A}^{(j)} \\ iH^{(j)}/c &= R_4^{(j)} = p_4^{(j)} + q^{(j)}A_4^{(j)} \end{aligned}. \tag{24}$$

If the j-th particle in the N particle system is in its stationary state, then the fourth component of the riding-wave momentum of the *j*-th particle is a constant. It is called as the Hamiltonian of the *j*-th particle $H^{(j)}$, or called as the energy of the *j*-th particle $E^{(j)}$, given by

$$\begin{aligned} R_4^{(j)} &= (1+\sum_k S^{(jk)}) p_4^{(k)} \\ R_4^{(j)} &= p_4^{(j)} + q^{(j)}A_4^{(j)} \equiv \frac{iH^{(j)}}{c} \equiv \frac{iE^{(j)}}{c} \end{aligned}, \tag{25}$$

where $A_4^{(j)}$ is the electric potential at where the j-th particle experiences. In the followings, we drop the superscript (j), substituting them into the total momentum formula

$$p^2 = p_1^2 + p_2^2 + p_3^2 + p_4^2 = -m^2c^2, \tag{26}$$

we get

$$(\mathbf{R}-q\mathbf{A})^2 - (H/c+iqA_4)^2 = -m^2c^2. \tag{27}$$

That is

$$H = c\sqrt{|\mathbf{R}-q\mathbf{A}|^2 + m^2c^2} - icqA_4 \tag{28}$$

This Hamiltonian is completely the same as that in usual textbooks of electrodynamics, under this square root, the canonical momentum of analytical mechanics is recognizable as **R**, rather than **p**. In quantum mechanics, a stationary state definitely means the energy (or the Hamiltonian) to be constant, so

$$R_4 \equiv \frac{iE^{(j)}}{c} = const. \tag{29}$$

Proof finished

Exercise:
1. It is easy to derive the Lagrangian of the j-th particle from its Hamiltonian

$$L = -m^2c^2\sqrt{1-v^2/c^2} + q\mathbf{v}\cdot\mathbf{A}/c - iqA_4 \tag{30}$$

Hint: look up electrodynamics textbook, e.g., see Ref.[18] (p.266,Eq.(10-57)).

2. It is also easy to derive the 4-vector force that acts on the j-th particle from its Lagrangian

$$f_\mu = m\frac{dp_\mu}{d\tau} = qF_{\mu\nu}u_\nu$$
$$F_{\mu\nu} = \frac{\partial A_\mu}{\partial x_\nu} - \frac{\partial A_\nu}{\partial x_\mu} \tag{31}$$

Hint: look up electrodynamics textbook, e.g., see Ref.[18] (p.264,Eq.(10-44)).

**2. *N*=2 particle system and SU(2) symmetry**

We have plenty of knowledge of the electromagnetism. In order to investigate the electromagnetic interaction, let us study a two-particle system: light particle 1 + massive particle 2, in which there is a pure electromagnetic interaction. For two particle system, the eight Gell-Mann matrices reduce into three independent Pauli Matrices, as follows

$$\begin{bmatrix} R^{(1)} \\ R^{(2)} \end{bmatrix} = (1+S)\begin{bmatrix} p^{(1)} \\ p^{(2)} \end{bmatrix} = (1+\alpha_1\sigma_1 + \alpha_2\sigma_2 + \alpha_3\sigma_3)\begin{bmatrix} p^{(1)} \\ p^{(2)} \end{bmatrix}. \tag{32}$$

Where, the Pauli matrices (SU(2) group) are given by (corresponding to $\lambda_1,\lambda_2,\lambda_3$ in SU(3) group)

$$\sigma_1 = \begin{bmatrix} 0 & 1 \\ 1 & 0 \end{bmatrix}, \quad \sigma_2 = \begin{bmatrix} 0 & -i \\ i & 0 \end{bmatrix}, \quad \sigma_3 = \begin{bmatrix} 1 & 0 \\ 0 & -1 \end{bmatrix}, \tag{33}$$

and $\alpha_1$, $\alpha_2$, $\alpha_3$ are three independent real first order small parameters.

To note that the matrix *S* has involved imaginary numbers, since we want study the pure electromagnetic interaction between real momentum $p^{(1)}$ and real momentum $p^{(2)}$, so that all entries of matrix *S* must be real numbers, thus we require $\alpha_2=0$. We discard the $\alpha_3$, because it represents the self-feedback-interaction: $p^{(1)} \sim \alpha_3 p^{(1)}$ whose explosive or decaying behaviors are not what we expect for stationary solutions. Thus we have

$$\begin{bmatrix} R^{(1)} \\ R^{(2)} \end{bmatrix} = (1+\alpha_1\sigma_1)\begin{bmatrix} p^{(1)} \\ p^{(2)} \end{bmatrix}, \tag{34}$$

it becomes algebra equations:

$$R^{(1)} = p^{(1)} + \alpha_1 p^{(2)} \Leftrightarrow \frac{1}{\hbar}\oint_{L^{(1)}} \mathbf{R}^{(1)} \cdot d\mathbf{x} = 2\pi n^{(1)} \quad n^{(1)} = 1,2,3...$$
$$R^{(2)} = p^{(2)} + \alpha_1 p^{(1)} \Leftrightarrow \frac{1}{\hbar}\oint_{L^{(2)}} \mathbf{R}^{(2)} \cdot d\mathbf{x} = 2\pi n^{(2)} \quad n^{(2)} = 1,2,3...$$

. (35)

The particle 1 is in a stationary state with a constant energy, so that

$$R_4^{(1)} = \frac{iE^{(1)}}{c} = const. \quad . \tag{36}$$

Substituting it into the quantization condition, that is

$$p_4^{(1)} + \alpha_1 p_4^{(2)} = \frac{iE^{(1)}}{c} = const. \quad . \tag{37}$$

The particle 2 is massive, approximately $u^{(2)}=(0,ic)$, thus

$$\frac{icm^{(1)}}{\sqrt{1-v^{(1)2}/c^2}} + \alpha_1 icm^{(2)} = \frac{iE^{(1)}}{c} = const. \quad . \tag{38}$$

We evaluate it step by step

$$\begin{aligned}
icm^{(1)}(1+\frac{v^{(1)2}}{2c^2}) + O(\frac{v^{(1)4}}{c^4}) + \alpha_1 icm^{(2)} &= \frac{iE^{(1)}}{c} = const. \\
icm^{(1)} + \frac{E_k^{(1)}}{c^2}ic + \alpha_1 icm^{(2)} + O(\frac{v^{(1)4}}{c^4}) &= \frac{iE^{(1)}}{c} = const. \\
icm^{(1)} + \frac{E_{classic} - E_p^{(1)}}{c^2}ic + \alpha_1 icm^{(2)} + O(\frac{v^{(1)4}}{c^4}) &= \frac{iE^{(1)}}{c} = const.
\end{aligned} \tag{39}$$

Where, $E_{classic}$ is the constant classical (non-relativistic) mechanical energy, $E_k^{(1)}$ is the kinetic energy of the particle 1, $E_p^{(1)}$ is the potential energy of the particle 1, given by

$$E_p^{(1)} = \frac{q^{(1)}q^{(2)}}{4\pi\varepsilon_0 r} \quad , \tag{40}$$

Where $q^{(1)}$ and $q^{(2)}$ are the two charges of the two particles, respectively, $r$ is the distance of the two particles. Neglecting higher order terms, we take variation on it, and get

$$\delta\left(-\frac{E_p^{(1)}}{c^2}ic + \alpha_1 icm^{(2)}\right) = 0 \implies \alpha_1 = \frac{E_p^{(1)}}{m^{(2)}c^2} = \frac{1}{4\pi\varepsilon_0 m^{(2)}c^2}\frac{q^{(1)}q^{(2)}}{r} \quad . \tag{41}$$

For the pure electromagnetic interaction of the two particle system, we have

$$\begin{bmatrix} R^{(1)} \\ R^{(2)} \end{bmatrix} = (1+\alpha_1\sigma_1)\begin{bmatrix} p^{(1)} \\ p^{(2)} \end{bmatrix} = (1+\frac{E_p^{(1)}}{m^{(2)}c^2}\begin{bmatrix} 0 & 1 \\ 1 & 0 \end{bmatrix})\begin{bmatrix} p^{(1)} \\ p^{(2)} \end{bmatrix} . \tag{42}$$

We know that the electromagnetic potential acting on the particles 1 is given by

$$A_\mu = \frac{q^{(2)}u_\mu^{(2)}}{4\pi\varepsilon_0 c^2 r} \quad \mu = 1,2,3,4 \quad . \tag{43}$$

We derive out

$$R^{(1)} = p^{(1)} + q^{(1)}A \quad . \tag{44}$$

Or we derive out

$$\begin{pmatrix} R_1^{(1)} \\ R_2^{(1)} \\ R_3^{(1)} \\ R_4^{(1)} \end{pmatrix} = \begin{pmatrix} p_1^{(1)} \\ p_2^{(1)} \\ p_3^{(1)} \\ p_4^{(1)} \end{pmatrix} + q^{(1)} \begin{pmatrix} A_1 \\ A_2 \\ A_3 \\ A_4 \end{pmatrix} . \quad (45)$$

In order to keep the Lorentz invariance for the matrix, we redefine

$$r^2 = |\mathbf{x}^{(1)} - \mathbf{x}^{(2)}|^2 - \frac{|\mathbf{x}^{(1)} - \mathbf{x}^{(2)}|^2}{c^2} . \quad (46)$$

It is the distance of the two particles with the measurement of retardation time, it is Lorentz-invariant, i.e., the equation holds in this mathematical form for any inertial reference frame.

Gravitational interaction between the sun and the earth shares the same formalism with the electromagnetic interaction.

As a simple example, consider a hydrogen atom, the nucleus provides the Coulomb potential for the electron which moves about the massive nucleus, the electron is the light particle. The relativistic matter wave of the electron (particle 1) is given by

$$\phi(x) = \exp\left(\frac{i}{\hbar} \int_{0(L)}^{x} R_\mu dx_\mu \right) = \exp\left(\frac{i}{\hbar} \int_{0(L)}^{x} (p_\mu + qA_\mu) dx_\mu \right) , \quad (47)$$

so the quantization condition for the electron in the hydrogen atom is given by

$$\frac{1}{\hbar} \oint_{L\_orbit} (\mathbf{p} + q\mathbf{A}) \cdot d\mathbf{x} = 2\pi a , \quad (48)$$

where, the integral path takes on the orbit of the electron.

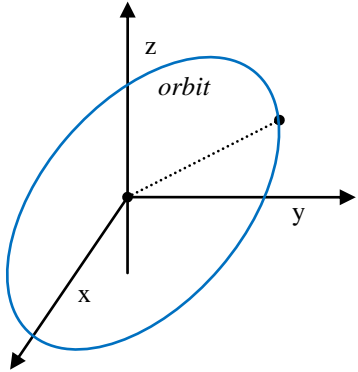

Fig.2 The electron moves around the nucleus in a hydrogen atom.

In a hydrogen atom as shown Fig. 2, the electron moves around the nucleus in an orbit $L$, satisfying the quantization condition, the integral path along the orbit $L$ spells out

$$\delta = \frac{1}{\hbar} \int_0^{2\pi} p_\varphi r \sin\theta d\varphi + \frac{1}{\hbar} \int_0^{2\pi} p_\theta r d\theta + \frac{2}{\hbar} \int_{r_1}^{r_2} p_r dr = 2\pi n; \quad n = 1, 2, \dots , \quad (49)$$

where $r_1$ and $r_2$ are the minimum radius and maximum radius for the orbit. Because the spherical coordinates are three independent coordinates, the general quantization equation can be divided into three independent quantization equations

$$\frac{1}{\hbar}\int_0^{2\pi} p_\varphi r\sin\theta d\varphi = 2\pi m; \quad m = 0, \pm 1, \ldots \quad , \tag{50}$$

$$\frac{1}{\hbar}\int_0^{2\pi} p_\theta r d\theta = 2\pi n_\theta; \quad n_\theta = 0, 1, 2, \ldots \quad , \tag{51}$$

$$\frac{2}{\hbar}\int_{r_1}^{r_2} p_r dr = 2\pi n_r; \quad n_r = 0, 1, 2, \ldots \quad . \tag{52}$$

The elliptic orbit must close up, usually

$$\frac{1}{\hbar}\int_0^{2\pi} p_\varphi r\sin\theta d\varphi = \frac{1}{\hbar}\int_0^{2\pi} J_z d\varphi = 2\pi m; \quad m = 0, \pm 1, \ldots \quad , \tag{53}$$

$$\frac{1}{\hbar}\int_0^{2\pi} p_\theta r d\theta = \frac{1}{\hbar}\int_0^{2\pi} \sqrt{J^2 - \frac{J_z^2}{\sin^2\theta}} d\theta = 2\pi n_\theta; \quad n_\theta = 0, 1, 2, \ldots \tag{54}$$

Substituting constant angular momentum into the above equations, we get

$$\begin{aligned} J_z &= p_\varphi r\sin\theta \\ \frac{1}{\hbar}\int_0^{2\pi} J_z d\varphi &= \frac{2\pi J_z}{\hbar} = 2\pi m \quad \Rightarrow \quad J_z = m\hbar \end{aligned} \tag{55}$$

The total angular momentum $J$ is also a constant, we get

$$\begin{aligned} p_\theta &= \sqrt{J^2/r^2 - p_\varphi^2} \\ \frac{1}{\hbar}\int_0^{2\pi} p_\theta r d\theta &= \frac{1}{\hbar}\int_0^{2\pi} \sqrt{J^2 - \frac{m^2\hbar^2}{\sin^2\theta}} d\theta = 2\pi n_\theta \end{aligned} \tag{56}$$

This integral was evaluated in [20], the result is

$$\begin{aligned} \frac{1}{\hbar}\int_0^{2\pi} p_\theta r d\theta &= 2\pi(J/\hbar - |m|) = 2\pi n_\theta \\ J &= (n_\theta + |m|)\hbar = j\hbar \end{aligned} \tag{57}$$

Where $j=k+|m|$, total angular momentum $J$ is quantized. The momentum in the radial axis is given by

$$p_r = \sqrt{-m_e^2 c^2 - J^2/r^2 - p_4^2} \quad , \tag{58}$$

where $m_e$ is the rest mass of the electron. The total energy $E$ is a constant, from

$$\frac{i}{\hbar}\int_0^t (p_4 + qA_4)dx_4 = \frac{-ct}{\hbar}(p_4 + qA_4) = -\frac{iEt}{\hbar} \quad , \tag{59}$$

using the Coulomb potential, we get

$$p_4 = \frac{iE}{c} - qA_4 = \frac{iE}{c} + ik_e \frac{e^2}{cr} . \tag{60}$$

Substituting it into the radial momentum, we get

$$\frac{2}{\hbar} \int_{r_1}^{r_2} p_r dr = \frac{2}{\hbar c} \int_{r_1}^{r_2} \sqrt{-m_e^2 c^4 - \frac{j^2 \hbar^2 c^2}{r^2} + (E + k_e \frac{e^2}{r})^2} dr = 2\pi n_r . \tag{61}$$

This integral was evaluated in [20], the result is

$$\frac{2\pi E \alpha}{\sqrt{m_e^2 c^4 - E^2}} - 2\pi \sqrt{j^2 - \alpha^2} = 2\pi n_r . \tag{62}$$

where $\alpha$ is the fine structure constant, we find it requires $j$=1,2,3,…

From it, we obtain the energy level formula for the hydrogen atom

$$E = m_e c^2 \left[ 1 + \frac{\alpha^2}{(\sqrt{j^2 - \alpha^2} + n_r)^2} \right]^{-1/2} . \tag{63}$$

It is completely the same as that in the Dirac wave equation for hydrogen atom, see Table 1.

Table 1. Energy level calculation of hydrogen atom

| Notations | | Energy levels (eV) | | |
|---|---|---|---|---|
| shell | Eq.(50-52) | this calculation | Standard values, Eq.(63) | error |
| n=1,l=0 | m=1,n_θ=0,n_r=0 | -13.605879 | -13.605879 | 0 |
| n=2,l=0 | m=2,n_θ=0,n_r=0 | -3.401436 | -3.401436 | 0 |
| n=2,l=1 | m=1,n_θ=0,n_r=1 | -3.401482 | -3.401482 | 0 |
| n=3,l=0 | m=3,n_θ=0,n_r=0 | -1.511746 | -1.511746 | 0 |
| n=3,l=1 | m=2,n_θ=0,n_r=1 | -1.511751 | -1.511751 | 0 |
| n=3,l=2 | m=1,n_θ=0,n_r=2 | -1.511765 | -1.511765 | 0 |
| n=4,l=0 | m=4,n_θ=0,n_r=0 | -0.850357 | -0.850357 | 0 |
| n=4,l=1 | m=3,n_θ=0,n_r=1 | -0.850358 | -0.850358 | 0 |
| n=4,l=2 | m=2,n_θ=0,n_r=2 | -0.850360 | -0.850360 | 0 |
| n=4,l=3 | m=1,n_θ=0,n_r=3 | -0.850365 | -0.850365 | 0 |
| n=5,l=0 | m=5,n_θ=0,n_r=0 | -0.544228 | -0.544228 | 0 |
| n=5,l=1 | m=4,n_θ=0,n_r=1 | -0.544229 | -0.544229 | 0 |
| n=5,l=2 | m=3,n_θ=0,n_r=2 | -0.544229 | -0.544229 | 0 |
| n=5,l=3 | m=2,n_θ=0,n_r=3 | -0.544230 | -0.544230 | 0 |
| n=5,l=4 | m=1,n_θ=0,n_r=4 | -0.544233 | -0.544233 | 0 |

Gravitational interaction between the sun and the earth shares the same quantization condition with electromagnetic interaction in a hydrogen atom, if we know the Planck constant that belongs to the earth.

## 3. *N*=3 particle system: the Pauli exclusion principle

*N*-particle system may contain many kinds of interactions among them, each interaction has its own performance. In a helium atom, the nucleus (particle 3) provides the Coulomb electric field for two electrons (particle 1 and particle 2) which move about the massive nucleus.

$$\begin{bmatrix} R^{(1)} \\ R^{(2)} \\ R^{(3)} \end{bmatrix} = (1+S) \begin{bmatrix} p^{(1)} \\ p^{(2)} \\ p^{(3)} \end{bmatrix} . \qquad (64)$$

The massive nucleus (particle 3) contribute the electromagnetic 4-vector potential *A*, can be separated from the particles 1 and 2, thus

$$\begin{bmatrix} R^{(1)} \\ R^{(2)} \end{bmatrix} = (1+S) \begin{bmatrix} p^{(1)} \\ p^{(2)} \end{bmatrix} + \begin{bmatrix} q^{(1)}A \\ q^{(2)}A \end{bmatrix} . \qquad (65)$$

If the helium atom was placed in an external electromagnetic field, the electromagnetic 4-vector potential *A* should include the external electromagnetic field. Nevertheless, in this case, the *N*=3 system reduce to *N*=2 system with an electromagnetic 4-vector potential *A*, and *S* is the combination of the Pauli matrices as follows

$$\begin{bmatrix} R^{(1)} \\ R^{(2)} \end{bmatrix} = (1+\alpha_1\sigma_1+\alpha_2\sigma_2+\alpha_3\sigma_3) \begin{bmatrix} p^{(1)} \\ p^{(2)} \end{bmatrix} + \begin{bmatrix} q^{(1)}A \\ q^{(2)}A \end{bmatrix} . \qquad (66)$$

Where $\alpha_1$, $\alpha_2$, $\alpha_3$ are three independent real first order small parameters for the SU(2) group.

To note that the matrix *S* has involved imaginary numbers, since we want study the pure electromagnetic interaction between real momentum $p^{(1)}$ and real momentum $p^{(2)}$, so that all entries of matrix *S* must be real numbers, thus we require $\alpha_2=0$. This time, we discard the $\alpha_1$ for the simple reasons: the $\sigma_1$ provides electromagnetic interaction that we have discussed in the preceding section, it should be involved into the electromagnetic 4-vector *A*, in other words: *A* absorbed up $\sigma_1$, thus we have

$$\begin{bmatrix} R^{(1)} \\ R^{(2)} \end{bmatrix} = (1+\alpha_3 \begin{bmatrix} 1 & 0 \\ 0 & -1 \end{bmatrix}) \begin{bmatrix} p^{(1)} \\ p^{(2)} \end{bmatrix} + \begin{bmatrix} q^{(1)}A \\ q^{(2)}A \end{bmatrix} , \qquad (67)$$

it becomes algebra equations:

$$\begin{aligned} R^{(1)} &= p^{(1)} + \alpha_3 p^{(1)} + q^{(1)}A \\ R^{(2)} &= p^{(2)} - \alpha_3 p^{(2)} + q^{(2)}A \end{aligned} \qquad (68)$$

The two particles are in a stationary state with a constant Hamiltonian, so that

$$p_4^{(1)} + p_4^{(2)} = iH/c = const. . \qquad (69)$$

Substituting it into the above equation, we get

$$p_4^{(1)} + \alpha_3 p_4^{(1)} + q^{(1)}A + p_4^{(2)} - \alpha_3 p_4^{(2)} + q^{(2)}A = \frac{iH}{c} , \qquad (70)$$

so

$$\alpha_3 = \frac{iH/c - (p_4^{(1)} + q^{(1)}A + p_4^{(2)} + q^{(2)}A)}{p_4^{(1)} - p_4^{(2)}} \quad , \tag{71}$$

or

$$\alpha_3 = \frac{iH/c - (p_4^{(1)} + q^{(1)}A + p_4^{(2)} + q^{(2)}A)}{(m^{(1)}c^2 + E_k^{(1)}) - (m^{(2)}c^2 + E_k^{(2)})} \quad , \tag{72}$$

where, $E_k^{(j)}$ is the kinetic energy of the *j*-th particle, the denominator may arises a singularity. If the two particles are identical particles, in a helium atom they are two electrons, then they will not be allowed to have the same kinetic energy, otherwise the SU(2) group parameter $\alpha_3$ will blow up! The coupling between the two electrons will blow up! This is just what the Pauli Exclusion Principle implies. In terms of the above formalism, the Pauli Exclusion Principle says that the two particles in a stationary state cannot share the same relativistic kinetic energy! i.e.

$$p_4^{(1)} \neq p_4^{(2)} \quad . \tag{73}$$

Applying the Virial theorem: $2<E_k> = -<E_p>$ to calculate the energy levels of the two electrons, it is easily understood that the Pauli Exclusion Principle actually bans the two electron to share the same energy level: $E = <E_k> + <E_p>$.

Now we discuss the influence of the SU(2) group parameter $\alpha_3$ in energy levels of the two electrons. According to the riding-wave momentum formula, the two particles about a massive nucleus evolves as

$$\begin{aligned} R^{(1)} &= p^{(1)} + \alpha_3 p^{(1)} + q^{(1)}A \\ R^{(2)} &= p^{(2)} - \alpha_3 p^{(2)} + q^{(2)}A \end{aligned} \quad . \tag{74}$$

The non-relativistic energy of the particle 1 is calculated by

$$E^{(1)} = \frac{1}{2m^{(1)}} \mathbf{p}^{(1)} \cdot \mathbf{p}^{(1)} + q^{(1)}V^{(1)} \quad . \tag{75}$$

Neglecting higher order terms, that is

$$\begin{aligned} E^{(1)} &= \frac{1}{2m^{(1)}} [\mathbf{R}^{(1)} - \alpha_3 \mathbf{R}^{(1)} - q^{(1)}\mathbf{A}] \cdot [\mathbf{R}^{(1)} - \alpha_3 \mathbf{R}^{(1)} - q^{(1)}\mathbf{A}] + q^{(1)}V^{(1)} \\ &= \frac{1}{2m^{(1)}} [\mathbf{R}^{(1)} \cdot \mathbf{R}^{(1)} - 2\alpha_3 \mathbf{R}^{(1)} \cdot \mathbf{R}^{(1)} - 2q^{(1)}\mathbf{R}^{(1)} \cdot \mathbf{A} - 2\alpha_3 q^{(1)}\mathbf{R}^{(1)} \cdot \mathbf{A} \\ &+ q^{(1)2}\mathbf{A} \cdot \mathbf{A}] + q^{(1)}V^{(1)} \end{aligned} \quad . \tag{76}$$

The particle 2 has the same formula for calculating its non-relativistic energy with minus $\alpha_3$.

Then single particle has common energy formula as

$$\begin{aligned} E^{(\pm)} &= \frac{1}{2m} [\mathbf{R} \mp \alpha_3 \mathbf{R} - q\mathbf{A}] \cdot [\mathbf{R} \mp \alpha_3 \mathbf{R} - q\mathbf{A}] + qV \\ &= \frac{1}{2m} [\mathbf{R} \cdot \mathbf{R} \mp 2\alpha_3 \mathbf{R} \cdot \mathbf{R} - 2q\mathbf{R} \cdot \mathbf{A} \pm 2\alpha_3 q\mathbf{R} \cdot \mathbf{A} + q^2 \mathbf{A} \cdot \mathbf{A}] + qV \end{aligned} \quad . \tag{77}$$

In there the plus sign is suitable for the particle 1, the minus sign for the particle 2. Why should we use the riding-wave momentum *R* to replace the true momentum *p*? Because the riding-wave

momentum *R* satisfies the quantization condition while the true momentum *p* not:

$$\frac{1}{\hbar}\oint_{L\_orbit} \mathbf{R}\cdot d\mathbf{x} = 2\pi n \quad n=1,2,3...$$
$$\frac{1}{\hbar}\oint_{L\_orbit} p_\mu dx_\mu = ? \quad (unknown, if\ A_\mu \neq 0)$$
(78)

Hence the riding-wave momentum *R* is a measureable quantity (locked by the quantization condition), rather than the true momentum *p* (unlocked) in experiment.

We put this helium atom in a place where the electric field is $V=cV_4/i$, including coulomb field from the nucleus, and the magnetic field B points in the direction of $x_3$-axis (z-axis), the 4-vector potential is given by.

$$V_4 = iV/c \iff V = cV_4/i$$
$$\mathbf{A} = \frac{1}{2}\mathbf{B}\times\mathbf{r} \iff \mathbf{A}\times\mathbf{r} = \frac{1}{2}r^2\mathbf{B} - \frac{1}{2}(rB_r)\mathbf{r}$$
(79)

Where **B** is perpendicular to the plane of electron motion, is perpendicular to the radius of electron motion. We invoke the following vector formulas

$$(\mathbf{A}\times\mathbf{B})\cdot(\mathbf{C}\times\mathbf{D}) = (\mathbf{A}\cdot\mathbf{C})(\mathbf{B}\cdot\mathbf{D}) - (\mathbf{A}\cdot\mathbf{D})(\mathbf{B}\cdot\mathbf{C})$$
or
$$(\mathbf{A}\cdot\mathbf{C})(\mathbf{B}\cdot\mathbf{D}) = (\mathbf{A}\times\mathbf{B})\cdot(\mathbf{C}\times\mathbf{D}) + (\mathbf{A}\cdot\mathbf{D})(\mathbf{B}\cdot\mathbf{C})$$
(80)

to find the angular momentum of the riding momentum *R* as follows

$$(\mathbf{r}\cdot\mathbf{r})(\mathbf{A}\cdot\mathbf{p}) = (\mathbf{r}\times\mathbf{A})\cdot(\mathbf{r}\times\mathbf{p}) + (\mathbf{A}\cdot\mathbf{r})(\mathbf{p}\cdot\mathbf{r})$$
$$= (\mathbf{r}\times\mathbf{A})\cdot(\mathbf{r}\times\mathbf{p}) + (\mathbf{A}\cdot\mathbf{r})(\mathbf{p}\cdot\mathbf{r})$$
(81)

Where, we have

$$\mathbf{L} = \mathbf{r}\times\mathbf{R}$$
$$(\mathbf{r}\times\mathbf{A}) = \frac{1}{2}r^2\mathbf{B} - \frac{1}{2}rB_r\mathbf{r} = \frac{1}{2}r^2\mathbf{B}$$
$$(\mathbf{A}\cdot\mathbf{r})(\mathbf{R}\cdot\mathbf{r}) = r^2 A_r R_r = 0$$
(82)

Therefore

$$\mathbf{A}\cdot\mathbf{R} = \frac{1}{2}\mathbf{B}\cdot\mathbf{L}$$
(83)

$$\mathbf{R}\cdot\mathbf{R} = \frac{1}{r^2}\mathbf{L}\cdot\mathbf{L} + R_r^2$$
(84)

That is

$$E^{(\pm)} = \frac{1}{2m}[\frac{1}{r^2}\mathbf{R}\cdot\mathbf{R} \mp 2\alpha_3 \frac{1}{r^2}\mathbf{L}\cdot\mathbf{L} \mp 2\alpha_3 R_r^2$$
$$-q\mathbf{B}\cdot(\mathbf{L}\mp\alpha_3\mathbf{L}) + q^2\mathbf{A}\cdot\mathbf{A}] + qV$$
(85)

The second term in the above equation represents the coupling effect between the angular momentum of the particle 1 and the angular momentum of the particle 1 itself. The total energy of

helium is given by

$$E = E^+ + E^- = \frac{1}{2m}[\frac{2}{r^2}\mathbf{R}\cdot\mathbf{R} - 2q\mathbf{B}\cdot\mathbf{L} + 2q^2\mathbf{A}\cdot\mathbf{A}] + 2qV \quad , \tag{86}$$

without involving the self-coupling.

**4. *N*=3 particle system: L-S coupling and spin concept**

Consider an alkali atom, the atomic core provides the Coulomb potential for the valence electron which moves about the massive core, the valence electron is the light particle. The riding-wave momentum of the valence electron (particle 1) and the atomic core (particle 2) are given by

$$\begin{bmatrix} R^{(1)} \\ R^{(2)} \end{bmatrix} = (1 + \alpha_1\sigma_1 + \alpha_2\sigma_2 + \alpha_3\sigma_3)\begin{bmatrix} p^{(1)} \\ p^{(2)} \end{bmatrix} + \begin{bmatrix} q^{(1)}A \\ q^{(2)}A \end{bmatrix} \quad , \tag{87}$$

where $\alpha_1$, $\alpha_2$, $\alpha_3$ are three independent real first order small parameters for the Pauli matrices (SU(2) group), the 4-vector potential *A* represents the electromagnetic interaction from the atomic core and external field. According to the discussion in the preceding sections, we know that $\alpha_1\sigma_1$ represents the electromagnetic interaction that can be absorbed into the 4-vector potential *A*, we should discard $\alpha_3\sigma_3$ term because it represents a smaller self-interaction that was recognized as the Pauli Exclusion Principle. Thus, we have

$$\begin{bmatrix} R^{(1)} \\ R^{(2)} \end{bmatrix} = (1 + \alpha_2\sigma_2)\begin{bmatrix} p^{(1)} \\ p^{(2)} \end{bmatrix} + \begin{bmatrix} q^{(1)}A \\ q^{(2)}A \end{bmatrix} = (1 + \alpha_2\begin{bmatrix} 0 & -i \\ i & 0 \end{bmatrix})\begin{bmatrix} p^{(1)} \\ p^{(2)} \end{bmatrix} + \begin{bmatrix} q^{(1)}A \\ q^{(2)}A \end{bmatrix} \quad . \tag{88}$$

We choose the center of mass as the reference frame, we have

$$\mathbf{p}^{(1)} + \mathbf{p}^{(2)} = 0 \quad . \tag{89}$$

Substituting it into the riding-wave equation, besides the fourth components, we get

$$\begin{bmatrix} \mathbf{R}^{(1)} \\ \mathbf{R}^{(2)} \end{bmatrix} = (1 + \alpha_2\begin{bmatrix} 0 & -i \\ i & 0 \end{bmatrix})\begin{bmatrix} \mathbf{p}^{(1)} \\ -\mathbf{p}^{(1)} \end{bmatrix} + \begin{bmatrix} q^{(1)}\mathbf{A} \\ q^{(2)}\mathbf{A} \end{bmatrix} \quad . \tag{90}$$

In this case, facing imaginary number, the valence electron has to consider itself having real-part momentum and imaginary-part momentum as follows

$$\mathbf{p}^{(1)} = \mathbf{p}^{\text{Re}(1)} + i\mathbf{p}^{\text{Im}(1)} \quad . \tag{91}$$

If only to observe the valence electron, we have

$$\begin{aligned} \mathbf{R}^{\text{Re}(1)} &= \mathbf{p}^{\text{Re}(1)} - \alpha_2\mathbf{p}^{\text{Im}(1)} + q^{(1)}\mathbf{A} \\ \mathbf{R}^{\text{Im}(1)} &= \mathbf{p}^{\text{Im}(1)} + \alpha_2\mathbf{p}^{\text{Re}(1)} \end{aligned} \quad . \tag{92}$$

It is interesting to regard the valence electron as dual particles: the real-part particle (with momentum $p^{\text{Re}(1)}$) and imaginary-part particle (with momentum $p^{\text{Im}(1)}$), obviously experiment can observe the real-part particle but unable to observe the imaginary-part particle. In matrix form, they are

$$\begin{bmatrix} \mathbf{R}^{Re(1)} \\ \mathbf{R}^{Im(2)} \end{bmatrix} = (1+\alpha_2) \begin{bmatrix} 0 & -1 \\ 1 & 0 \end{bmatrix} \begin{bmatrix} \mathbf{p}^{Re(1)} \\ \mathbf{p}^{Im(2)} \end{bmatrix} + \begin{bmatrix} q^{(1)}\mathbf{A} \\ 0 \end{bmatrix} . \tag{93}$$

The imaginary-momentum particle did not has the term $q^{(1)}A$, indicating that the imaginary-momentum particle has not energy transformation with the massive particle 2 or with the real-momentum particle, i.e. without channel, i.e. it has not energy variation. Consider the particle in Bohr's circular orbit. The their non-relativistic energies are given by

$$\begin{aligned} E^{Re} &= \frac{1}{2m}\mathbf{p}^{Re}\cdot\mathbf{p}^{Re} + qV^{Re} \\ &= \frac{1}{2}(\mathbf{R}^{Re}+\alpha_2\mathbf{p}^{Im}-q\mathbf{A})\cdot(\mathbf{R}^{Re}+\alpha_2\mathbf{p}^{Im}-q\mathbf{A}) + qV^{Re} \\ &= \frac{1}{2m}\left(\mathbf{R}^{Re}\cdot\mathbf{R}^{Re}+2\alpha_2\mathbf{R}^{Re}\cdot\mathbf{p}^{Im}-2q\mathbf{R}^{Re}\cdot\mathbf{A}-2\alpha_2 q\mathbf{p}^{Im}\cdot\mathbf{A}+q^2\mathbf{A}\cdot\mathbf{A}\right) + qV^{Re} \end{aligned} \tag{94}$$

Why should we use the riding-wave momentum $R$ to replace the true momentum $p$? Because the riding-wave momentum $R$ satisfies the quantization condition while the true momentum $p$ not:

$$\begin{aligned} &\frac{1}{\hbar}\oint_{L\_orbit} \mathbf{R}\cdot d\mathbf{x} = 2\pi n \quad n=1,2,3... \\ &\frac{1}{\hbar}\oint_{L\_orbit} p_\mu dx_\mu = ? \quad (unknown, if\ A_\mu \neq 0) \end{aligned} \tag{95}$$

Hence the riding-wave momentum $R$ is a measureable quantity (locked by the quantization condition), rather than the true momentum $p$ (unlocked) in experiment. The angular moment is defined in terms of the riding-wave momentum as

$$\mathbf{L} = \mathbf{r}\times\mathbf{R} , \tag{96}$$

$$(\mathbf{A}\cdot\mathbf{C})(\mathbf{B}\cdot\mathbf{D}) = (\mathbf{A}\times\mathbf{B})\cdot(\mathbf{C}\times\mathbf{D}) + (\mathbf{A}\cdot\mathbf{D})(\mathbf{B}\cdot\mathbf{C})$$
$$\mathbf{R}\cdot\mathbf{R} = \frac{1}{r^2}\mathbf{L}\cdot\mathbf{L} \quad (circular\ obit) \tag{97}$$

$$\mathbf{A}\cdot\mathbf{R} = \frac{1}{2}\mathbf{B}\cdot\mathbf{L} . \tag{98}$$

Therefore the energy is given by

$$E^{Re} = \frac{1}{2m}\left(\mathbf{R}^{Re}\cdot\mathbf{R}^{Re}+2\alpha_2\frac{1}{r^{Re}r^{Im}}[\mathbf{L}^{Re}\cdot(\mathbf{r}^{Im}\times\mathbf{p}^{Im})]-\mathbf{B}\cdot(\mathbf{L}^{Re}+\alpha_2(\mathbf{r}^{Im}\times\mathbf{p}^{Im}))\right) \tag{99}$$
$$+q^2\mathbf{A}\cdot\mathbf{A}) + qV^{Re}$$

For the imaginary-part particle, we have the almost same energy formula

$$E^{Im} = \frac{1}{2m}\left(\mathbf{R}^{Im}\cdot\mathbf{R}^{Im}-2\alpha_2\frac{1}{r^{Re}r^{Im}}[\mathbf{L}^{Im}\cdot(\mathbf{r}^{Re}\times\mathbf{p}^{Re})]-\mathbf{B}\cdot(\mathbf{L}^{Im}+\alpha_2(\mathbf{r}^{Re}\times\mathbf{p}^{Re}))\right)$$
$$+q^2\mathbf{A}\cdot\mathbf{A})$$

(100)

For the dual particle matrix, the second term in the brackets is the coupling energies of the real-part particle with the imaginary-part particle.

$$E^{coupling} = \frac{\alpha_2}{m} \frac{1}{r^{Re} r^{Im}} [\mathbf{L}^{Re} \cdot (\mathbf{r}^{Im} \times \mathbf{p}^{Im})] \ . \tag{101}$$

Now, let us find a way to determine the $\alpha_2$. The imaginary-momentum particle did not has the potential term $qA^{(1)}$, indicating that the imaginary-momentum particle has not energy transformation with the massive particle 2 or the real-momentum particle, i.e., without channel, i.e. it has not energy variation. The system energy only distributes within the real-momentum particle and massive particle 2. Omitting the energy variation of the massive particle 2, the particle 1 must be in a stationary state with a constant Hamiltonian energy $H^{Re}$, do not take into account of the imaginary-momentum particle energy.

$$\frac{iH^{Re}}{c} = R_4^{Re} = p_4^{Re} - \alpha_2 p_4^{Im} + qA_4 = const. \tag{102}$$

That is approximately

$$\frac{iH^{Re}}{c} = R_4^{Re} = icm + \frac{iE_k^{Re}}{c} - \alpha_2 icm + qA_4 = const. \tag{103}$$

Taking variation over the above equation, we get

$$\alpha_2 = \frac{E_k^{Re} - icqA_4}{mc^2} = \frac{E_k^{Re} + qV}{mc^2} \ . \tag{104}$$

From the Virial theorem, we know $E_k^{Re} + qV = qV/2$ for circular orbits, so that

$$\alpha_2 = \frac{qV}{2mc^2} = \frac{qq^{(core)}}{8\pi\varepsilon_0 r^{Re} mc^2} \ . \tag{105}$$

The imaginary=part particle leads a life as

$$\begin{aligned} R^{Im(1)} &= p^{Im(1)} + \alpha_2 p^{Re(1)} \\ &= p^{Im(1)} + \frac{qV}{2mc^2} p^{Re(1)} \end{aligned} \ . \tag{106}$$

Or

$$\begin{aligned} \mathbf{R}^{Im} &= \mathbf{p}^{Im} + \frac{qV}{2mc^2} \frac{m\mathbf{v}^{Re}}{\sqrt{1 - v^{Re2}/c^2}} \\ R_4^{Im} &= p_4^{Im} + \frac{qV}{2mc^2} \frac{imc}{\sqrt{1 - v^{Re2}/c^2}} \end{aligned} \ . \tag{107}$$

Approximately, we have

$$\begin{aligned} \mathbf{R}^{Im} &= \mathbf{p}^{Im} \\ R_4^{Im} &= p_4^{Im} + \frac{iqV}{2c} \end{aligned} \ . \tag{108}$$

The imaginary-part particle moves in an equivalent electric field V/2, comparing with the real-part particle which moves in the electric field V, so that the Bohr's radius of the imaginary-part particle

is given by
$$r^{Im} = 2r^{Re} \quad . \tag{109}$$
Total Angular momentum must be zero, thus this coupling effect is

$$\begin{aligned}
E^{(coupling)} &= \frac{qq^{(core)}}{8\pi\varepsilon_0 m^2 c^2 r^{Re3}} \frac{1}{2}[\mathbf{L}^{Re} \cdot (\mathbf{r}^{Im} \times \mathbf{p}^{Im})] \\
&\simeq \frac{qq^{(core)}}{8\pi\varepsilon_0 m^2 c^2 r^{Re3}} \frac{1}{2}(\mathbf{L}^{Re} \cdot \mathbf{L}^{Im}) \\
&= \frac{qq^{(core)}}{8\pi\varepsilon_0 m^2 c^2 r^{Re3}} \frac{1}{2}(s\hbar^2) \quad s = \pm 1, \pm 2, ...
\end{aligned} \tag{110}$$

For the minimal $s=+1$ or $s=-1$, it is completely the same with the formulae in the Schrodinger wave equation for alkali orbit-spin coupling effect, in the later the valence electron has the angular momentum $m\hbar$ and spin $\hbar/2$.

Remark: we introduced 1/2 concept for spin by $r^{Im}=2r^{Re}$, we eventually objected to $L_{spin} = \hbar/2$. Fundamentally, The half Bohr' angular momentum is unable to be quantized in physics, it must be rejected by the advanced physics.

## 5.  $N=2$ particle system: Fermi's Golden Rule under SU(2) symmetry

Quantum electrodynamics (QED) is the theory describing electromagnetic interaction between high energy particles, like electrons, leptons and quarks. Some of physicist's favorite QED processes are $e^+e^- \to \mu^+\mu^-$, $eq \to eq$, $\gamma \to (e^+e^-)$ and so on, where q denotes a quark.

Consider two particles, their interaction matrix $S$ is the combination of three Pauli matrices, according the preceding sections, the first Pauli matrix $\alpha_1\sigma_1$ represents the electromagnetic interaction, understood as basic interaction; the second Pauli matrix $\alpha_2\sigma_2$ arises the spin effect for electron; the third Pauli matrix $\alpha_3\sigma_3$ serves fermions for which it represents a kind of self-interaction that is recognized as the Pauli Exclusion Principle. In this section we investigate the scattering of electrons off a proton, calculate the differential cross section in their collision, ignoring the spin effect ($\alpha_2=0$) and the Pauli exclusive interaction ($\alpha_3=0$).

In high energy collision for two fermions 1 and 2, the $N=2$ system has the riding-wave momenta given by

$$\begin{bmatrix} R^{(1)} \\ R^{(2)} \end{bmatrix} = (1+\alpha_1\sigma_1)\begin{bmatrix} p^{(1)} \\ p^{(2)} \end{bmatrix} = (1+\alpha_1\begin{bmatrix} 0 & 1 \\ 1 & 0 \end{bmatrix})\begin{bmatrix} p^{(1)} \\ p^{(2)} \end{bmatrix} \quad . \tag{111}$$

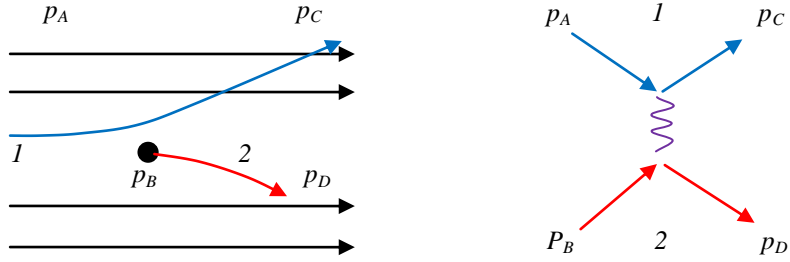

Fig.3 Particle 1 scatters off particle 2.

The interaction of the electron and the proton is the Coulomb potential energy $V(r)$ which appears in the group constant $\alpha_1$ in section 3, given by

$$\alpha_1 = \frac{1}{4\pi\varepsilon_0}\frac{q^{(1)}q^{(2)}}{rm^{(2)}c^2} = \frac{1}{m^{(2)}c^2}V(r)$$

$$A_\mu = \frac{1}{4\pi\varepsilon_0}\frac{q^{(2)}u_\mu^{(2)}}{c^2 r} = \frac{\alpha_1 p_\mu^{(2)}}{q^{(1)}}$$

(112)

where $q^{(1)}$ and $q^{(2)}$ are the charges of the incident particle and the target particle respectively. The single incident particle 1 is governed by

$$-i\hbar\frac{\partial \phi^{(1)}}{\partial x_\mu} = R_\mu^{(1)}\phi^{(1)} = (p_\mu^{(1)} + \alpha_1 p_\mu^{(2)})\phi^{(1)} \ . \tag{113}$$

Considering a small volume element $dV$ in where there are N particles 1 at a time, their average quantities (labeled AVE) satisfy

$$-i\hbar\frac{\partial \phi^{(1\_AVE)}}{\partial x_\mu} = (p_\mu^{(1\_AVG)} + \alpha_1 p_\mu^{(2)})\phi^{(1\_AVE)};$$

$$\phi^{(1\_AVE)} = \sum_{j=1}^{N}\phi^{(1\_j)}/N;$$

$$p_\mu^{(1\_AVE)} = \sum_{j=1}^{N}p_\mu^{(1\_j)}\phi^{(1\_j)}/\sum_{j=1}^{N}\phi^{(1\_j)}$$

(114)

Suppose that the incident particles 1 have the same initial state $i$ with the momentum $p_i$, since in high energy collision 99% incident particles go almost straightforward by the target particle, only about 1% incident particles would be apparently scattered by the target particle, their average quantities well approximately satisfy

$$p_\mu^{(1\_AVE)} = 0.99 p_\mu^{(i)} + 0.01 p_\mu^{(1\_other)} \simeq p_\mu^{(i)}$$

$$-i\hbar\frac{\partial \phi^{(1\_AVE)}}{\partial x_\mu} = (p_\mu^{(i)} + \alpha_1 p_\mu^{(2)})\phi^{(1\_AVE)}$$

(115)

The fewer scattered partition, the more accurate expectation. The wave function of the particle 1 can be expressed in terms of its free plane waves as

$$\phi^{(1\_AVE)} = \sum_{n=1}^{\infty} a_n(t)\exp(\frac{ip_n^{(1)}\cdot x}{\hbar}) \ . \tag{116}$$

To find the unknown coefficient $a_n$, we have to study its time-dependent equation. We know in the space-time $(x_1,x_2,x_3,x_4=-ct)$, the fourth component expresses its evolution, substituting it into the left side of the fourth component equation,

$$-i\hbar \frac{\partial \phi^{(1\_AVE)}}{\partial x_4} = (p_4^{(1)} + \alpha_1 p_4^{(2)}) \phi^{(1\_AVE)} \quad . \tag{117}$$

we get

$$-i\hbar \sum_{n=1}^{\infty} (\frac{\partial a_n}{\partial x_4} + \frac{a_n i p_{n4}^{(1)}}{\hbar}) \exp(\frac{i p_n^{(1)} \cdot x}{\hbar}) = \sum_{n=1}^{\infty} a_n (p_{i4}^{(1)} + \alpha_1 p_4^{(2)}) \exp(\frac{i p_n^{(1)} \cdot x}{\hbar}) \quad . \tag{118}$$

It reduces to

$$-i\hbar \sum_{n=1}^{\infty} (\frac{\partial a_n}{\partial x_4}) \exp(\frac{i p_n^{(1)} \cdot x}{\hbar}) = \sum_{n=1}^{\infty} a_n (p_{i4}^{(1)} - p_{n4}^{(1)}) \exp(\frac{i p_n^{(1)} \cdot x}{\hbar})$$

$$+ \sum_{n=1}^{\infty} a_n (\alpha_1 p_4^{(2)}) \exp(\frac{i p_n^{(1)} \cdot x}{\hbar}) \tag{119}$$

Multiplying by another eigen state $\phi_f^*$, integrating over the whole volume $V$, and using the orthonorrmality relation of eigen states, using the initial condition $a_{ni}=\delta_{ni}$, it leads to

$$\frac{\partial a_f}{\partial x_4} = \frac{1}{V} \int dx^3 (\frac{i\alpha_1 p_4^{(2)}}{\hbar}) \exp(\frac{i(p_i^{(1)} - p_f^{(1)}) \cdot x}{\hbar})$$

$$\frac{i\alpha_1 p_4^{(2)}}{\hbar} = \frac{iV(r)}{m^{(2)}c^2 \hbar} m^{(2)} ic = -\frac{V(r)}{c\hbar} \tag{120}$$

**Hint**: this formula is the same as that in Schrodinger's time-dependent perturbation theory, if the particle 2 is at rest. The particle 1 at the initial time t=-T/2=-∞, is in the initial state *i* as shown in Fig.3, we have

$$a_n(t = -\frac{T}{2}) = \delta_{ni} \quad . \tag{121}$$

At a later time *t*, we have

$$a_f(t) = \frac{1}{V} \int_{-T/2}^{t} dx_4 \int dx^3 (\frac{i\alpha_1 p_4^{(2)}}{\hbar}) \exp(\frac{i(p_i^{(1)} - p_f^{(1)}) \cdot x}{\hbar}) \quad . \tag{122}$$

At a final departure time t=T/2=∞, we have

$$a_f(t = \frac{T}{2}) = \frac{1}{V} \int dx^4 (\frac{i\alpha_1 p_4^{(2)}}{\hbar}) \exp(\frac{i(p_i^{(1)} - p_f^{(1)}) \cdot x}{\hbar}) \quad . \tag{123}$$

We define the $a(t=\infty)$ as the **transition amplitude** that the particle has scattered from an initial state *i* to a final state *f*, whose formula divides into two parts as

$$a_f(t=\frac{T}{2}) = \frac{1}{V}\int dx_4 \exp(\frac{i(p_i^{(1)} - p_f^{(1)}) \cdot x_4}{\hbar})\int dx^3(\frac{i\alpha_1 p_4^{(2)}}{\hbar})\exp(\frac{i(p_i^{(1)} - p_f^{(1)}) \cdot x}{\hbar})$$

$$= \int d(ict)\exp(\frac{i(E_f^{(1)} - E_i^{(1)}) \cdot t}{\hbar})\frac{1}{V}\int dx^3(\frac{i\alpha_1 p_4^{(2)}}{\hbar})\exp(\frac{i(\mathbf{p}_i^{(1)} - \mathbf{p}_f^{(1)}) \cdot \mathbf{x}}{\hbar})$$

$$= (2\pi\hbar)\delta(E_f^{(1)} - E_i^{(1)})\Gamma_{if}$$

$$\Gamma_{if} \equiv \frac{1}{V}\int dx^3(\frac{-c\alpha_1 p_4^{(2)}}{\hbar})\exp(\frac{i(\mathbf{p}_i^{(1)} - \mathbf{p}_f^{(1)}) \cdot \mathbf{x}}{\hbar})$$

(124)

And we define the **transition probability per unit time** as

$$W = \lim_{T\to\infty}\frac{|a(t=\infty)|^2}{T} \quad . \tag{125}$$

Then we have

$$\begin{aligned}
W &= \lim_{T\to\infty}\frac{|a_{if}(t=T/2)|^2}{T} \\
&= \lim_{T\to\infty}\frac{|\Gamma_{if}|^2}{T}(2\pi\hbar)^2\delta(E_f - E_i)^2 \\
&= \lim_{T\to\infty}\frac{|\Gamma_{if}|^2}{T}(2\pi\hbar)\delta(E_f - E_i)\int_{-T/2}^{T/2}\exp(\frac{i(E_f - E_i)}{\hbar})dt \\
&= \lim_{T\to\infty}\frac{|\Gamma_{if}|^2}{T}(2\pi\hbar)\delta(E_f - E_i)\int_{-T/2}^{T/2}1\,dt \quad (\text{because}\quad E_f - E_i \approx 0) \\
&= \lim_{T\to\infty}\frac{|\Gamma_{if}|^2}{T}(2\pi\hbar)\delta(E_f - E_i)T \\
&= 2\pi\hbar|\Gamma_{if}|^2\delta(E_f - E_i)
\end{aligned} \tag{126}$$

In practice physics, we usually deal with situations where we start with a specified initial state and end up in one of a set of final states. Let $\rho(E)$ be the density of final states; that is, $\rho(E)dE$ is the number of states in the energy interval $E$ to $E + dE$. We integrate over this density, we obtain the transition rate

$$\begin{aligned}
W_{if} &= 2\pi\hbar\int|\Gamma_{if}|^2\delta(E_f - E_i)\rho(E)dE_f \\
&= 2\pi\hbar|\Gamma_{if}|^2\rho(E_i)
\end{aligned} \tag{127}$$

This is the Fermi's Golden Rule.

The incident particle 1 after the collision changes its state from A to C, as shown in Fig.3. If the incident particle current density is $j$, then the particle density is $n=j/v_A$. the number of the particles outgoing into an element of solid angle $d\Omega$ about $\mathbf{p}_C$ per unit time is given by $dN=jD\,d\Omega$, where the $D$ is defined as the **differential cross section**, which is a quantity easily measured in the laboratory: the detector accepts particles scattering into a solid angle $d\Omega$, we simply count the number recorded per unit time, divided by $d\Omega$, and normalize to the current density of the incident particles.

$$D = \frac{dN}{jd\Omega} . \tag{128}$$

At the same time we know that due to the Fermi's Golden Rule, the scattered number is given by

$$dN = nV(2\pi\hbar)|\Gamma_{AC}|^2 \rho(E_C) . \tag{129}$$

Thus the differential cross section is given by

$$D = \frac{dN}{jd\Omega} = \frac{nV(2\pi\hbar)|\Gamma_{AC}|^2 \rho(E_C)}{jd\Omega} = \frac{V(2\pi\hbar)|\Gamma_{AC}|^2 \rho(E_C)}{v_A d\Omega} . \tag{130}$$

The density of final states is determined by box model: one particle in a box of volume $V$ has

$$\rho(E_C)dE_C = \frac{Vd^3 p_C}{(2\pi\hbar)^3} = \frac{V|\mathbf{p}_C|^2 d|\mathbf{p}_C|d\Omega}{(2\pi\hbar)^3} . \tag{131}$$

Using the relation $dE_C = |\mathbf{v}_C| d|\mathbf{p}_C|$, we get

$$\rho(E_C) = \frac{V|\mathbf{p}_C|^2 d\Omega}{(2\pi\hbar)^3 |\mathbf{v}_C|} . \tag{132}$$

Thus we obtain the differential cross section

$$D = \frac{V^2 |\mathbf{p}_C|^2}{(2\pi\hbar)^2 (|\mathbf{v}_A||\mathbf{v}_C|)} |\Gamma_{AC}|^2 = \frac{V^2 m^{(1)2} |\mathbf{p}_C|}{(2\pi\hbar)^2 |\mathbf{p}_A|} |\Gamma_{AC}|^2 . \tag{133}$$

6. **Born approximation scattering formula under SU(2) symmetry**

Suppose potential V(r) is localized about $x_0=0$. The incident particle 1 is electron; the target particle 2 is approximately at rest, as shown in Fig.4.

$$\begin{aligned}\alpha_1 &= \frac{1}{4\pi\varepsilon_0} \frac{q^{(1)}q^{(2)}}{rm^{(2)}c^2} = \frac{1}{m^{(2)}c^2} V(r) \\ A_\mu &= \frac{1}{4\pi\varepsilon_0} \frac{q^{(2)} u_\mu^{(2)}}{c^2 r} = \frac{\alpha_1 p_\mu^{(2)}}{q^{(1)}}\end{aligned} . \tag{134}$$

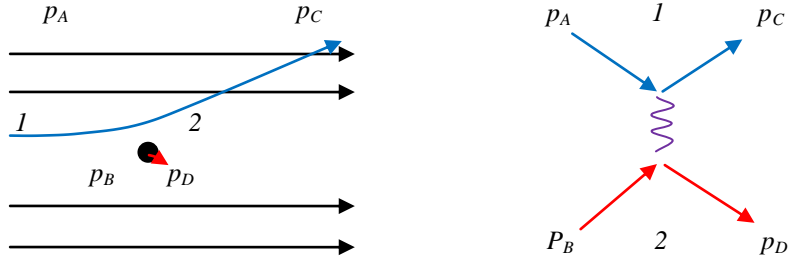

Fig.4 Particle 1 scatters off particle 2 at rest

Defining

$$q = q_A - q_C; \quad \mathbf{q} = \mathbf{q}_A - \mathbf{q}_C \ . \tag{135}$$

Ignoring the magnetic effect of $A_\mu$, the transition amplitude from A to C is given by

$$\begin{aligned}
\Gamma_{AC} &= \frac{1}{V}\int dx^3 (\frac{-c\alpha_1 p_4^{(2)}}{\hbar}) \exp(\frac{i(\mathbf{p}_A^{(1)} - \mathbf{p}_C^{(1)})\cdot \mathbf{x}}{\hbar}) \\
&= \frac{1}{V}\int dx^3 (-\frac{iV(r)}{\hbar}) \exp\left(\frac{i\mathbf{q}\cdot\mathbf{x}}{\hbar}\right) \\
&= \frac{-i}{V\hbar}\int d^3 x V(r) \exp\left(\frac{i\mathbf{q}\cdot\mathbf{x}}{\hbar}\right)
\end{aligned} \tag{136}$$

Then from it, the differential cross section is given by

$$\frac{d\sigma}{d\Omega} = D = \frac{V^2 m^{(1)2}|\mathbf{p}_C|}{(2\pi\hbar)^2 |\mathbf{p}_A|} \left[\frac{-i}{V\hbar}\int d^3 x V(r)\exp\left(\frac{i\mathbf{q}\cdot\mathbf{x}}{\hbar}\right)\right]^2 \ . \tag{137}$$

For elastic collision, we know $|\mathbf{p}_A| = |\mathbf{p}_C|$, thus

$$\frac{d\sigma}{d\Omega} = D = \frac{m^{(1)2}}{(2\pi\hbar)^2 \hbar^2}\left[\int d^3 x V(r)\exp\left(\frac{i\mathbf{q}\cdot\mathbf{x}}{\hbar}\right)\right]^2 \ . \tag{138}$$

This is the Born approximation scattering formula.

It is important to point out that the scattering formula is hold for the condition that the first Pauli matrix works and other Pauli matrices idle:

$$\begin{bmatrix} R^{(1)} \\ R^{(2)} \end{bmatrix} = (1 + \alpha_1 \sigma_1 + 0\cdot \sigma_2 + 0\cdot \sigma_3) \begin{bmatrix} p^{(1)} \\ p^{(2)} \end{bmatrix} \ . \tag{139}$$

If the second and third Pauli matrices work, the collision becomes a complicated process fully engaged with the SU(2) group symmetry. Thinking about mesons, leptons, quarks, spin, Pauli exclusion, SU(N) structures, etc, we expect the engagement to expose further.

## 7. Rutherford scattering formula under SU(2) symmetry

As shown in Fig.5, the differential cross section is given by

$$\frac{d\sigma}{d\Omega} = D = \frac{m^{(1)2}}{(2\pi\hbar)^2 \hbar^2} \left[ \int d^3 x V(r) \exp\left(\frac{i\mathbf{q}\cdot\mathbf{x}}{\hbar}\right) \right]^2 . \tag{140}$$

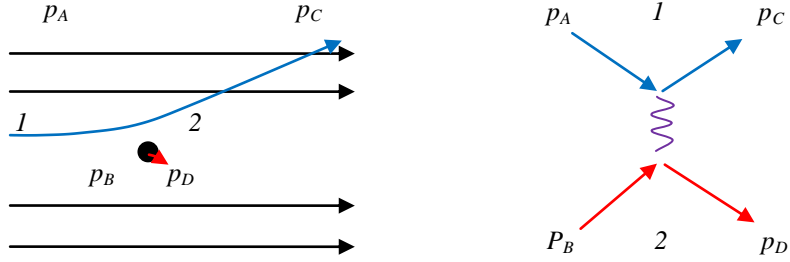

Fig.5 Particle 1 scatters off particle 2 at rest.

Applying the following mathematical formula

$$\int (\nabla^2 F) \exp\left(\frac{i\mathbf{q}\cdot\mathbf{x}}{\hbar}\right) d^3 x = -\frac{|\mathbf{q}|^2}{\hbar^2} \int F \exp\left(\frac{i\mathbf{q}\cdot\mathbf{x}}{\hbar}\right) d^3 x . \tag{141}$$

to the differential cross section, we obtain

$$\frac{d\sigma}{d\Omega} = \frac{m^{(1)2}}{(2\pi\hbar)^2 \hbar^2} \left[ \frac{\hbar^2}{|\mathbf{q}|^2} \int d^3 x \nabla^2 V(r) \exp\left(\frac{i\mathbf{q}\cdot\mathbf{x}}{\hbar}\right) \right]^2 . \tag{142}$$

For the Coulomb potential, we know

$$\nabla^2 V(r) = \frac{Z^{(1)} Z^{(2)} e^2}{4\pi\varepsilon_0} [4\pi\delta(\mathbf{r})] = \frac{Z^{(1)} Z^{(2)} e^2}{\varepsilon_0} \delta(\mathbf{r}) . \tag{143}$$

Substituting it into the differential cross section, we get

$$\frac{d\sigma}{d\Omega} = \frac{Z^{(1)2} Z^{(2)2} e^4 m^{(1)2}}{\varepsilon_0^2 (2\pi)^2} \frac{1}{|\mathbf{q}|^4} . \tag{144}$$

For the elastic collision, the transfer momentum is given by

$$|\mathbf{q}| = 2|\mathbf{p}_A| \sin(\theta/2) . \tag{145}$$

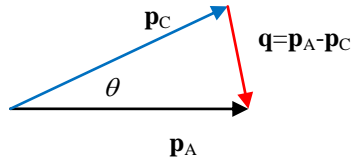

Fig.6 definition of the transfer momentum

As shown in Fig.6, the $\theta$ is the deflection angle of the particle 1, therefore, we obtain

$$\frac{d\sigma}{d\Omega} = \frac{Z^{(1)2}Z^{(2)2}e^4 m^{(1)2}}{16\varepsilon_0^2 (2\pi)^2 |\mathbf{p}_A|^4 \sin^4(\theta/2)} \quad . \tag{146}$$

This is the Rutherford scattering formula.

## 8. Mott scattering formula under SU(2) symmetry

As shown in Fig.7, the differential cross section is given by

$$\frac{d\sigma}{d\Omega} = D = \frac{m^{(1)2} |\mathbf{p}_C|}{(2\pi\hbar)^2 |\mathbf{p}_A|} \left[ \int d^3 x V(r) \exp\left(\frac{i\mathbf{q}\cdot\mathbf{x}}{\hbar}\right) \right]^2 \quad . \tag{147}$$

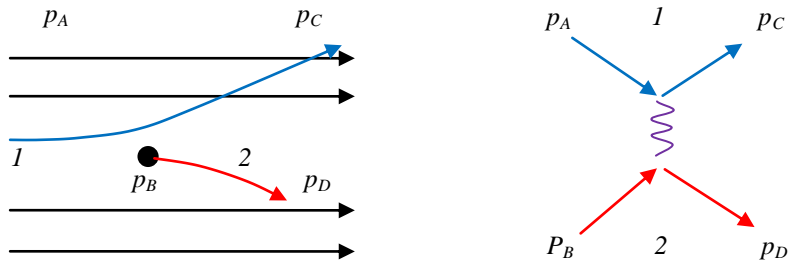

Fig.7 Particle 1 scatters off particle 2 with SU(2) symmetry.

Applying the following mathematical formula

$$\int (\nabla^2 F) \exp(\frac{i\mathbf{q}\cdot\mathbf{x}}{\hbar}) d^3 x = -\frac{|\mathbf{q}|^2}{\hbar^2} \int F \exp(\frac{i\mathbf{q}\cdot\mathbf{x}}{\hbar}) d^3 x \quad . \tag{148}$$

to the differential cross section, we obtain

$$\frac{d\sigma}{d\Omega} = \frac{m^{(1)2}|\mathbf{p}_C|}{(2\pi\hbar)^2|\mathbf{p}_A|}\left[\frac{\hbar^2}{|\mathbf{q}|^2}\int d^3x \nabla^2 V(r)\exp\left(\frac{i\mathbf{q}\cdot\mathbf{x}}{\hbar}\right)\right]^2 . \quad (149)$$

For the Coulomb potential, we know

$$\nabla^2 V(r) = \frac{Z^{(1)}Z^{(2)}e^2}{4\pi\varepsilon_0}[-4\pi\delta(\mathbf{r})] = -\frac{Z^{(1)}Z^{(2)}e^2}{\varepsilon_0}\delta(\mathbf{r}) . \quad (150)$$

Substituting it into the differential cross section, we get

$$\frac{d\sigma}{d\Omega} = \frac{Z^{(1)2}Z^{(2)2}e^4 m^{(1)2}}{\varepsilon_0^2(2\pi)^2}\frac{|\mathbf{p}_C|}{|\mathbf{p}_A|}\frac{1}{|\mathbf{q}|^4} . \quad (151)$$

For inelastic collision, we know $|\mathbf{p}_A| > |\mathbf{p}_C|$, the transfer momentum is given by

$$|\mathbf{q}|^2 \simeq 4|\mathbf{p}_A||\mathbf{p}_C|\sin^2(\theta/2) . \quad (152)$$

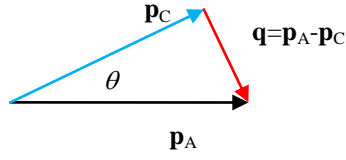

Fig.8 definition of the transfer momentum

As shown in Fig.8, the $\theta$ is the deflection angle of the particle 1. In relativistic limit, we consider $\gamma = |\mathbf{p}|/E$ as a constant, because in $p^2 = \mathbf{p}^2 - E^2/c^2 = -m^2c^2$, if the momentum is large enough we can neglect the particle mass, then $\gamma = |\mathbf{p}|/E \simeq 1/c$.

$$\begin{aligned}
q^2 &= (p_A - p_C)^2 \simeq |\mathbf{q}|^2 = 4|\mathbf{p}_A|\cdot|\mathbf{p}_C|\sin^2(\theta/2) = 4\gamma^2 E_A E_C \sin^2(\theta/2) \\
(q + p_B)^2 &= p_D^2 \Rightarrow q^2 - 2m^{(2)}(E_A - E_C) + p_B^2 = p_D^2
\end{aligned}, \quad (153)$$

where $p_B^2 = p_D^2 = -m^{(2)2}c^2$, $p_B = (0,0,0,m^{(2)}ic)$, we work out

$$E_C = \frac{E_A}{1 + 2\dfrac{E_A\gamma^2}{m^{(2)}}\sin^2(\theta/2)} . \quad (154)$$

The differential cross section is

$$\frac{d\sigma}{d\Omega} = \frac{Z^{(1)2}Z^{(2)2}e^4 m^{(1)2}}{\varepsilon_0^2 (2\pi)^2} \frac{|\mathbf{p}_C|}{|\mathbf{p}_A|} \frac{1}{|\mathbf{q}|^4} = \frac{Z^{(1)2}Z^{(2)2}e^4 m^{(1)2}}{\varepsilon_0^2 (2\pi)^2} \frac{E_C}{E_A} \frac{1}{|\mathbf{q}|^4}$$
$$= \frac{Z^{(1)2}Z^{(2)2}e^4 m^{(1)2}}{\varepsilon_0^2 (2\pi)^2} \frac{1}{|\mathbf{q}|^4} \left(1 - 2\frac{E_A \gamma^2}{m^{(2)}} \sin^2(\theta/2)\right)$$
(155)

Or, we rewrite it as

$$\frac{d\sigma}{d\Omega} = \left(\frac{d\sigma}{d\Omega}\right)_{Rutherford} \cdot \left(1 - 2\frac{E_A \gamma^2}{m^{(2)}} \sin^2(\theta/2)\right) .$$
(156)

This is the Mott scattering formula for inelastic collision.

## 9. Rosenbluth scattering formula under SU(2) symmetry

The interaction between incident particle and target particle is given by

$$\alpha_1 = \frac{1}{4\pi\varepsilon_0} \frac{q^{(1)} q^{(2)}}{rm^{(2)} c^2} = \frac{1}{m^{(2)} c^2} V(r)$$
$$A_\mu = \frac{1}{4\pi\varepsilon_0} \frac{q^{(2)} u_\mu^{(2)}}{c^2 r} = \frac{\alpha_1 p_\mu^{(2)}}{q^{(1)}}$$
(157)

In the Mott model and Rutherford model, we have calculated the Coulomb potential ($A_4$), but neglected the magnetic components ($A_1$, $A_2$, $A_3$) coming from the target particle which has a recoil during collision. In this section we discuss a model which contains the first three components of the 4-vector potential $A$.

Remember that we have proposed an assertion: the relativistic de Broglie matter wave is a high resolution wave with respect to old traditional quantum mechanics. In this section we will demonstrate what the assertion means by deriving the Rosenbluth scattering formula, which requires a higher precision than the Born approximation.

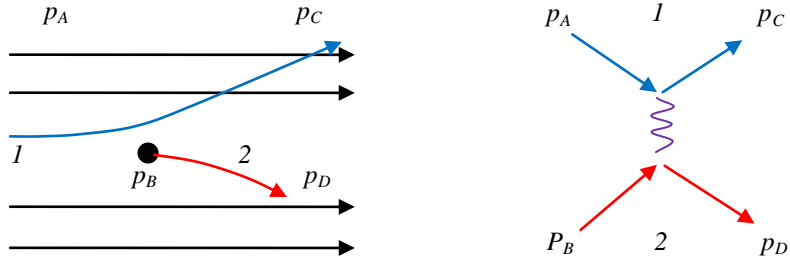

Fig.9 Particle 1 scatters off particle 2 with recoil.

As shown in Fig.9, let us begin with its origins. In high energy collision for two fermions 1 and 2, the N=2 system has the riding-wave momenta given by

$$\begin{bmatrix} R^{(1)} \\ R^{(2)} \end{bmatrix} = (1+\alpha_1\sigma_1)\begin{bmatrix} p^{(1)} \\ p^{(2)} \end{bmatrix} = (1+\alpha_1\begin{bmatrix} 0 & 1 \\ 1 & 0 \end{bmatrix})\begin{bmatrix} p^{(1)} \\ p^{(2)} \end{bmatrix} \quad . \tag{158}$$

The single incident particle 1 is governed by

$$-i\hbar\frac{\partial \phi^{(1)}}{\partial x_\mu} = R_\mu^{(1)}\phi^{(1)} = (p_\mu^{(1)} + \alpha_1 p_\mu^{(2)})\phi^{(1)} \quad . \tag{159}$$

Considering a small volume element *dV* in where there are N particles 1 at a time, their average quantities (labeled *AVE*) satisfy

$$-i\hbar\frac{\partial \phi^{(1\_AVE)}}{\partial x_\mu} = (p_\mu^{(1\_AVE)} + \alpha_1 p_\mu^{(2)})\phi^{(1\_AVE)};$$

$$\phi^{(1\_AVE)} = \sum_{j=1}^{N}\phi^{(1\_j)} / N; \tag{160}$$

$$p_\mu^{(1\_AVE)} = \sum_{j=1}^{N} p_\mu^{(1\_j)}\phi^{(1\_j)} / \sum_{j=1}^{N}\phi^{(1\_j)}$$

Suppose that the incident particles 1 have the same initial state *i* with the momentum $p_i$, since in high energy collision 99% incident particles go almost straightforward by the target particle, only about 1% incident particles would be apparently scattered by the target particle, their average quantities well approximately satisfy

$$p_\mu^{(1\_AVE)} = 0.99 p_\mu^{(1\_i)} + 0.01 p_\mu^{(1\_other)} \simeq p_\mu^{(1\_i)}$$

$$-i\hbar\frac{\partial \phi^{(1\_AVE)}}{\partial x_\mu} = (p_\mu^{(1\_i)} + \alpha_1 p_\mu^{(2)})\phi^{(1\_AVE)} \tag{161}$$

The fewer scattered partition, the more accurate expectation. The wave function of the incident particles are given by

$$\phi^{(1\_AVE)} = \frac{1}{\sqrt{V}}\exp\left(\frac{i}{\hbar}\int_{x_0}^{x}R_\mu^{(1)}dx_\mu\right)$$

$$= \frac{1}{\sqrt{V}}\exp\left(\frac{i}{\hbar}p_\mu^{(1\_i)}\cdot x + \frac{i}{\hbar}\int_{x_0}^{x}(\alpha_1 p_\mu^{(2)})dx_\mu\right) \quad . \tag{162}$$

Where, the wave function has normalized to one particle in the Volume *V*. The integral paths take on any mathematical paths in their 4-vector velocity fields (integral path independency). It is interesting to look at the following integral.

$$\frac{i}{\hbar}\int_{x_0}^{x}(\alpha_1 p_\mu^{(2)})dx_\mu = \frac{i}{\hbar}\int_{t_0}^{t}(\alpha_1 p_4^{(2)})dx_4 + \frac{i}{\hbar}\int_{\mathbf{x}_0}^{\mathbf{x}}(\alpha_1 \mathbf{p}^{(2)})\cdot d\mathbf{x} \quad . \tag{163}$$

The integral path takes on the z-axis in which the particle 1 comes, as shown in Fig.9. We have

$$\frac{i}{\hbar}\int_{x_0}^{x}(\alpha_1 p_\mu^{(2)})dx_\mu = \frac{i}{\hbar}\int_{t_0}^{t}(\alpha_1 p_4^{(2)})dx_4 + \frac{i}{\hbar}\int_{\mathbf{x}_0}^{\mathbf{x}}(\alpha_1\mathbf{p}^{(2)})\cdot d\mathbf{z}$$
$$= \frac{i}{\hbar}\int_{t_0}^{t}(\alpha_1 p_4^{(2)})dx_4 + \frac{i}{\hbar}\int_{\mathbf{x}_0}^{\mathbf{x}}\alpha_1 p_z^{(2)}dz$$
$$= \frac{i}{\hbar}\int_{t_0}^{t}(\alpha_1 p_4^{(2)})dx_4 + \frac{i}{\hbar}\int_{\tau_0}^{\tau}\alpha_1 p_z^{(2)}\frac{m^{(2)}dz}{m^{(2)}d\tau^{(2)}}d\tau^{(2)} \qquad (164)$$
$$= \frac{i}{\hbar}\int_{t_0}^{t}(\alpha_1 p_4^{(2)})dx_4 + \frac{i}{\hbar}\int_{z_0}^{\tau}\alpha_1 p_z^{(2)}\frac{p_z^{(2)}}{m^{(2)}}d\tau^{(2)}$$

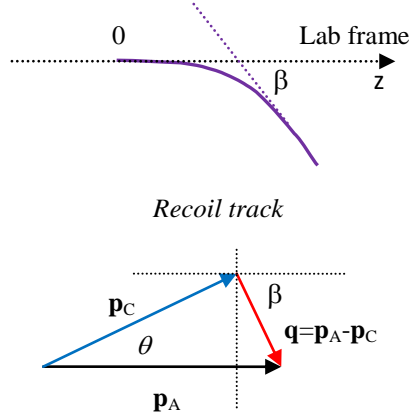

Fig.10 Recoil and momentum transfer.

In fact, we know that the target particle $p_z^{(2)}$ increases from zero to $q_z = |\mathbf{q}|\cos\beta$ in the lab frame during the collision, as shown in Fig.10, we have to take its average for the calculation, we have

$$\frac{i}{\hbar}\int_{x_0}^{x}(\alpha_1 p_\mu^{(2)})dx_\mu = \frac{i}{\hbar}\int_{t_0}^{t}(\alpha_1 p_4^{(2)})dx_4 + \frac{i}{\hbar}\int_{z_0}^{\tau}\alpha_1 p_z^{(2)}\frac{p_z^{(2)}}{m^{(2)}}d\tau^{(2)}$$
$$= \frac{i}{\hbar}\int_{t_0}^{t}(\alpha_1 p_4^{(2)})dx_4 + \frac{i}{\hbar}\int_{\tau_0}^{\tau}\frac{\alpha_1}{m^{(2)}}(\frac{|\mathbf{q}|\cos\beta+0}{2})^2 d\tau^{(2)} \qquad (165)$$

We know that $\cos\beta = \sin(\theta/2)$, $d\tau^{(2)} = icm^{(2)}dt/p_4^{(2)}$, thus we get

$$\frac{i}{\hbar}\int_{x_0}^{x}(\alpha_1 p_\mu^{(2)})dx_\mu = \frac{i}{\hbar}\int_{t_0}^{t}(\alpha_1 p_4^{(2)})dx_4 + \frac{i}{\hbar}\int_{t_0}^{t}\frac{\alpha_1}{m^{(2)}}\frac{|\mathbf{q}|^2\cos^2(\theta/2)}{4}\frac{icm^{(2)}dt}{p_4^{(2)}}$$
$$= \frac{i}{\hbar}\int_{t_0}^{t}(\alpha_1 p_4^{(2)})dx_4 + \frac{i}{\hbar}\int_{t_0}^{t}\alpha_1\frac{|\mathbf{q}|^2\cos^2(\theta/2)}{4 p_4^{(2)}}dx_4 \qquad (166)$$

Now the wave function of the incident particles are given by

$$\phi^{(1\_AVE)} = \frac{1}{\sqrt{V}}\exp\left(\frac{i}{\hbar}\int_{x_0}^{x} R_\mu^{(1)}dx_\mu\right)$$
$$= \frac{1}{\sqrt{V}}\exp\left(\frac{i}{\hbar}p_\mu^{(i)}\cdot x + \frac{i}{\hbar}\int_{t_0}^{t}(\alpha_1 p_4^{(2)})dx_4 + \frac{i}{\hbar}\int_{t_0}^{t}\alpha_1\frac{|\mathbf{q}|^2\cos^2(\theta/2)}{4 p_4^{(2)}}dx_4\right) \qquad (167)$$

Actually, we have moved the magnetic components ($A_1$, $A_2$, $A_3$) coming from the target particle to the time-axis effect, thus the wave function of the incident particles satisfy

$$-i\hbar \frac{\partial \phi^{(1\_AVE)}}{\partial x_k} = (p_k^{(1\_i)} + 0)\phi^{(1\_AVE)}; \quad k=1,2,3$$
$$-i\hbar \frac{\partial \phi^{(1\_AVE)}}{\partial x_4} = (p_4^{(1\_i)} + \alpha_1 p_4^{(2)} + \alpha_1 \frac{|\mathbf{q}|^2 \cos^2(\theta/2)}{4p_4^{(2)}})\phi^{(1\_AVE)};$$
(168)

The wave function can also be expressed in terms of its free plane waves as (the Fourier expansion)

$$\phi^{(1\_AVE)} = \sum_{n=1}^{\infty} a_n(t) \exp(\frac{ip_n^{(1)} \cdot x}{\hbar}) .$$
(169)

And we know in the space-time ($x_1,x_2,x_3,x_4$=-ct), the fourth component expresses its evolution. For our purpose, substituting it into the left side of the fourth component equation, we get

$$-i\hbar \sum_{n=1}^{\infty} (\frac{\partial a_n}{\partial x_4} + \frac{a_n ip_{n4}^{(1)}}{\hbar}) \exp(\frac{ip_n^{(1)} \cdot x}{\hbar})$$
$$= \frac{1}{\sqrt{V}} (p_{i4}^{(1)} + \alpha_1 p_4^{(2)} + \alpha_1 \frac{|\mathbf{q}|^2 \cos^2(\theta/2)}{4p_4^{(2)}}) \exp\left(\frac{i}{\hbar} p_\mu^{(i)} \cdot x\right)$$
(170)

Multiplying by another plane wave eigen state $\phi_f^*$, integrating over the whole volume $V$, and using the orthonorrmality relation of eigen states, using the initial condition $a_{ni}=\delta_{ni}$, it leads to

$$\frac{\partial a_f}{\partial x_4} = \frac{i}{V\hbar} \int dx^3 (\alpha_1 p_4^{(2)} + \alpha_1 \frac{|\mathbf{q}|^2 \cos^2(\theta/2)}{4p_4^{(2)}}) \exp\left(\frac{i(p_i^{(1)} - p_f^{(1)}) \cdot x}{\hbar}\right) .$$
(171)

We have

$$a_n(t=-\frac{T}{2}) = \delta_{ni} .$$
(172)

At a later time $t$, we have

$$a_f(t) = \frac{i}{V\hbar} \int_{-T/2}^{t} dx_4 \int dx^3 (\alpha_1 p_4^{(2)} + \alpha_1 \frac{|\mathbf{q}|^2 \cos^2(\theta/2)}{4p_4^{(2)}}) \exp\left(\frac{i(p_i^{(1)} - p_f^{(1)}) \cdot x}{\hbar}\right) .$$
(173)

At a final departure time t=T/2=∞, we have

$$a_f(t=\frac{T}{2}) = \frac{i}{V\hbar} \int dx^4 (\alpha_1 p_4^{(2)} + \alpha_1 \frac{|\mathbf{q}|^2 \cos^2(\theta/2)}{4p_4^{(2)}}) \exp\left(\frac{i(p_i^{(1)} - p_f^{(1)}) \cdot x}{\hbar}\right) . \quad (174)$$

We define the $a(t=\infty)$ as the **transition amplitude** that the particle has scattered from an initial state $i$ to a final state $f$, whose formula divides into two parts as

$$a_f(t=\frac{T}{2}) = \int dx_4 \exp\left(\frac{i(p_i^{(1)} - p_f^{(1)}) \cdot x_4}{\hbar}\right)$$
$$\cdot \frac{i}{V\hbar} \int dx^3 (\alpha_1 p_4^{(2)} + \alpha_1 \frac{|\mathbf{q}|^2 \cos^2(\theta/2)}{4p_4^{(2)}}) \exp\left(\frac{i(\mathbf{p}_i^{(1)} - \mathbf{p}_f^{(1)}) \cdot \mathbf{x}}{\hbar}\right)$$
(175)

We have

$$a_f(t=\frac{T}{2}) = (2\pi\hbar)\delta(E_f^{(1)} - E_i^{(1)} + w)\Gamma_{if}$$

(176)

$$\Gamma_{if} \equiv \frac{i}{V\hbar}\int dx^3(\alpha_1 p_4^{(2)} + \alpha_1 \frac{|\mathbf{q}|^2 \cos^2(\theta/2)}{4p_4^{(2)}})\exp\left(\frac{i(\mathbf{p}_i^{(1)} - \mathbf{p}_f^{(1)})\cdot\mathbf{x}}{\hbar}\right)$$

In practice physics, we usually deal with the transition rate

$$W_{if} = 2\pi\hbar\int |\Gamma_{if}|^2 \delta(E_f - E_i)\rho(E)dE_f$$
$$= 2\pi\hbar|\Gamma_{if}|^2 \rho(E_i)$$

(177)

This is the Fermi's Golden Rule. The incident particle 1 after the collision changes its state from A to C, as shown in Fig.9. Due to the Fermi's Golden Rule, we obtain the differential cross section

$$D = \frac{V^2|\mathbf{p}_C|^2}{(2\pi\hbar)^2(|\mathbf{v}_A||\mathbf{v}_C|)}|\Gamma_{AC}|^2 = \frac{V^2 m^{(1)2}|\mathbf{p}_C|}{(2\pi\hbar)^2|\mathbf{p}_A|}|\Gamma_{AC}|^2 \ .$$

(178)

Applying the following formula

$$\int(\nabla^2 F)\exp(\frac{i\mathbf{q}\cdot\mathbf{x}}{\hbar})d^3x = -\frac{|\mathbf{q}|^2}{\hbar^2}\int F\exp(\frac{i\mathbf{q}\cdot\mathbf{x}}{\hbar})d^3x \ .$$

(179)

to the transition amplitude, we obtain

$$\Gamma_{AC} = \frac{-i\hbar}{V|\mathbf{q}|^2}\int d^3x \nabla^2(\alpha_1 p_4^{(2)} + \alpha_1\frac{|\mathbf{q}|^2 \sin^2(\theta/2)}{4p_4^{(2)}})\exp\left(\frac{i\mathbf{q}\cdot\mathbf{x}}{\hbar}\right)$$

(180)

for the Coulomb electric potential, assume that $p_4^{(2)}$ is smooth function, we have

$$\Gamma_{AC} = -\frac{i\hbar}{V|\mathbf{q}|^2}\int d^3x \nabla^2(\alpha_1)(p_4^{(2)} + \frac{|\mathbf{q}|^2 \sin^2(\theta/2)}{4p_4^{(2)}})\exp\left(\frac{i\mathbf{q}\cdot\mathbf{x}}{\hbar}\right)$$
$$= \frac{i\hbar}{V|\mathbf{q}|^2}\int d^3x \frac{Z^{(1)}Z^{(2)}e^2}{\varepsilon_0 m^{(2)}c^2}\delta(\mathbf{r})(p_4^{(2)} + \frac{|\mathbf{q}|^2 \sin^2(\theta/2)}{4p_4^{(2)}})\exp\left(\frac{i\mathbf{q}\cdot\mathbf{x}}{\hbar}\right)$$
$$= \frac{i\hbar}{V|\mathbf{q}|^2}\frac{Z^{(1)}Z^{(2)}e^2}{\varepsilon_0 m^{(2)}c^2}(p_4^{(2)} + \frac{|\mathbf{q}|^2 \sin^2(\theta/2)}{4p_4^{(2)}})$$

(181)

We take $p_4^{(2)} \simeq icm^{(2)}, |\mathbf{q}|^2 \simeq -q^2$, the differential cross section is given by

$$\frac{d\sigma}{d\Omega} = D = \frac{V^2 m^{(1)2} |\mathbf{p}_C|}{(2\pi\hbar)^2 |\mathbf{p}_A|} |\Gamma_{AC}|^2$$

$$= \frac{V^2 m^{(1)2} |\mathbf{p}_C|}{(2\pi\hbar)^2 |\mathbf{p}_A|} \left| \frac{i\hbar}{V|\mathbf{q}|^2} \frac{Z^{(1)}Z^{(2)}e^2}{\varepsilon_0 m^{(2)} c^2} [p_4^{(2)} - \frac{q^2 \sin^2(\theta/2)}{4 p_4^{(2)}}] \right|^2 \quad (182)$$

$$= \frac{Z^{(1)2} Z^{(2)2} e^4 m^{(1)2}}{\varepsilon_0^2 (2\pi)^2} \frac{|\mathbf{p}_C|}{|\mathbf{p}_A| |\mathbf{q}|^4} [1 - \frac{q^2 \sin^2(\theta/2)}{4 m^{(2)2} c^2}]^2$$

$$= \left(\frac{d\sigma}{d\Omega}\right)_{Mott} [1 - \frac{q^2}{4 m^{(2)2} c^2} \sin^2(\theta/2)]^2$$

In the lowest order, we have

$$\frac{d\sigma}{d\Omega} = \left(\frac{d\sigma}{d\Omega}\right)_{Mott} [1 - \frac{q^2}{2 m^{(2)2} c^2} \sin^2(\theta/2)] \quad (183)$$

This is the Rosenbluth scattering formula.

Some fundamental problems still remain there in this rough model.
(1) we have used several kinematic approximations in the derivation, some modification may be suggested to modifying the primitive Rosenbluth scattering formula as follows

$$\frac{d\sigma}{d\Omega} = \left(\frac{d\sigma}{d\Omega}\right)_{Mott} [F_1(\theta, E) - \frac{q^2}{2 m^{(2)} c^2} F_2(\theta, E) \sin^2(\theta/2)] \quad (184)$$

Where $F_1$ and $F_2$ are the form factors, as we know, the form factors work very well in experiments of high energy particle physics. The first form factor originates from electric interaction; the second form factor originates from magnetic interaction.
(2) in our line, we doubt the temporary theory that the form factors represent charge distribution of target particle, or there are partons and gluons in the proton, we think that some physicists had misread the form factors. In the following chapters, we will discuss the standard model in a new theory: factory of prosthesis (rehab technique).
(3) if we modify the scattering formula as follows

$$\frac{d\sigma}{d\Omega} = \left(\frac{d\sigma}{d\Omega}\right)_{Mott} [F_1(\theta, E) + \frac{q^2}{2 m^{(2)} c^2} F_2(\theta, E) \tan^2(\theta/2)] \quad (185)$$

It is also acceptable, since our model is so primitive that many aspects could be improved.

## 10. Scattering off magnetic moment under SU(2) symmetry

If the target particle has a magnetic moment which has a magnetic field B through a circular area with radius $r_m$, as shown in Fig.11, let 4-vector potential $a_\mu$ denote the magnetic field, then the interaction between incident particle and target particle is given by

$$\alpha_1 = \frac{1}{4\pi\varepsilon_0} \frac{q^{(1)}q^{(2)}}{rm^{(2)}c^2} = \frac{1}{m^{(2)}c^2} V(r)$$

$$A_\mu = \frac{1}{4\pi\varepsilon_0} \frac{q^{(2)} u_\mu^{(2)}}{c^2 r} + a_\mu = \frac{\alpha_1 p_\mu^{(2)}}{q^{(1)}} + a_\mu \tag{186}$$

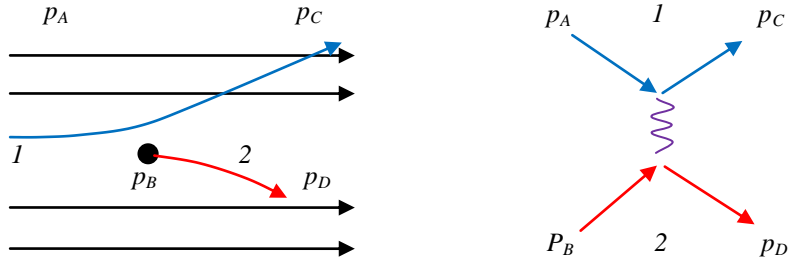

Fig.11 Particle 1 scatters off particle 2 with magnetic flux.

Actually, the magnetic field is regarded as a uniform magnetic field which occupies a cylinder through circular area in diameter $r_m$. In textbook, its 4-vector potential $a_\mu$ is expressed as

$$\mathbf{a} = \frac{1}{2} \mathbf{B} \times \mathbf{r} \cos\varphi \quad (r\cos\varphi < r_m)$$
$$a_4 = 0 \tag{187}$$

Where $\varphi$ is the angle between the track plane and the incident plan to which the magnetic field B is perpendicular, see Fig.12.

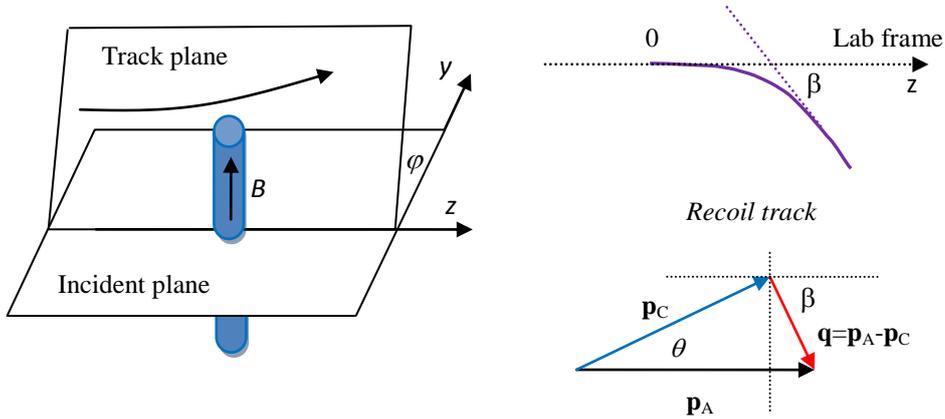

Fig.12 The orientation of the magnetic flux

Suppose that the incident particles 1 have the same initial state $i$ with the momentum $p_i$, the wave function of the incident particles are given by

$$\phi^{(1\_AVE)} = \frac{1}{\sqrt{V}} \exp\left(\frac{i}{\hbar} \int_{x_0}^{x} R_{\mu}^{(1)} dx_{\mu}\right)$$
$$= \frac{1}{\sqrt{V}} \exp\left(\frac{i}{\hbar} p_{\mu}^{(1\_i)} \cdot x + \frac{i}{\hbar} \int_{x_0}^{x} (\alpha_1 p_{\mu}^{(2)}) dx_{\mu} + \frac{i}{\hbar} \int_{x_0}^{x} (q^{(1)}\mathbf{a}) \cdot d\mathbf{x}\right) \quad . \tag{188}$$

Where, the wave function has normalized to one particle in the Volume *V*. The integral paths take on any mathematical paths in their 4-vector velocity fields (integral path independency). The integral path takes on the z-axis in which the particle 1 comes, as shown in Fig.9, the target particle is at rest, we have

$$\frac{i}{\hbar} \int_{x_0}^{x} (\alpha_1 p_{\mu}^{(2)}) dx_{\mu} + \frac{i}{\hbar} \int_{x_0}^{x} (q^{(1)}\mathbf{a}) \cdot d\mathbf{x} = \frac{i}{\hbar} \int_{t_0}^{t} (\alpha_1 p_4^{(2)}) dx_4 + \frac{i}{\hbar} \int_{x_0}^{x} (q^{(1)}\mathbf{a}) \cdot d\mathbf{z} \quad . \tag{189}$$

We need first to calculate the line integral in the above equation, in lab reference frame. Here we only calculate the interaction of the magnetic moment of the target particle, other terms we have calculated in the preceding section in Rosenbluth formula.

$$\frac{i}{\hbar} \int_{x_0}^{x} (q^{(1)}\mathbf{a}) \cdot d\mathbf{x} = \frac{i}{\hbar} \int_{z_0}^{z} q^{(1)} \mathbf{a} \cdot d\mathbf{z}$$
$$= \frac{i}{\hbar} \int_{z_0}^{z} q^{(1)} (\frac{1}{2} \mathbf{B} \times \mathbf{r} \cos\varphi) \cdot d\mathbf{z} \tag{190}$$
$$= \frac{i}{\hbar} \int_{z_0}^{z} q^{(1)} \mathbf{B} \cdot (\frac{1}{2} d\mathbf{z} \times \mathbf{r} \cos\varphi)$$

Where $\frac{1}{2} d\mathbf{z} \times \mathbf{r} \cos\varphi$ is the area element $d\mathbf{s}$ swept out by the radius vector in the incident plane considered, if the incident particle goes outside the magnetic cylinder, as shown in Fig.13, then we have

$$\frac{i}{\hbar} \int_{x_0}^{x} (q^{(1)}\mathbf{a}) \cdot d\mathbf{x} = \frac{i}{\hbar} \int_{z_0}^{z} q^{(1)} \mathbf{B} \cdot d\mathbf{s} = \frac{i}{\hbar} \int_{0}^{\delta} q^{(1)} B \pi r_m^2 d(\frac{\delta}{\pi})$$
$$= \frac{i}{\hbar} \int_{0}^{\delta} q^{(1)} B r_m^2 d\delta = \frac{i}{\hbar} \int_{0}^{\delta} q^{(1)} B r_m^2 \left(\frac{m^{(1)} r^2 d\delta}{d\tau^{(1)}}\right) \frac{d\tau^{(1)}}{m^{(1)} r^2} \tag{191}$$
$$= \frac{i}{\hbar} \int_{0}^{\delta} q^{(1)} B r_m^2 J^{(1)} \frac{d\tau^{(1)}}{m^{(1)} r^2}$$

Where $J^{(1)}$ is the angular momentum of the incident particle which is a constant and expressed in terms of impact parameter *y* as

$$J^{(1)} = y p_A \tag{192}$$

This equals to the initial angular momentum. Then we have

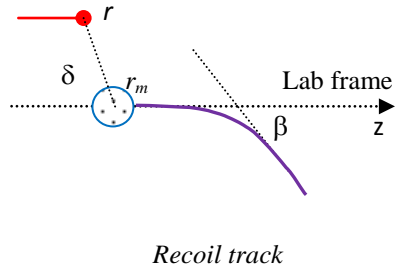

*Recoil track*

Fig.13 Recoil track

$$\frac{i}{\hbar}\int_{x_0}^{x}(q^{(1)}\mathbf{a})\cdot d\mathbf{x}=\frac{i}{\hbar}\int_{0}^{\delta}q^{(1)}Br_m^{\,2}p_A\frac{y}{m^{(1)}r^2}d\tau^{(1)} \qquad (193)$$

We know that $d\tau^{(1)}=icm^{(1)}dt/p_4^{(1)}$, thus we get

$$\frac{i}{\hbar}\int_{x_0}^{x}(q^{(1)}\mathbf{a})\cdot d\mathbf{x}=\frac{i}{\hbar}\int_{0}^{\delta}q^{(1)}Br_m^{\,2}p_A\frac{icy}{p_4^{(1)}r^2}dt=\frac{i}{\hbar}\int_{0}^{\delta}q^{(1)}Br_m^{\,2}p_A\frac{ic}{p_4^{(1)}r}\left(\frac{y}{r}\right)dt \qquad (194)$$

During the collision, the $y/r$ changes from 0 to 1 then back to 0, having an average value of 1/2. Taking this approximation, regarding the incident particle inside the magnetic cylinder as outside the magnetic cylinder, we have

$$\frac{i}{\hbar}\int_{x_0}^{x}(q^{(1)}\mathbf{a})\cdot d\mathbf{x}=\frac{i}{\hbar}\int_{0}^{\delta}\frac{q^{(1)}Br_m^{\,2}p_A}{2p_4^{(1)}}\frac{1}{r}dx_4 \qquad (195)$$

Now the wave function of the incident particles are given by

$$\phi^{(1\_AVE)}=\frac{1}{\sqrt{V}}\exp\left(\frac{i}{\hbar}\int_{x_0}^{x}R_\mu^{(1)}dx_\mu\right)$$
$$=\frac{1}{\sqrt{V}}\exp\left(\frac{i}{\hbar}p_\mu^{(i)}\cdot x+\frac{i}{\hbar}\int_{t_0}^{t}(\alpha_1 p_4^{(2)})dx_4+\frac{i}{\hbar}\int_{0}^{\delta}\frac{q^{(1)}Br_m^{\,2}p_A}{2p_4^{(1)}}\frac{1}{r}dx_4\right) \qquad (196)$$

Actually, we have moved the magnetic components ($A_1$, $A_2$, $A_3$) coming from the target particle to the time-axis effect, thus the wave function of the incident particles satisfy

$$-i\hbar\frac{\partial\phi^{(1\_AVE)}}{\partial x_k}=(p_k^{(1\_i)}+0)\phi^{(1\_AVE)};\quad k=1,2,3$$
$$-i\hbar\frac{\partial\phi^{(1\_AVE)}}{\partial x_4}=(p_4^{(1\_i)}+\alpha_1 p_4^{(2)}+\frac{q^{(1)}Br_m^{\,2}p_A}{2p_4^{(1)}}\frac{1}{r})\phi^{(1\_AVE)}; \qquad (197)$$

The wave function can also be expressed in terms of its free plane waves as (the Fourier expansion )

$$\phi^{(1\_AVG)}=\sum_{n=1}^{\infty}a_n(t)\exp(\frac{ip_n^{(1)}\cdot x}{\hbar}) \quad. \qquad (198)$$

And we know in the space-time $(x_1, x_2, x_3, x_4=-ct)$, the fourth component expresses its evolution. For our purpose, substituting it into the left side of the fourth component, we get

$$-i\hbar \sum_{n=1}^{\infty} \left( \frac{\partial a_n}{\partial x_4} + \frac{a_n i p_{n4}^{(1)}}{\hbar} \right) \exp\left( \frac{i p_n^{(1)} \cdot x}{\hbar} \right)$$
$$= \frac{1}{\sqrt{V}} (p_{i4}^{(1)} + \alpha_1 p_4^{(2)} + \frac{q^{(1)} B r_m^2 p_A}{2 p_4^{(1)}} \frac{1}{r}) \exp\left( \frac{i}{\hbar} p_\mu^{(i)} \cdot x \right) \quad (199)$$

Multiplying by another plane wave eigen state $\phi_f^*$, integrating over the whole volume $V$, and using the orthonorrmality relation of eigen states, using the initial condition $a_{ni}=\delta_{ni}$, it leads to

$$\frac{\partial a_f}{\partial x_4} = \frac{i}{V\hbar} \int dx^3 (\alpha_1 p_4^{(2)} + \frac{q^{(1)} B r_m^2 p_A}{2 p_4^{(1)}} \frac{1}{r}) \exp\left( \frac{i(p_i^{(1)} - p_f^{(1)}) \cdot x}{\hbar} \right) . \quad (200)$$

We have

$$a_n(t = -\frac{T}{2}) = \delta_{ni} . \quad (201)$$

At a later time $t$, we have

$$a_f(t) = \frac{i}{V\hbar} \int_{-T/2}^{t} dx_4 \int dx^3 (\alpha_1 p_4^{(2)} + \frac{q^{(1)} B r_m^2 p_A}{2 p_4^{(1)}} \frac{1}{r}) \exp\left( \frac{i(p_i^{(1)} - p_f^{(1)}) \cdot x}{\hbar} \right) . \quad (202)$$

At a final departure time $t=T/2=\infty$, we have

$$a_f(t = \frac{T}{2}) = \frac{i}{V\hbar} \int dx^4 (\alpha_1 p_4^{(2)} + \frac{q^{(1)} B r_m^2 p_A}{2 p_4^{(1)}} \frac{1}{r}) \exp\left( \frac{i(p_i^{(1)} - p_f^{(1)}) \cdot x}{\hbar} \right) . \quad (203)$$

We define the $a(t=\infty)$ as the **transition amplitude** that the particle has scattered from an initial state $i$ to a final state $f$, whose formula divides into two parts as

$$a_f(t = \frac{T}{2}) = \int dx_4 \exp\left( \frac{i(p_i^{(1)} - p_f^{(1)}) \cdot x_4}{\hbar} \right)$$
$$\cdot \frac{i}{V\hbar} \int dx^3 (\alpha_1 p_4^{(2)} + \frac{q^{(1)} B r_m^2 p_A}{2 p_4^{(1)}} \frac{1}{r}) \exp\left( \frac{i(\mathbf{p}_i^{(1)} - \mathbf{p}_f^{(1)}) \cdot \mathbf{x}}{\hbar} \right) . \quad (204)$$

We have

$$a_f(t = \frac{T}{2}) = (2\pi\hbar)\delta(E_f^{(1)} - E_i^{(1)} + w)\Gamma_{if}$$
$$\Gamma_{if} \equiv \frac{i}{V\hbar} \int dx^3 (\alpha_1 p_4^{(2)} + \frac{q^{(1)} B r_m^2 p_A}{2 p_4^{(1)}} \frac{1}{r}) \exp\left( \frac{i(\mathbf{p}_i^{(1)} - \mathbf{p}_f^{(1)}) \cdot \mathbf{x}}{\hbar} \right) \quad (205)$$

In practice physics, we usually deal with the transition rate

$$W_{if} = 2\pi\hbar \int |\Gamma_{if}|^2 \delta(E_f - E_i) \rho(E) dE_f$$
$$= 2\pi\hbar |\Gamma_{if}|^2 \rho(E_i) \quad (206)$$

This is the Fermi's Golden Rule. The incident particle 1 after the collision changes its state from A to C, as shown in Fig.9. Due to the Fermi's Golden Rule, we obtain the differential cross section

$$D = \frac{V^2 |\mathbf{p}_C|^2}{(2\pi\hbar)^2 (|\mathbf{v}_A||\mathbf{v}_C|)} |\Gamma_{AC}|^2 = \frac{V^2 m^{(1)2} |\mathbf{p}_C|}{(2\pi\hbar)^2 |\mathbf{p}_A|} |\Gamma_{AC}|^2 \quad . \tag{207}$$

The transition amplitude is given by

$$\Gamma_{AC} = \frac{i}{V\hbar} \int d^3x \left[ \alpha_1 p_4^{(2)} + \frac{q^{(1)} Br_m^2 p_A}{2 p_4^{(1)}} \frac{1}{r} \right] \cdot \exp\left( \frac{i(\mathbf{p}_A - \mathbf{p}_C) \cdot \mathbf{x}}{\hbar} \right) \tag{208}$$

Applying the following formula

$$\int (\nabla^2 F) \exp(\frac{i\mathbf{q} \cdot \mathbf{x}}{\hbar}) d^3x = -\frac{|\mathbf{q}|^2}{\hbar^2} \int F \exp(\frac{i\mathbf{q} \cdot \mathbf{x}}{\hbar}) d^3x \quad . \tag{209}$$

to the transition amplitude, for the Coulomb electric potential, assume that $p_4^{(2)}$ is smooth function, we have

$$\begin{aligned}
\Gamma_{AC} &= -\frac{i\hbar}{V|\mathbf{q}|^2} \int d^3x \nabla^2 (\alpha_1 p_4^{(2)} + \frac{q^{(1)} Br_m^2 p_A}{2 p_4^{(1)}} \frac{1}{r}) \exp\left( \frac{i\mathbf{q} \cdot \mathbf{x}}{\hbar} \right) \\
&= \frac{i\hbar}{V|\mathbf{q}|^2} \int d^3x \left( \frac{Z^{(1)} Z^{(2)} e^2}{\varepsilon_0 m^{(2)} c^2} \delta(\mathbf{r}) p_4^{(2)} + \frac{q^{(1)} Br_m^2 p_A}{2 p_4^{(1)}} 4\pi \delta(\mathbf{r}) \right) \exp\left( \frac{i\mathbf{q} \cdot \mathbf{x}}{\hbar} \right) \\
&= \frac{i\hbar}{V|\mathbf{q}|^2} \left( \frac{Z^{(1)} Z^{(2)} e^2}{\varepsilon_0 m^{(2)} c^2} p_4^{(2)} + \frac{q^{(1)} Br_m^2 p_A}{2 p_4^{(1)}} 4\pi \right) \\
&= \frac{i\hbar}{V|\mathbf{q}|^2} \frac{Z^{(1)} Z^{(2)} e^2}{\varepsilon_0 m^{(2)} c^2} p_4^{(2)} \left( 1 + \frac{2\pi q^{(1)} Br_m^2 p_A \varepsilon_0 m^{(2)} c^2}{p_4^{(1)2} Z^{(1)} Z^{(2)} e^2} \right) \\
&= \frac{i\hbar}{V|\mathbf{q}|^2} \frac{Z^{(1)} Z^{(2)} e^2}{\varepsilon_0 m^{(2)} c^2} p_4^{(2)} \left( 1 + \frac{2\pi q^{(1)} Br_m^2 p_A \varepsilon_0 m^{(2)} c^2}{p_4^{(1)2} Z^{(1)} Z^{(2)} e^2} \right)
\end{aligned} \tag{210}$$

We take $p_4^{(2)} \simeq icm^{(2)}$, the differential cross section is given by

$$\begin{aligned}
\frac{d\sigma}{d\Omega} &= D = \frac{V^2 m^{(1)2} |\mathbf{p}_C|}{(2\pi\hbar)^2 |\mathbf{p}_A|} |\Gamma_{AC}|^2 \\
&= \frac{Z^{(1)2} Z^{(2)2} e^4 m^{(1)2}}{\varepsilon_0^2 (2\pi)^2} \frac{|\mathbf{p}_C|}{|\mathbf{p}_A|} \frac{1}{|\mathbf{q}|^4} \left( 1 + \frac{2\pi q^{(1)} Br_m^2 p_A \varepsilon_0 m^{(2)} c^2}{p_4^{(1)2} Z^{(1)} Z^{(2)} e^2} \right)^2 \\
&= \left( \frac{d\sigma}{d\Omega} \right)_{Mott} \left( 1 + \frac{2\pi q^{(1)} Br_m^2 p_A \varepsilon_0 m^{(2)} c^2}{p_4^{(1)2} Z^{(1)} Z^{(2)} e^2} \right)^2
\end{aligned} \tag{211}$$

Where $Br_m^2$ is the magnetic flux in the magnetic cylinder of the target particle. Therefore, the high energy scattering experiments can detect the magnetic flux of target particle rather than its magnetic moment. We do not know how to define anomalous magnetic moment of elementary particle theoretically and experimentally.

## 11. The influence of temperature on scattering

As shown in Fig.14, consider an electron beam bombarding crystal lattice, interaction between incident particle and target particle is given by

$$\alpha_1 = \frac{1}{4\pi\varepsilon_0} \frac{q^{(1)}q^{(2)}}{rm^{(2)}c^2} = \frac{1}{m^{(2)}c^2}V(r)$$

$$A_\mu = \frac{1}{4\pi\varepsilon_0} \frac{q^{(2)}u_\mu^{(2)}}{c^2 r} = \frac{\alpha_1 p_\mu^{(2)}}{q^{(1)}}$$

(212)

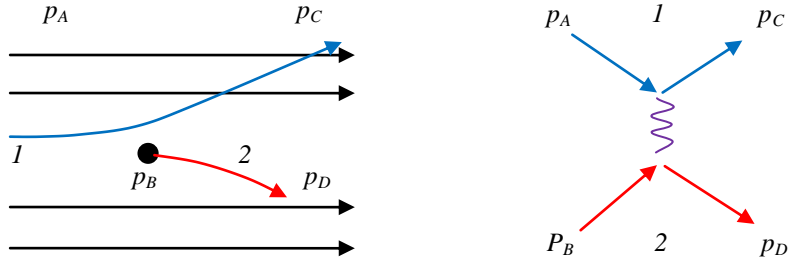

Fig.14 Particle 1 scatters off particle 2 at a temperature.

Suppose that the incident particles 1 have the same initial state $i$ with the momentum $p_i$, the wave function of the incident particles are given by

$$\phi^{(1\_AVE)} = \frac{1}{\sqrt{V}} \exp\left(\frac{i}{\hbar}\int_{x_0}^{x} R_\mu^{(1)} dx_\mu\right)$$

$$= \frac{1}{\sqrt{V}} \exp\left(\frac{i}{\hbar} p_\mu^{(1\_i)} \cdot x + \frac{i}{\hbar}\int_{x_0}^{x}(\alpha_1 p_\mu^{(2)})dx_\mu\right)$$

(213)

Where, the wave function has normalized to one particle in the Volume *V*. The integral paths take on any mathematical paths in their 4-vector velocity fields (integral path independency). The integral path takes on the z-axis in which the particle 1 comes, as shown in Fig.14, the target particle is at its equilibrium position at a temperature *T*, we have

$$\frac{i}{\hbar}\int_{x_0}^{x}(\alpha_1 p_\mu^{(2)})dx_\mu = \frac{i}{\hbar}\int_{t_0}^{t}(\alpha_1 p_4^{(2)})dx_4 + \frac{i}{\hbar}\int_{x_0}^{x}(\alpha_1 m^{(2)}\mathbf{u}^{(2)}) \cdot d\mathbf{z}$$

$$= \frac{i}{\hbar}\int_{t_0}^{t}(\alpha_1 p_4^{(2)})dx_4 + \frac{i}{\hbar}\int_{t_0}^{t}(\alpha_1 m^{(2)} \frac{dz^{(2)}}{d\tau}) \cdot \frac{dz}{dt}dt$$

$$= \frac{i}{\hbar}\int_{t_0}^{t}(\alpha_1 p_4^{(2)})dx_4 + \frac{i}{\hbar}\int_{t_0}^{t}(2\alpha_1)(\frac{1}{2}m^{(2)}v_z^2) \cdot \frac{dt}{d\tau}dt$$

$$= \frac{i}{\hbar}\int_{t_0}^{t}(\alpha_1 p_4^{(2)})dx_4 + \frac{i}{\hbar}\int_{t_0}^{t}(2\alpha_1)(\frac{1}{2}m^{(2)}v_z^2) \cdot \frac{p_4^{(2)}}{m^{(2)}ic}dt$$

(214)

According to the equipartition of kinetic energy theorem for the target particle, we have

$$\frac{i}{\hbar}\int_{x_0}^{x}(\alpha_1 p_\mu^{(2)})dx_\mu = \frac{i}{\hbar}\int_{t_0}^{t}(\alpha_1 p_4^{(2)})dx_4 + \frac{i}{\hbar}\int_{t_0}^{t}(2\alpha_1)(\frac{1}{2}kT)\cdot\frac{p_4^{(2)}}{m^{(2)}ic}dt$$
$$= \frac{i}{\hbar}\int_{t_0}^{t}(\alpha_1 p_4^{(2)})dx_4 - \frac{i}{\hbar}\int_{t_0}^{t}(\alpha_1 p_4^{(2)})\cdot\frac{kT}{m^{(2)}c^2}dx_4 \qquad (215)$$
$$= \frac{i}{\hbar}(1-\frac{kT}{m^{(2)}c^2})\int_{t_0}^{t}(\alpha_1 p_4^{(2)})dx_4$$

Now the wave function of the incident particles are given by

$$\phi^{(1\_AVE)} = \frac{1}{\sqrt{V}}\exp\left(\frac{i}{\hbar}\int_{x_0}^{x}R_\mu^{(1)}dx_\mu\right)$$
$$= \frac{1}{\sqrt{V}}\exp\left(\frac{i}{\hbar}p_\mu^{(i)}\cdot x + \frac{i}{\hbar}(1-\frac{kT}{m^{(2)}c^2})\int_{t_0}^{t}(\alpha_1 p_4^{(2)})dx_4\right) \qquad (216)$$

Actually, we have moved the magnetic components ($A_1$, $A_2$, $A_3$) coming from the target particle to the time-axis effect, thus the wave function of the incident particles satisfy

$$-i\hbar\frac{\partial\phi^{(1\_AVE)}}{\partial x_k} = (p_k^{(1\_i)} + 0)\phi^{(1\_AVE)}; \quad k=1,2,3$$
$$-i\hbar\frac{\partial\phi^{(1\_AVE)}}{\partial x_4} = \left(p_4^{(1\_i)} + (1-\frac{kT}{m^{(2)}c^2})\alpha_1 p_4^{(2)}\right)\phi^{(1\_AVE)}; \qquad (217)$$

The wave function can also be expressed in terms of its free plane waves as (the Fourier expansion)

$$\phi^{(1\_AVE)} = \sum_{n=1}^{\infty}a_n(t)\exp(\frac{ip_n^{(1)}\cdot x}{\hbar}) \quad . \qquad (218)$$

And we know in the space-time ($x_1,x_2,x_3,x_4=-ct$), the fourth component expresses its evolution. For our purpose, substituting it into the left side of the fourth component, we get

$$-i\hbar\sum_{n=1}^{\infty}(\frac{\partial a_n}{\partial x_4} + \frac{a_n ip_{n4}^{(1)}}{\hbar})\exp(\frac{ip_n^{(1)}\cdot x}{\hbar})$$
$$= \frac{1}{\sqrt{V}}\left(p_4^{(1\_i)} + (1-\frac{kT}{m^{(2)}c^2})\alpha_1 p_4^{(2)}\right)\exp\left(\frac{i}{\hbar}p_\mu^{(i)}\cdot x\right) \qquad (219)$$

Multiplying by another plane wave eigen state $\phi_f^*$, integrating over the whole volume $V$, and using the orthonorrmality relation of eigen states, using the initial condition $a_{ni}=\delta_{ni}$, it leads to

$$\frac{\partial a_f}{\partial x_4} = \frac{i}{V\hbar}\int dx^3\left((1-\frac{kT}{m^{(2)}c^2})\alpha_1 p_4^{(2)}\right)\exp\left(\frac{i(p_i^{(1)}-p_f^{(1)})\cdot x}{\hbar}\right) \quad . \qquad (220)$$

We have

$$a_n(t=-\frac{T}{2}) = \delta_{ni} \quad . \qquad (221)$$

At a later time $t$, we have

$$a_f(t) = \frac{i}{V\hbar}\int_{-T/2}^{t}dx_4\int dx^3\left((1-\frac{kT}{m^{(2)}c^2})\alpha_1 p_4^{(2)}\right)\exp\left(\frac{i(p_i^{(1)}-p_f^{(1)})\cdot x}{\hbar}\right) \quad . \qquad (222)$$

At a final departure time t=T/2=∞, we have

$$a_f(t = \frac{T}{2}) = \frac{i}{V\hbar} \int dx^4 \left( (1 - \frac{kT}{m^{(2)}c^2}) \alpha_1 p_4^{(2)} \right) \exp\left( \frac{i(p_i^{(1)} - p_f^{(1)}) \cdot x}{\hbar} \right) . \quad (223)$$

We define the $a(t=\infty)$ as the **transition amplitude** that the particle has scattered from an initial state $i$ to a final state $f$, whose formula divides into two parts as

$$a_f(t = \frac{T}{2}) = \int dx_4 \exp\left( \frac{i(p_i^{(1)} - p_f^{(1)}) \cdot x_4}{\hbar} \right)$$
$$\cdot \frac{i}{V\hbar} \int dx^3 \left( (1 - \frac{kT}{m^{(2)}c^2}) \alpha_1 p_4^{(2)} \right) \exp\left( \frac{i(\mathbf{p}_i^{(1)} - \mathbf{p}_f^{(1)}) \cdot \mathbf{x}}{\hbar} \right) \quad (224)$$

We have

$$a_f(t = \frac{T}{2}) = (2\pi\hbar)\delta(E_f^{(1)} - E_i^{(1)} + w)\Gamma_{if}$$

$$\Gamma_{if} \equiv \frac{i}{V\hbar} \int dx^3 \left( (1 - \frac{kT}{m^{(2)}c^2}) \alpha_1 p_4^{(2)} \right) \exp\left( \frac{i(\mathbf{p}_i^{(1)} - \mathbf{p}_f^{(1)}) \cdot \mathbf{x}}{\hbar} \right) \quad (225)$$

In practice physics, we usually deal with the transition rate

$$W_{if} = 2\pi\hbar \int |\Gamma_{if}|^2 \delta(E_f - E_i)\rho(E)dE_f$$
$$= 2\pi\hbar |\Gamma_{if}|^2 \rho(E_i) \quad (226)$$

This is the Fermi's Golden Rule. The incident particle 1 after the collision changes its state from A to C, as shown in Fig.14. Due to the Fermi's Golden Rule, we obtain the differential cross section

$$D = \frac{V^2 |\mathbf{p}_C|^2}{(2\pi\hbar)^2 (|\mathbf{v}_A||\mathbf{v}_C|)} |\Gamma_{AC}|^2 = \frac{V^2 m^{(1)2} |\mathbf{p}_C|}{(2\pi\hbar)^2 |\mathbf{p}_A|} |\Gamma_{AC}|^2 . \quad (227)$$

**Chapter 2    Factory of prosthesis**

The weak interaction and strong interaction are investigated by using the relativistic de Broglie matter wave with SU(n) symmetry.

**12. An electron interacting with an electromagnetic field**

In the author's early paper[19], we have derived out the relativistic matter wave from the Newton's second law as follows

$$\phi = \exp\left( \frac{i}{\hbar} \int_{x_0}^{x} (mu_\mu + qA_\mu)dx_\mu \right) = \exp\left( \frac{i}{\hbar} \int_{x_0}^{x} R_\mu dx_\mu \right)$$
$$R_\mu = mu_\mu + qA_\mu \quad (228)$$

where the integral takes from the initial point $x_0$ to the final point $x$ by an arbitrary mathematical path in the velocity field in which the particle has uncertain positions duo to the quantum hidden variable. $A_\mu$ is the electromagnetic potential; $R_\mu$ is the riding-wave momentum.

As shown in Fig.1, in bound states, due to periodic boundary condition, the relativistic matter wave is quantized for a closed orbit $L$ by

$$\frac{1}{\hbar}\oint_L (m\mathbf{u}+q\mathbf{A})\cdot d\mathbf{x} = 2\pi n \quad n=1,2,3... \quad . \tag{229}$$

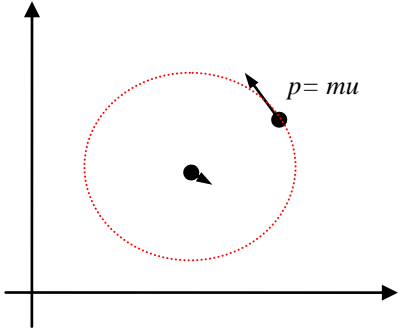

Fig.1 Two particle system.

As a simplest case, consider a hydrogen atom in a uniform magnetic field **B**, as in a Zeeman experiment. As we know, the Coulomb's potential energy V and the magnetic potential are

$$qA_4 = \frac{iV}{c}$$
$$\mathbf{A} = \frac{1}{2}\mathbf{B}\times\mathbf{r}, \quad \mathbf{B} = \frac{2}{r^2}(\mathbf{r}\times\mathbf{A}) \tag{230}$$

The electronic matter wave is given by

$$\begin{aligned}
\phi &= \exp\left(\frac{i}{\hbar}\int_{x_0}^x (mu_\mu + qA_\mu)dx_\mu\right) \\
&= \exp\left(\frac{i}{\hbar}\int_{\mathbf{x}_0}^{\mathbf{x}} (m\mathbf{u}+q\mathbf{A})\cdot d\mathbf{x} + \frac{i}{\hbar}\int_{\mathbf{x}_0}^{\mathbf{x}} (mp_4 + qA_4)\cdot dx_4\right) \\
&= \exp\left(\frac{i}{\hbar}\int_{\mathbf{x}_0}^{\mathbf{x}} (m\mathbf{u})\cdot d\mathbf{x} + \frac{i}{\hbar}\int_{\mathbf{x}_0}^{\mathbf{x}} (q\mathbf{A})\cdot\frac{d\mathbf{x}}{d\tau}d\tau + \frac{i}{\hbar}\int_{x_0}^{x} (mp_4 + \frac{iV}{c})\cdot dx_4\right) \\
&= \exp\left(\frac{i}{\hbar}\int_{\mathbf{x}_0}^{\mathbf{x}} (m\mathbf{u})\cdot d\mathbf{x} + \frac{i}{\hbar}\int_{\tau_0}^{\tau} (q\mathbf{A}\cdot\mathbf{u})d\tau + \frac{i}{\hbar}\int_{t_0}^{t} (icmp_4 - V)dt\right)
\end{aligned} \tag{231}$$

We invoke the following vector formula

$$(\mathbf{A}\times\mathbf{B})\cdot(\mathbf{C}\times\mathbf{D}) = (\mathbf{A}\cdot\mathbf{C})(\mathbf{B}\cdot\mathbf{D}) - (\mathbf{A}\cdot\mathbf{D})(\mathbf{B}\cdot\mathbf{C}) \tag{232}$$

Then

$$\phi = \exp\left(\frac{i}{\hbar}\int_{\mathbf{x}_0}^{\mathbf{x}}(m\mathbf{u})\cdot d\mathbf{x} + \frac{i}{\hbar}\int_{\tau_0}^{\tau}\frac{q}{r^2}(\mathbf{r}\cdot\mathbf{r})(\mathbf{A}\cdot\mathbf{u})d\tau + \frac{i}{\hbar}\int_{t_0}^{t}(icmp_4 - V)dt\right)$$

$$= \exp\left(\frac{i}{\hbar}\int_{\mathbf{x}_0}^{\mathbf{x}}(m\mathbf{u})\cdot d\mathbf{x} + \frac{i}{\hbar}\int_{\tau_0}^{\tau}\frac{q}{r^2}[(\mathbf{r}\times\mathbf{A})\cdot(\mathbf{r}\times\mathbf{u}) + (\mathbf{r}\cdot\mathbf{u})(\mathbf{A}\cdot\mathbf{r})]d\tau + \frac{i}{\hbar}\int_{t_0}^{t}(icmp_4 - V)dt\right)$$

$$= \exp\left(\frac{i}{\hbar}\int_{\mathbf{x}_0}^{\mathbf{x}}(m\mathbf{u})\cdot d\mathbf{x} + \frac{i}{\hbar}\int_{\tau_0}^{\tau}\frac{q}{r^2}[(\mathbf{r}\times\mathbf{A})\cdot(\mathbf{r}\times\mathbf{u}) + 0]d\tau + \frac{i}{\hbar}\int_{t_0}^{t}(icmp_4 - V)dt\right)$$

(233)

We have taken average $<\mathbf{r}\cdot\mathbf{u}> = <ru_r> = 0$. Using the angular momentum $L$, the matter wave is given by

$$\phi = \exp\left(\frac{i}{\hbar}\int_{\mathbf{x}_0}^{\mathbf{x}}(m\mathbf{u})\cdot d\mathbf{x} + \frac{i}{\hbar}\int_{t_0}^{t}\frac{q}{2m}(\mathbf{B}\cdot\mathbf{L})\frac{dt}{\sqrt{1-v^2/c^2}} + \frac{i}{\hbar}\int_{t_0}^{t}(icmp_4 - V)dt\right)$$

$$= \exp\left(\frac{i}{\hbar}\int_{\mathbf{x}_0}^{\mathbf{x}}(m\mathbf{u})\cdot d\mathbf{x} + \frac{i}{\hbar}\int_{t_0}^{t}(icmp_4 - V + \frac{q(\mathbf{B}\cdot\mathbf{L})}{2m\sqrt{1-v^2/c^2}})dt\right)$$

(234)

$$= \exp\left(\frac{i}{\hbar}\int_{\mathbf{x}_0}^{\mathbf{x}}(m\mathbf{u})\cdot d\mathbf{x} + \frac{i}{\hbar}\int_{t_0}^{t}[-mc^2 - \frac{1}{2}mv^2 - V + \frac{q(\mathbf{B}\cdot\mathbf{L})}{2m} + O(v^2/c^2)]dt\right)$$

$$= \exp\left(\frac{i}{\hbar}\int_{\mathbf{x}_0}^{\mathbf{x}}(m\mathbf{u})\cdot d\mathbf{x} + \frac{i}{\hbar}[-mc^2 - \frac{1}{2}mv^2 - V + \frac{q(\mathbf{B}\cdot\mathbf{L})}{2m}]t\right)$$

From the last line, we obtain the electronic energy $E$ that is

$$E = mc^2 + \frac{1}{2}mv^2 + V - \frac{q(\mathbf{B}\cdot\mathbf{L})}{2m} \tag{235}$$

It contains the kinematic energy, the Coulomb's electric energy and the magnetic energy. The electronic matter wave is given by

$$\phi = \exp\left(\frac{i}{\hbar}\int_{\mathbf{x}_0}^{\mathbf{x}}(m\mathbf{u})\cdot d\mathbf{x} + \frac{iEt}{\hbar}\right) \tag{236}$$

This result indicates that the interaction energy $E^{(12)}$ between two particles arises from the following integral

$$E^{(12)} = \int_{x_0}^{x}(q^{(1)}A_\mu^{(2)})dx_\mu^{(1)} = \int_{\tau_0}^{\tau}(q^{(1)}A_\mu^{(2)})u_\mu^{(1)}d\tau \tag{237}$$

which appears in the matter wave as an addition term to the momentum line integral.

$$\phi^{(1)} = \exp\left(\frac{i}{\hbar}\int_{x_0}^{x}(m^{(1)}u_\mu^{(1)} + q^{(1)}A_\mu)dx_\mu^{(1)}\right) \tag{238}$$

Recalling that for structure-less point particle we have

$$A_\mu = \frac{1}{4\pi\varepsilon_0}\frac{q^{(2)}u_\mu^{(2)}}{c^2 r} = \frac{\alpha_1 p_\mu^{(2)}}{q^{(1)}} \tag{239}$$

the interaction energy becomes

$$E^{(12)} = \int_{x_0}^{x} (q^{(1)} A_\mu^{(2)}) dx_\mu^{(1)} = \int_{\tau_0}^{\tau} \frac{\alpha_1}{m^{(1)}} p_\mu^{(2)} p_\mu^{(1)} d\tau \tag{240}$$

It transcends that the interaction energy should be in the from

$$E^{(12)} \propto p_\mu^{(2)} p_\mu^{(1)} \propto j_\mu^{(2)} j_\mu^{(1)} \tag{241}$$

This form is the jet-jet interaction which is very prevalent in high energy collisions, for example, V-A interaction of weak currents proposed by Fermi in 1932 with introducing the famous Fermi's constant G for $\beta$-decay.

### 13. The jet-jet interaction in electron-muon scattering

Consider two particle system in an inertial reference frame, as shown in Fig.2, their riding-wave momenta are given in terms of the Pauli matrices by

$$\begin{bmatrix} R^{(1)} \\ R^{(2)} \end{bmatrix} = (1+S) \begin{bmatrix} p^{(1)} \\ p^{(2)} \end{bmatrix} = (1 + \alpha_1 \sigma_1 + \alpha_2 \sigma_2 + \alpha_3 \sigma_3) \begin{bmatrix} p^{(1)} \\ p^{(2)} \end{bmatrix} \tag{242}$$

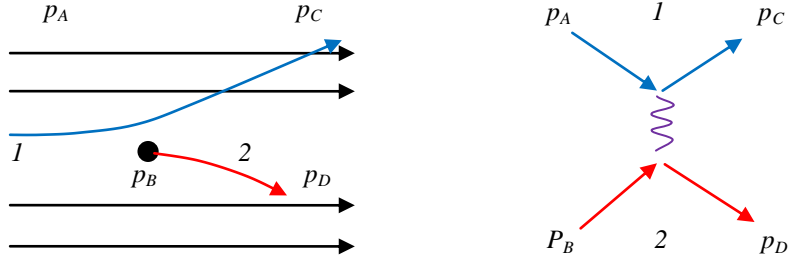

Fig.2 Particle 1 scatters off particle 2.

Where, the Pauli matrices (SU(2) group) are given by (corresponding to $\lambda_1, \lambda_2, \lambda_3$ in SU(3) group)

$$\sigma_1 = \begin{bmatrix} 0 & 1 \\ 1 & 0 \end{bmatrix}, \quad \sigma_2 = \begin{bmatrix} 0 & -i \\ i & 0 \end{bmatrix}, \quad \sigma_3 = \begin{bmatrix} 1 & 0 \\ 0 & -1 \end{bmatrix}, \tag{243}$$

and $\alpha_1$, $\alpha_2$, $\alpha_3$ are three independent real first order small parameters. Their interaction matrix $S$ is the combination of three Pauli matrices, according the preceding sections, the first Pauli matrix $\alpha_1 \sigma_1$ represents the electromagnetic interaction, understood as basic interaction; the second Pauli matrix $\alpha_2 \sigma_2$ arises the spin effect for electron; the third Pauli matrix $\alpha_3 \sigma_3$ serves fermions for which it represents a kind of self-interaction that is associated with the Pauli Exclusion Principle.

At the first, we let $\alpha_1 \sigma_1$ to work while $\alpha_2 \sigma_2$ and $\alpha_3 \sigma_3$ to idle, in a moment we will add the $\alpha_2 \sigma_2$ and $\alpha_3 \sigma_3$ effects. The $N=2$ system has the riding-wave momenta given by

$$\begin{bmatrix} R^{(1)} \\ R^{(2)} \end{bmatrix} = (1 + \alpha_1 \sigma_1) \begin{bmatrix} p^{(1)} \\ p^{(2)} \end{bmatrix} = (1 + \alpha_1 \begin{bmatrix} 0 & 1 \\ 1 & 0 \end{bmatrix}) \begin{bmatrix} p^{(1)} \\ p^{(2)} \end{bmatrix} . \tag{244}$$

The single incident particle 1 is governed by

$$-i\hbar\frac{\partial \phi^{(1)}}{\partial x_\mu} = R_\mu^{(1)}\phi^{(1)} = (p_\mu^{(1)} + \alpha_1 p_\mu^{(2)})\phi^{(1)} \quad . \tag{245}$$

The incident particle 1 after the collision changes its state from A to C, the target particle from C to D, as shown in Fig.2. The matter wave is given by

$$\phi^{(1)} = \frac{1}{\sqrt{V}}\exp\left(\frac{i}{\hbar}p_\mu^{(A)}\cdot x + \frac{i}{\hbar}\int_{x_0}^{x}(\alpha_1 p_\mu^{(B)})dx_\mu\right) \quad . \tag{246}$$

The integral paths take on the incident line (integral path independency). It is interesting to look at the following integral.

$$\frac{i}{\hbar}\int_{x_0}^{x}(\alpha_1 p_\mu^{(B)})dx_\mu = \frac{i}{\hbar}\int_{\tau_0}^{\tau}\frac{1}{m^{(1)}}\alpha_1 p_\mu^{(B)} p_\mu^{(A)} d\tau^{(A)} \quad . \tag{247}$$

We define $\alpha_1 = g_1/r$, where $g_1$ is a proportional coefficient, then

$$\frac{i}{\hbar}\int_{x_0}^{x}(\alpha_1 p_\mu^{(B)})dx_\mu = \frac{i}{\hbar}\int_{t_0}^{t}\frac{icg_1}{p_4^{(A)}r} p_\mu^{(B)} p_\mu^{(A)} dt \quad . \tag{248}$$

Now the wave function of the incident particles are given by

$$\phi^{(1)} = \frac{1}{\sqrt{V}}\exp\left(\frac{i}{\hbar}p_\mu^{(i)}\cdot x + \frac{i}{\hbar}\int_{t_0}^{t}\frac{icg_1}{p_4^{(A)}r} p_\mu^{(B)} p_\mu^{(A)} dt\right) \quad . \tag{249}$$

Actually, we have moved the magnetic-like components (like $A_1, A_2, A_3$) coming from the target particle to the time-axis effect, thus the wave function of the incident particles satisfy

$$\begin{aligned}-i\hbar\frac{\partial \phi^{(1)}}{\partial x_k} &= (p_k^{(A)} + 0)\phi^{(1)}; \quad k=1,2,3 \\ -i\hbar\frac{\partial \phi^{(1)}}{\partial x_4} &= \left(p_4^{(A)} + \frac{g_1}{p_4^{(A)}r} p_\mu^{(B)} p_\mu^{(A)}\right)\phi^{(1)}\end{aligned} \quad . \tag{250}$$

Therefore the interaction energy between the two particles is given by

$$E^{(12)} = \frac{cg_1}{ip_4^{(A)}r} p_\mu^{(B)} p_\mu^{(A)} \quad \Leftrightarrow \quad -i\hbar\frac{\partial \phi^{(1)}}{\partial x_4} = \left(p_4^{(A)} + \frac{iE^{(12)}}{c}\right)\phi^{(1)} \quad . \tag{251}$$

The wave function for the two particles can also be expressed in terms of their free plane waves as (the Fourier expansion)

$$\phi^{(1)} = \sum_{n=1}^{\infty} a_n(t)\exp\left(\frac{ip_n^{(1)}\cdot x}{2\hbar}\right) \quad . \tag{252}$$

And we know in the space-time $(x_1,x_2,x_3,x_4=ict)$, the fourth component expresses its evolution. For our purpose, substituting it into the left side of the fourth component equation, we get

$$\begin{aligned}&-i\hbar\sum_{n=1}^{\infty}(\frac{\partial a_n}{\partial x_4} + \frac{a_n ip_{n4}^{(1)}}{\hbar})\exp\left(\frac{ip_n^{(1)}\cdot x}{\hbar}\right) \\ &= \frac{1}{\sqrt{V}}(p_4^{(A)} + \frac{g_1}{p_4^{(A)}r} p_\mu^{(B)} p_\mu^{(A)})\exp\left(\frac{ip_n^{(1)}\cdot x}{\hbar}\right)\end{aligned} \quad . \tag{253}$$

Multiplying by another plane wave eigen state $\phi_C^*$ with the momentum $p_C$, integrating over the whole volume $V$, and using the orthonorrmality relation of eigen states, using the initial condition $a_n=\delta_{nA}$, it leads to

$$\frac{\partial a_C}{\partial x_4} = \frac{i}{V\hbar}\int dx^3 (\frac{g_1}{p_4^{(A)}r} p_\mu^{(B)} p_\mu^{(A)})\exp\left(\frac{i(p^{(A)} - p^{(C)})\cdot x}{\hbar}\right) . \tag{254}$$

We have

$$a_C(t = -\frac{T}{2}) = \delta_{AC} . \tag{255}$$

At a later time $t$, we have

$$a_C(t) = \frac{i}{V\hbar}\int_{-T/2}^{t} dx_4 \int dx^3 (\frac{g_1}{p_4^{(A)}r} p_\mu^{(B)} p_\mu^{(A)})\exp\left(\frac{i(p^{(A)} - p^{(C)})\cdot x}{\hbar}\right) . \tag{256}$$

At a final departure time $t=T/2=\infty$, we have

$$a_C(t = \frac{T}{2}) = \frac{i}{V\hbar}\int dx^4 (\frac{g_1}{p_4^{(A)}r} p_\mu^{(B)} p_\mu^{(A)})\exp\left(\frac{i(p^{(A)} - p^{(C)})\cdot x}{\hbar}\right) . \tag{257}$$

We define the $a(t=\infty)$ as the **transition amplitude** that the particle has scattered from an initial state $i$ to a final state $f$, whose formula is

$$a_C(t = \frac{T}{2}) = \frac{i}{V\hbar}\int dx^4 (\frac{g_1}{p_4^{(A)}r} p_\mu^{(B)} p_\mu^{(A)})\exp\left(\frac{i(p^{(A)} - p^{(C)})\cdot x}{\hbar}\right)$$

$$= \frac{i}{V\hbar}\int dx^4 (\frac{g_1}{p_4^{(A)}r} p_\mu^{(B)} p_\mu^{(A)})\exp\left(\frac{iq\cdot x}{\hbar}\right) \tag{258}$$

$$q \equiv p^{(A)} - p^{(C)}$$

we have

$$a_C(t = \frac{T}{2}) = (2\pi\hbar)\delta(E^{(A)} - E^{(C)})\Gamma_{AC}$$

$$\Gamma_{AC} \equiv \frac{i}{V\hbar}\int dx^3 (\frac{g_1}{p_4^{(A)}r} p_\mu^{(B)} p_\mu^{(A)})\exp\left(\frac{i\mathbf{q}\cdot\mathbf{x}}{\hbar}\right) . \tag{259}$$

$$\mathbf{q} \equiv \mathbf{p}^{(A)} - \mathbf{p}^{(C)}$$

In practice physics, we usually deal with the transition rate

$$W_{AC} = 2\pi\hbar\int |\Gamma_{AC}|^2 \delta(E^{(A)} - E^{(C)})\rho(E^{(C)})dE^{(C)}$$

$$= 2\pi\hbar |\Gamma_{AC}|^2 \rho(E^{(C)}) \tag{260}$$

This is the Fermi's Golden Rule, from which we obtain the differential cross section

$$D = \frac{V^2 |\mathbf{p}_C|^2}{(2\pi\hbar)^2 (|\mathbf{v}_A||\mathbf{v}_C|)}|\Gamma_{AC}|^2 = \frac{V^2 m^{(1)2} |\mathbf{p}_C|}{(2\pi\hbar)^2 |\mathbf{p}_A|}|\Gamma_{AC}|^2 . \tag{261}$$

In order to eliminate the $1/r$ term in $\Gamma_{AC}$, applying the following formula

$$\int (\nabla^2 F)\exp(\frac{i\mathbf{q}\cdot\mathbf{x}}{\hbar})d^3x = -\frac{|\mathbf{q}|^2}{\hbar^2}\int F\exp(\frac{i\mathbf{q}\cdot\mathbf{x}}{\hbar})d^3x . \tag{262}$$

to the transition amplitude, we obtain

$$\Gamma_{AC} = \frac{i}{V\hbar} \frac{-\hbar^2}{|\mathbf{q}|^2} \int d^3x \nabla^2 (\frac{g_1}{p_4^{(A)} r} p_\mu^{(B)} p_\mu^{(A)}) \exp\left(\frac{i\mathbf{q}\cdot\mathbf{x}}{\hbar}\right)$$

$$= \frac{i}{V\hbar} \frac{-\hbar^2 g_1}{p_4^{(A)} |\mathbf{q}|^2} p_\mu^{(B)} p_\mu^{(A)} \int d^3x \nabla^2 (\frac{1}{r}) \exp\left(\frac{i\mathbf{q}\cdot\mathbf{x}}{\hbar}\right) \quad (263)$$

$$= \frac{-i4\pi\hbar g_1}{V p_4^{(A)} |\mathbf{q}|^2} p_\mu^{(B)} p_\mu^{(A)}$$

It is in the jet-jet form. Actually, the collision process can be divided into two stages: coming stage and departure stage, we need equally treat the two stages. The simplest approach is to take average over the coming stage and departure stage for the transition amplitude, this leads to

$$\Gamma_{AC} = \frac{-i4\pi\hbar g_1}{V p_4^{(A)} |\mathbf{q}|^2} \frac{p_\mu^{(B)} + p_\mu^{(D)}}{2} \frac{p_\mu^{(A)} + p_\mu^{(C)}}{2}$$

$$= \frac{-i\pi\hbar g_1}{V p_4^{(A)} |\mathbf{q}|^2} (p_\mu^{(B)} + p_\mu^{(D)})(p_\mu^{(A)} + p_\mu^{(C)}) \quad (264)$$

The differential cross section is given by

$$\frac{d\sigma}{d\Omega} = D = \frac{V^2 m^{(1)2} |\mathbf{p}_C|}{(2\pi\hbar)^2 |\mathbf{p}_A|} |\Gamma_{AC}|^2$$

$$= \frac{m^{(1)2} |\mathbf{p}_C|}{(2\pi\hbar)^2 |\mathbf{p}_A|} \left| \frac{-i\pi\hbar g_1}{V p_4^{(A)} |\mathbf{q}|^2} (p_\mu^{(B)} + p_\mu^{(D)})(p_\mu^{(A)} + p_\mu^{(C)}) \right|^2 \quad (265)$$

Now we arrive at a place where we can consider the second Pauli matrix $\alpha_2\sigma_2$ and the third Pauli matrix $\alpha_3\sigma_3$ for the collision. Current research suggests that the $\alpha_2\sigma_2$ contributes the spin effect as it was done in the high energy physics, ignoring the $\alpha_3\sigma_3$ Pauli exclusion effect for the collision.

This differential cross section formula agrees well with experiments of electron-muon scattering.

### 14. The jet-jet interaction of two particle beams in collision

Let $\alpha_1\sigma_1$ to work while $\alpha_2\sigma_2$ and $\alpha_3\sigma_3$ to idle, then regarding $\alpha_2\sigma_2$ and $\alpha_3\sigma_3$ as additive effects for the collision, this approach seems unfair for the Hertian interaction matrix $S$. Therefore we need rebuild our calculation. Consider two particle system in an inertial reference frame, as shown in Fig.3, their riding-wave momenta are given in terms of the Pauli matrices by

$$\begin{bmatrix} R^{(1)} \\ R^{(2)} \end{bmatrix} = (1+S) \begin{bmatrix} p^{(1)} \\ p^{(2)} \end{bmatrix} = (1 + \alpha_1\sigma_1 + \alpha_2\sigma_2 + \alpha_3\sigma_3) \begin{bmatrix} p^{(1)} \\ p^{(2)} \end{bmatrix} \quad (266)$$

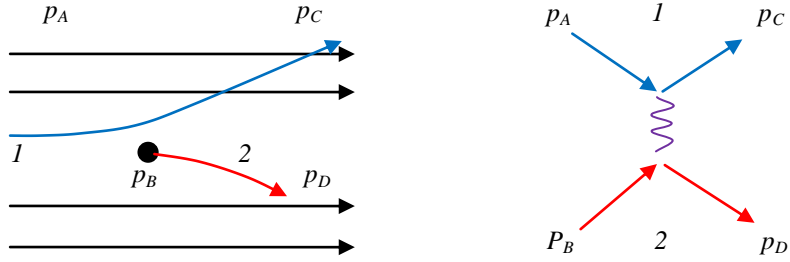

Fig.3 Particle 1 scatters off particle 2.

The single incident particle 1 is governed by

$$-i\hbar \frac{\partial \phi^{(1)}}{\partial x_\mu} = R_\mu^{(1)} \phi^{(1)} = (p_\mu^{(1)} + S^{(1k)} p_\mu^{(k)})\phi^{(1)} = (p_\mu^{(1)} + \alpha_a \sigma_a^{(1k)} p_\mu^{(k)})\phi^{(1)} \quad . \quad (267)$$
$$a = 1,2,3; \quad k = 1,2$$

The repeated indexes imply summation over them. The incident particle 1 after the collision changes its state from A to C, the target particle from C to D, as shown in Fig.3. Considering a small volume element *dV* in the incident beam in where there are N particles 1 at a time, their average quantities (labeled *AVE*) satisfy

$$-i\hbar \frac{\partial \phi^{(1\_AVE)}}{\partial x_\mu} \simeq (p_\mu^{(1\_AVG)} + \alpha_a \sigma_a^{(1k)} p_\mu^{(k\_AVG)})\phi^{(1\_AVE)};$$
$$\phi^{(1\_AVE)} = \sum_{j=1}^N \phi^{(1\_j)} / N; \quad (268)$$
$$p_\mu^{(1\_AVE)} = \sum_{j=1}^N p_\mu^{(1\_j)} \phi^{(1\_j)} / \sum_{j=1}^N \phi^{(1\_j)}$$

with a hope for a well approximation. Since in high energy collision 99% incident particles go almost straightforward by the target particle, only about 1% incident particles would be apparently scattered by the target particle, their average quantities well approximately satisfy

$$p_\mu^{(1\_AVE)} = 0.99 p_\mu^{(A)} + 0.01 p_\mu^{(1\_other)} \simeq p_\mu^{(A)}$$
$$-i\hbar \frac{\partial \phi^{(1\_AVE)}}{\partial x_\mu} = (p_\mu^{(A)} + \alpha_a \sigma_a^{(1k)} p_\mu^{(k\_AVG)})\phi^{(1\_AVE)} \quad . \quad (269)$$

The fewer scattered partition, the more accurate expectation. The wave function of the incident particles are given by

$$\phi^{(1\_AVE)} = \frac{1}{\sqrt{V}} \exp\left(\frac{i}{\hbar} \int_{x_0}^x R_\mu^{(1)} dx_\mu\right)$$
$$= \frac{1}{\sqrt{V}} \exp\left(\frac{i}{\hbar} p_\mu^{(A)} \cdot x + \frac{i}{\hbar} \int_{x_0}^x (\alpha_a \sigma_a^{(1k)} p_\mu^{(k\_AVE)}) dx_\mu\right) \quad . \quad (270)$$

Where, the wave function has normalized to one particle in the volume *V*. The integral paths take on the incident line (integral path independency). It is interesting to look at the following integral.

$$\frac{i}{\hbar}\int_{x_0}^{x}(\alpha_a\sigma_a^{(1k)}p_\mu^{(k\_AVG)})dx_\mu = \frac{i}{\hbar}\int_{\tau_0}^{\tau}\frac{1}{m^{(1)}}\alpha_a\sigma_a^{(1k)}p_\mu^{(k\_AVE)}p_\mu^{(A)}d\tau^{(A)} \quad . \tag{271}$$

Since we know $\alpha_1 \sim 1/r$, then we naturally assume $\alpha_a = g_a/r$, where $g_a$ are proportional coefficients, then the wave function of the incident particles are given by

$$\phi^{(1\_AVE)} = \frac{1}{\sqrt{V}}\exp\left(\frac{i}{\hbar}p_\mu^{(A)}\cdot x + \frac{i}{\hbar}\int_{t_0}^{t}\frac{ic}{p_4^{(i)}r}g_a\sigma_a^{(1k)}p_\mu^{(k\_AVE)}p_\mu^{(A)}dt\right) \quad . \tag{272}$$

Actually, we have moved the magnetic-like components (like $A_1, A_2, A_3$) coming from the target particle to the time-axis effect, thus the wave function of the incident particles satisfy

$$\begin{aligned}-i\hbar\frac{\partial\phi^{(1\_AVE)}}{\partial x_k} &= (p_k^{(A)}+0)\phi^{(1\_AVE)}; \quad k=1,2,3 \\ -i\hbar\frac{\partial\phi^{(1\_AVE)}}{\partial x_4} &= \left(p_4^{(A)}+\frac{1}{p_4^{(A)}r}g_a\sigma_a^{(1k)}p_\mu^{(k\_AVG)}p_\mu^{(A)}\right)\phi^{(1\_AVE)}; \end{aligned} \tag{273}$$

Therefore the interaction energy between the two particles is given by

$$\begin{aligned} E^{(12)} &= \frac{c}{ip_4^{(A)}r}g_a\sigma_a^{(1k)}p_\mu^{(k\_AVG)}p_\mu^{(A)} \\ -i\hbar\frac{\partial\phi^{(1\_AVE)}}{\partial x_4} &= \left(p_4^{(A)}+\frac{iE^{(12)}}{c}\right)\phi^{(1\_AVE)} \end{aligned} \tag{274}$$

The wave function can also be expressed in terms of its free plane waves as (the Fourier expansion)

$$\phi^{(1\_AVE)} = \sum_{n=1}^{\infty}a_n(t)\exp\left(\frac{ip_n^{(1)}\cdot x}{\hbar}\right) \quad . \tag{275}$$

And we know in the space-time $(x_1,x_2,x_3,x_4=ict)$, the fourth component expresses its evolution. For our purpose, substituting it into the left side of the fourth component equation, we get

$$\begin{aligned} &-i\hbar\sum_{n=1}^{\infty}(\frac{\partial a_n}{\partial x_4}+\frac{a_n ip_{n4}^{(1)}}{\hbar})\exp\left(\frac{ip_n^{(1)}\cdot x}{\hbar}\right) \\ &= \frac{1}{\sqrt{V}}(p_4^{(A)}+\frac{1}{p_4^{(A)}r}g_a\sigma_a^{(1k)}p_\mu^{(k\_AVE)}p_\mu^{(A)})\exp\left(\frac{ip_n^{(1)}\cdot x}{\hbar}\right) \end{aligned} \tag{276}$$

Multiplying by another plane wave eigen state $\phi_C^*$ with the momentum $p_C$, integrating over the whole volume $V$, and using the orthonorrmality relation of eigen states, using the initial condition $a_n=\delta_{nA}$, it leads to

$$\frac{\partial a_C}{\partial x_4} = \frac{i}{V\hbar}\int dx^3(\frac{1}{p_4^{(A)}r}g_a\sigma_a^{(1k)}p_\mu^{(k\_AVG)}p_\mu^{(A)})\exp\left(\frac{i(p^{(A)}-p^{(C)})\cdot x}{\hbar}\right) \quad . \tag{277}$$

We have

$$a_n(t=-\frac{T}{2}) = \delta_{nA} \quad . \tag{278}$$

At a later time $t$, we have

$$a_C(t) = \frac{i}{V\hbar}\int_{-T/2}^{t}dx_4\int dx^3(\frac{1}{p_4^{(A)}r}g_a\sigma_a^{(1k)}p_\mu^{(k\_AVG)}p_\mu^{(A)})\exp\left(\frac{i(p^{(A)}-p^{(C)})\cdot x}{\hbar}\right) \quad .$$

At a final departure time $t=T/2=\infty$, we have

$$a_C(t=\frac{T}{2}) = \frac{i}{V\hbar}\int dx^4 (\frac{1}{p_4^{(A)}r} g_a \sigma_a^{(1k)} p_\mu^{(k\_AVG)} p_\mu^{(A)}) \exp\left(\frac{i(p^{(A)}-p^{(C)})\cdot x}{\hbar}\right) \quad . \tag{280}$$

We define the $a(t=\infty)$ as the **transition amplitude** that the particle has scattered from an initial state $A$ to a final state $C$, whose formula is

$$\begin{aligned}
a_C(t=\frac{T}{2}) &= \frac{i}{V\hbar}\int dx^4 (\frac{1}{p_4^{(A)}r} g_a \sigma_a^{(1k)} p_\mu^{(k\_AVG)} p_\mu^{(A)}) \exp\left(\frac{i(p^{(A)}-p^{(C)})\cdot x}{\hbar}\right) \\
&= \frac{i}{V\hbar}\int dx^4 (\frac{1}{p_4^{(A)}r} g_a \sigma_a^{(1k)} p_\mu^{(k\_AVG)} p_\mu^{(A)}) \exp\left(\frac{iq\cdot x}{\hbar}\right) \\
q &\equiv p^{(A)} - p^{(C)}
\end{aligned} \tag{281}$$

we have

$$\begin{aligned}
a_C(t=\frac{T}{2}) &= (2\pi\hbar)\delta(E^{(A)}-E^{(C)})\Gamma_{AC} \\
\Gamma_{AC} &\equiv \frac{i}{V\hbar}\int dx^3 (\frac{1}{p_4^{(A)}r} g_a \sigma_a^{(1k)} p_\mu^{(k\_AVG)} p_\mu^{(A)}) \exp\left(\frac{i\mathbf{q}\cdot\mathbf{x}}{\hbar}\right) \quad . \\
\mathbf{q} &\equiv \mathbf{p}^{(A)} - \mathbf{p}^{(C)}
\end{aligned} \tag{282}$$

In practice physics, we usually deal with the transition rate

$$\begin{aligned}
W_{AC} &= 2\pi\hbar \int |\Gamma_{AC}|^2 \delta(E^{(A)}-E^{(C)})\rho(E^{(C)})dE^{(C)} \\
&= 2\pi\hbar |\Gamma_{AC}|^2 \rho(E^{(C)})
\end{aligned} \tag{283}$$

This is the Fermi's Golden Rule, from which we obtain the differential cross section

$$D = \frac{V^2 |\mathbf{p}_C|^2}{(2\pi\hbar)^2 (|\mathbf{v}_A||\mathbf{v}_C|)}|\Gamma_{AC}|^2 = \frac{V^2 m^{(1)2} |\mathbf{p}_C|}{(2\pi\hbar)^2 |\mathbf{p}_A|}|\Gamma_{AC}|^2 \quad . \tag{284}$$

In order to eliminate the $1/r$ term in $\Gamma_{AC}$, applying the following formula

$$\int (\nabla^2 F)\exp(\frac{i\mathbf{q}\cdot\mathbf{x}}{\hbar})d^3x = -\frac{|\mathbf{q}|^2}{\hbar^2}\int F \exp(\frac{i\mathbf{q}\cdot\mathbf{x}}{\hbar})d^3x \quad . \tag{285}$$

to the transition amplitude, we obtain

$$\begin{aligned}
\Gamma_{AC} &= \frac{-i\hbar}{V|\mathbf{q}|^2}\int d^3x \nabla^2 (\frac{1}{p_4^{(A)}r} g_a \sigma_a^{(1k)} p_\mu^{(k\_AVG)} p_\mu^{(A)}) \exp\left(\frac{i\mathbf{q}\cdot\mathbf{x}}{\hbar}\right) \\
&= \frac{-i\hbar}{V|\mathbf{q}|^2}\frac{1}{p_4^{(A)}} g_a \sigma_a^{(1k)} p_\mu^{(k\_AVG)} p_\mu^{(A)} \int d^3x \nabla^2 (\frac{1}{r}) \exp\left(\frac{i\mathbf{q}\cdot\mathbf{x}}{\hbar}\right) \\
&= \frac{-i\hbar}{V|\mathbf{q}|^2}\frac{4\pi}{p_4^{(A)}} g_a \sigma_a^{(1k)} p_\mu^{(k\_AVG)} p_\mu^{(A)}
\end{aligned} \tag{286}$$

It is in the form of jet-jet weak interaction, with a proportion constant can be identified as the Fermi's constant for weak β-decay. The differential cross section is given by

$$\frac{d\sigma}{d\Omega} = D = \frac{V^2 m^{(1)2} |\mathbf{p}_C|}{(2\pi\hbar)^2 |\mathbf{p}_A|} |\Gamma_{AC}|^2$$

$$= \frac{V^2 m^{(1)2} |\mathbf{p}_C|}{(2\pi\hbar)^2 |\mathbf{p}_A|} \left| \frac{4\pi}{p_4^{(A)}} g_a \sigma_a^{(1k)} p_\mu^{(k\_AVG)} p_\mu^{(A)} \right|^2 . \quad (287)$$

It was found this differential cross section formula to be un-practicable, the reason arises from the second Pauli matrix which contains imaginary numbers as

$$\sigma_1 = \begin{bmatrix} 0 & 1 \\ 1 & 0 \end{bmatrix}, \quad \sigma_2 = \begin{bmatrix} 0 & -i \\ i & 0 \end{bmatrix}, \quad \sigma_3 = \begin{bmatrix} 1 & 0 \\ 0 & -1 \end{bmatrix} . \quad (288)$$

It brings us with a trouble: require redefining particle momentum to be imaginary number (with respect to the first three components, the fourth one is on time-axis).

$$\sigma_2 = \begin{bmatrix} 0 & -i \\ i & 0 \end{bmatrix} \to demands \to \quad p = p^{(Re)} + ip^{(Im)} . \quad (289)$$

Recall that the second Pauli matrix $\alpha_2 \sigma_2$ arises the spin effect for electron; the third Pauli matrix $\alpha_3 \sigma_3$ serves fermions for which it represents a kind of self-interaction that is associated with the Pauli Exclusion Principle. In practice, we fall back to the old place where we let $\alpha_1 \sigma_1$ to work while $\alpha_2 \sigma_2$ and $\alpha_3 \sigma_3$ to idle, in a moment we will add the $\alpha_2 \sigma_2$ spin effects, ignoring the Pauli exclusion effect for the collision.

But its merit exists, waiting for us to explore, for example, to find a real matrix β to replace the second Pauli matrix $\sigma_2$ if the real matrix β can yield the spin effects completely or partially for the problem under investigation. Without redefining momentum, we have

$$\begin{bmatrix} R^{(1)} \\ R^{(2)} \end{bmatrix} = (1+S) \begin{bmatrix} p^{(1)} \\ p^{(2)} \end{bmatrix} = (1 + \alpha_1 \sigma_1 + \alpha_2 \beta + \alpha_3 \sigma_3) \begin{bmatrix} p^{(1)} \\ p^{(2)} \end{bmatrix} . \quad (290)$$

This real matrix β as substitution is called as the prosthesis matrix. The differential cross section is given by

$$\frac{d\sigma}{d\Omega} = D = \frac{V^2 m^{(1)2} |\mathbf{p}_C|}{(2\pi\hbar)^2 |\mathbf{p}_A|} |\Gamma_{AC}|^2$$

$$= \frac{V^2 m^{(1)2} |\mathbf{p}_C|}{(2\pi\hbar)^2 |\mathbf{p}_A|} \left| \frac{4\pi}{p_4^{(A)}} rS^{(1k)} p_\mu^{(k\_AVG)} p_\mu^{(A)} \right|^2 . \quad (291)$$

where the interaction matrix $S$ contains the prosthesis matrix, $rS$ being independent of $r$. It agrees well with experiment.

### 15. Glashow-Weinberg-Salam weak interaction

Leptons and quarks participate in weak interactions in a form of doublet states:

$$\begin{pmatrix} \nu_e \\ e \end{pmatrix}, \quad \begin{pmatrix} \nu_\mu \\ \mu \end{pmatrix}, \quad \begin{pmatrix} \nu_\tau \\ \tau \end{pmatrix}, \quad (292)$$

$$\begin{pmatrix} u \\ d \end{pmatrix}, \quad \begin{pmatrix} c \\ s \end{pmatrix}, \quad \begin{pmatrix} t \\ b \end{pmatrix}, \quad . \tag{293}$$

We have used these conventional notations in the common textbooks[18]. Of cause, these doublets are the best samples for calculating their weak interactions as a *N*=2 particle system by using the concept of riding-wave momentum, we have

$$\begin{bmatrix} R^{(1)} \\ R^{(2)} \end{bmatrix} = (1+S) \begin{bmatrix} p^{(1)} \\ p^{(2)} \end{bmatrix} = (1+\alpha_1\sigma_1 + \alpha_2\sigma_2 + \alpha_3\sigma_3) \begin{bmatrix} p^{(1)} \\ p^{(2)} \end{bmatrix} . \tag{294}$$

In the reference frame of center of mass, we assume that

$$\alpha_1 = gg_1/r, \quad \alpha_2 = gg_2/r, \quad \alpha_3 = gg_3/r \ , \tag{295}$$

Where $r$ is the radius distance to the center of mass for the particle 1 considered, $g_1, g_2$ and $g_3$ are the coupling strengths, respectively corresponding to electromagnetic interaction, spin interaction and Pauli exclusive interaction, while $g$ is the common coupling strength. Substituting the Pauli matrices into the interaction matrix, we obtain

$$\begin{bmatrix} R^{(1)} \\ R^{(2)} \end{bmatrix} = (1 + \frac{g}{r} \begin{bmatrix} g_3 & g_1 - ig_2 \\ g_1 + ig_2 & -g_3 \end{bmatrix}) \begin{bmatrix} p^{(1)} \\ p^{(2)} \end{bmatrix} . \tag{296}$$

We immediately find that we are getting in trouble with the term $\alpha_2\sigma_2$, which demands the particle momentum to be complex number (with respect to the first three components, the fourth one is on time-axis) as follows

$$\sigma_2 = \begin{bmatrix} 0 & -i \\ i & 0 \end{bmatrix} \rightarrow demands \rightarrow \quad p = p^{(Re)} + ip^{(Im)} \ . \tag{297}$$

After examination, we conclude that we are not able to deal with a momentum in complex number in their riding-wave momentum equations.

To cope with this failure, we propose a modification that consists of two ideas: (1) the particle momenta are conventionally real numbers (with respect to the first three components); (2) smearing off the term $\alpha_2\sigma_2$ and compensating with a real matrix $\alpha_4\sigma_4$ which play the equivalent function of the dismissed term $\alpha_2\sigma_2$, we call the $\sigma_4$ as the prosthesis matrix.

$$\begin{bmatrix} R^{(1)} \\ R^{(2)} \end{bmatrix} = (1+S) \begin{bmatrix} p^{(1)} \\ p^{(2)} \end{bmatrix} = (1+\alpha_1\sigma_1 + 0 + \alpha_3\sigma_3) \begin{bmatrix} p^{(1)} \\ p^{(2)} \end{bmatrix} + (\alpha_4\sigma_4) \begin{bmatrix} p^{(1)} \\ p^{(2)} \end{bmatrix} . \tag{298}$$

Analyzing experimental data, it is fortunate for us to find that the prosthesis matrix is at least approximately equal to the unit matrix, that is

$$\sigma_4 = \begin{bmatrix} 1 & 0 \\ 0 & 1 \end{bmatrix}, \quad \alpha_4 = gg_4/r \ . \tag{299}$$

Thus the interaction matrix *S* becomes a real matrix, and the associated momenta are real numbers under the condition that $\alpha_4\sigma_4$ works as full-function prosthesis. Actually, we eventually find that the $g_1$ is connected with charge while the $g_4$ is connected with hypercharge of the particle 1, they all are with U(1)$_Y$ group symmetry.

As we know that the $\alpha_1\sigma_1+\alpha_2\sigma_2+\alpha_3\sigma_3$ matrix is with SU(2) group symmetry, so that we have an impression that weak interactions belong to SU(2)xU(1)$_Y$ group symmetry. Smearing off the second Pauli matrix has confronted with a more complicate situation: the SU(2) group symmetry is not complete while the U(1)$_Y$ group symmetry still exists for the hypercharge of the particle considered. Finally, the weak interaction is expressed in the matrix equations

$$\begin{bmatrix} R^{(1)} \\ R^{(2)} \end{bmatrix} = (1 + \frac{g}{r}\begin{bmatrix} g_3 + g_4 & g_1 \\ g_1 & -g_3 + g_4 \end{bmatrix})\begin{bmatrix} p^{(1)} \\ p^{(2)} \end{bmatrix}. \tag{300}$$

(1) A defect of the Glashow-Weinberg-Salam weak interaction model

In the Glashow-Weinberg-Salam weak interaction, the interaction is given by

$$\begin{bmatrix} R^{(1)} \\ R^{(2)} \end{bmatrix} = (1 + \alpha_1\sigma_1 + \alpha_2\sigma_2 + \alpha_3\sigma_3)\begin{bmatrix} p^{(1)} \\ p^{(2)} \end{bmatrix} + (\alpha_4\sigma_4)\begin{bmatrix} p^{(1)} \\ p^{(2)} \end{bmatrix}. \tag{301}$$

Or

$$\begin{bmatrix} R^{(1)} \\ R^{(2)} \end{bmatrix} = (1 + \frac{g}{r}\begin{bmatrix} g_3 + g_4 & g_1 - ig_2 \\ g_1 + ig_2 & -g_3 + g_4 \end{bmatrix})\begin{bmatrix} p^{(1)} \\ p^{(2)} \end{bmatrix}. \tag{302}$$

By defining

$$\begin{aligned} W^- &= g_1 - ig_2 \\ W^+ &= g_1 + ig_2 \end{aligned}. \tag{303}$$

The interaction becomes

$$\begin{bmatrix} R^{(1)} \\ R^{(2)} \end{bmatrix} = (1 + \frac{g}{r}\begin{bmatrix} g_3 + g_4 & W^- \\ W^+ & -g_3 + g_4 \end{bmatrix})\begin{bmatrix} p^{(1)} \\ p^{(2)} \end{bmatrix}. \tag{304}$$

It is right, but throughout all particle physics textbooks by searching for the form $p=p^{(Re)}+ip^{(Im)}$, it was found that physicists treated the $W^-$ and $W^+$ as two real numbers, that is equivalent to smearing off the $\alpha_2\sigma_2$. This smearing is reversible deceitful defect in the Glashow-Weinberg-Salam weak interaction model, so that the SU(2) group symmetry is not complete for the model. In honesty, it accounts for the fact that early physicists are not able to confront $\alpha_2\sigma_2$ with imaginary entries in an accurate way.

(2) Welding of the prosthesis: Weinberg angle

We return to the weak interaction expressed in the matrix equations

$$\begin{bmatrix} R^{(1)} \\ R^{(2)} \end{bmatrix} = (1 + \frac{g}{r}\begin{bmatrix} g_3 + g_4 & g_1 \\ g_1 & -g_3 + g_4 \end{bmatrix})\begin{bmatrix} p^{(1)} \\ p^{(2)} \end{bmatrix}, \tag{305}$$

where the $g_4\sigma_4$ represents the prosthesis duo to cutting off the $\sigma_2$, in a hope that the prosthesis works well with an expected precision. Analyzing experimental data, after installed a prosthesis, ones find that the fields must mix in such a way as

$$\begin{aligned} A &= g_3 \sin\theta_w + g_4 \cos\theta_w \\ Z &= g_3 \cos\theta_w - g_4 \sin\theta_w \end{aligned}, \tag{306}$$

Or

$$g_3 = A\sin\theta_w + Z\cos\theta_w$$
$$g_4 = A\cos\theta_w - Z\sin\theta_w \quad , \tag{307}$$

where $\theta_w$ is called the Weinberg angle or weak mixing angle (although Glashow was the first to introduce the idea). From the analyses of data for inclusive and exclusive neutrino-nucleon processes, we have $\sin^2\theta_w=0.234$. We may write the weak interaction as

$$\begin{bmatrix} R^{(1)} \\ R^{(2)} \end{bmatrix} = (1+S)\begin{bmatrix} p^{(1)} \\ p^{(2)} \end{bmatrix}$$
$$= (1 + \frac{g}{r}\begin{bmatrix} (A\sin\theta_w + Z\cos\theta_w) + (A\cos\theta_w - Z\sin\theta_w) & g_1 \\ g_1 & -(A\sin\theta_w + Z\cos\theta_w) + (A\cos\theta_w - Z\sin\theta_w) \end{bmatrix})\begin{bmatrix} p^{(1)} \\ p^{(2)} \end{bmatrix}. \tag{308}$$

The differential cross section is given by

$$\frac{d\sigma}{d\Omega} = D = \frac{V^2 m^{(1)2} |\mathbf{p}_C|}{(2\pi\hbar)^2 |\mathbf{p}_A|}|\Gamma_{AC}|^2$$
$$= \frac{V^2 m^{(1)2} |\mathbf{p}_C|}{(2\pi\hbar)^2 |\mathbf{p}_A|}\left|\frac{4\pi}{p_4^{(A)}} rS^{(1k)} p_\mu^{(k\_AVG)} p_\mu^{(A)}\right|^2. \tag{309}$$

where the interaction matrix $S$ contains the prosthesis matrix, $rS$ being independent of $r$.

The observed lifetimes of weak interaction are considerably longer than those of strong interaction, it is found that

$$\pi^- \to \mu^- \nu_\mu^* \quad \text{with } 2.6 \times 10^{-8} \text{ sec}$$
$$\mu^- \to e^- \nu_e^* \nu_\mu \quad \text{with } 2.2 \times 10^{-6} \text{ sec} \tag{310}$$

whereas particles decay by strong interactions in about $10^{-23}$ sec and through electromagnetic interaction in about $10^{-16}$ sec. The longer lifetime in weak process implies the radius $r$ to be almost hold at a constant for a longer time, as shown in Fig.4, thus we can softly assume the radius to be a constant in our calculation for weak interaction as

$$G = \frac{g}{r} = cons. \quad , \tag{311}$$

Where the constant number G represent the coupling now---Fermi's constant. We may write the weak interaction as

$$\begin{bmatrix} R^{(1)} \\ R^{(2)} \end{bmatrix} = (1+S)\begin{bmatrix} p^{(1)} \\ p^{(2)} \end{bmatrix}$$
$$= (1 + G\begin{bmatrix} (A\sin\theta_w + Z\cos\theta_w) + (A\cos\theta_w - Z\sin\theta_w) & g_1 \\ g_1 & -(A\sin\theta_w + Z\cos\theta_w) + (A\cos\theta_w - Z\sin\theta_w) \end{bmatrix})\begin{bmatrix} p^{(1)} \\ p^{(2)} \end{bmatrix}. \tag{312}$$

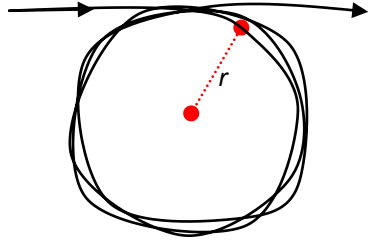

Fig.4 The radius seems to be a constant in stationary state.

In classical physics, $p^2/2m$ represents kinetic energy; in theory of relativity, $p_\mu p_\mu/m = -mc^2$ represents relativistic rest energy; then, what energy means the product of $R_\mu R_\mu/m$ for riding-wave momentum? Obviously, it means the interaction energy. The particle 2 is almost at rest, the interacting energy is given by

$$E^{(12)} = \left(-\frac{1}{m^{(1)}} R_\mu^{(1)} R_\mu^{(1)} - m^{(1)}c^2\right) + \left(-\frac{1}{m^{(2)}} R_\mu^{(2)} R_\mu^{(2)} - m^{(2)}c^2\right)$$
$$\simeq \left(-\frac{1}{m^{(1)}} R_\mu^{(1)} R_\mu^{(1)} - m^{(1)}c^2\right) + 0 \tag{313}$$

Or, the energy of the system for weak process is

$$H \simeq -\frac{1}{m^{(1)}} R_\mu^{(1)} R_\mu^{(1)} \tag{314}$$

It agrees well with experiment. The Fermi's constant is $G_F = 1.166 \times 10^{-5} \text{GeV}^{-2}$.

(3) Welding of the prosthesis: Cabibbo angle

In describing the charged current weak interactions of the quarks, after installed a prosthesis, ones found that the quarks must mix in such a way as

$$\begin{bmatrix} d' \\ s' \end{bmatrix} = \begin{bmatrix} \cos_c & \sin_c \\ -\sin_c & \cos_c \end{bmatrix} \begin{bmatrix} d \\ s \end{bmatrix}, \tag{315}$$

where $\theta_c$ is called the Cabibbo angle. We may write the weak interaction as

$$\begin{bmatrix} R^{(1)} \\ R^{(2)} \end{bmatrix} = (1 + \frac{g}{r}\begin{bmatrix} (A\sin\theta_w + Z\cos\theta_w) + (A\cos\theta_w - Z\sin\theta_w) & g_1 \\ g_1 & -(A\sin\theta_w + Z\cos\theta_w) + (A\cos\theta_w - Z\sin\theta_w) \end{bmatrix} \begin{bmatrix} \cos_c & \sin_c \\ -\sin_c & \cos_c \end{bmatrix})\begin{bmatrix} d \\ s \end{bmatrix}$$
. (316)

The differential cross section is given by

$$\frac{d\sigma}{d\Omega} = D = \frac{V^2 m^{(1)2} |\mathbf{p}_C|}{(2\pi\hbar)^2 |\mathbf{p}_A|} |\Gamma_{AC}|^2$$
$$= \frac{V^2 m^{(1)2} |\mathbf{p}_C|}{(2\pi\hbar)^2 |\mathbf{p}_A|} \left|\frac{4\pi}{p_4^{(A)}} rS^{(1k)} p_\mu^{(k\_AVG)} p_\mu^{(A)}\right|^2 . \tag{317}$$

where the interaction matrix $S$ contains the prosthesis matrix, $rS$ being independent of $r$.

The observed lifetimes of weak interaction are considerably longer than those of strong

interaction, thus we can softly assume the radius to be a constant in our calculation for weak interaction as

$$G = \frac{g}{r} = cons. \quad , \tag{318}$$

Where the constant number G represent the coupling now---Fermi's constant. We may write the weak interaction as

$$\begin{bmatrix} R^{(1)} \\ R^{(2)} \end{bmatrix} = (1+G) \begin{bmatrix} (A\sin\theta_w + Z\cos\theta_w) + (A\cos\theta_w - Z\sin\theta_w) & g_1 \\ g_1 & -(A\sin\theta_w + Z\cos\theta_w) + (A\cos\theta_w - Z\sin\theta_w) \end{bmatrix} \begin{bmatrix} \cos_c & \sin_c \\ -\sin_c & \cos_c \end{bmatrix} \begin{bmatrix} d \\ s \end{bmatrix}$$

In classical physics, $p^2/2m$ represents kinetic energy; in theory of relativity, $p_\mu p_\mu/m = -mc^2$ represents relativistic rest energy; then, what energy means the product of $R_\mu R_\mu/m$ for riding-wave momentum? Obviously, it means the interaction energy. The particle 2 is almost at rest, the interacting energy is given by

$$E^{(12)} = \left( -\frac{1}{m^{(1)}} R_\mu^{(1)} R_\mu^{(1)} - m^{(1)} c^2 \right) + \left( -\frac{1}{m^{(2)}} R_\mu^{(2)} R_\mu^{(2)} - m^{(2)} c^2 \right)$$
$$\simeq \left( -\frac{1}{m^{(1)}} R_\mu^{(1)} R_\mu^{(1)} - m^{(1)} c^2 \right) + 0 \tag{319}$$

Or, the energy of the system for weak process is

$$H \simeq -\frac{1}{m^{(1)}} R_\mu^{(1)} R_\mu^{(1)} \tag{320}$$

It agrees well with experiment.
(4) Welding of the prosthesis: Cabibbo-Kobayashi-Maskawa matrix

By analogy with two particle system, after installed a prosthesis, three quark system must mix in a way that is the Cabibbo-Kobayashi-Maskawa matrix

$$\begin{bmatrix} d' \\ s' \\ b' \end{bmatrix} = \begin{bmatrix} V_{ud} & V_{us} & V_{ub} \\ V_{cd} & V_{cs} & V_{cb} \\ V_{td} & V_{tb} & V_{tb} \end{bmatrix} \begin{bmatrix} d \\ s \\ b \end{bmatrix} . \tag{321}$$

It is called the Kobayashi-Maskawa model for its hidden prosthesis.

**16. The Higgs mechanism**

What is the vacuum made of? Everyone has his or her owning answer, among them which answers are imposed with SU(2) symmetry? Only few geniuses gave out physically-interesting answer.

Imagining vacuum composes with two scale fields $\varphi(x)^{(1)}$ and $\varphi(x)^{(2)}$, they satisfy two conditions:(1) their wave functions are quantized and given by

$$\psi^{(1)} = \exp\left(\frac{i}{\hbar} R^{(1)}\right)$$
$$\psi^{(2)} = \exp\left(\frac{i}{\hbar} R^{(2)}\right), \tag{322}$$

$$\begin{bmatrix} R^{(1)} \\ R^{(1)} \end{bmatrix} = [1 + G(g_1\sigma_1 + g_2\sigma_2 + g_3\sigma_3)] \begin{bmatrix} \varphi^{(1)} \\ \varphi^{(2)} \end{bmatrix}. \tag{323}$$

This implies that the scale field doublet is with SU(2) symmetry. Where, G is the coupling constant. (2) the scale field doublet has energy density as follows

$$E = R^{(1)*}R^{(1)} + R^{(2)*}R^{(2)} = \begin{bmatrix} R^{(1)} \\ R^{(2)} \end{bmatrix}^{+} \begin{bmatrix} R^{(1)} \\ R^{(2)} \end{bmatrix}. \tag{324}$$

Thus the scale fields $\varphi(x)^{(1)}$ and $\varphi(x)^{(2)}$ have sqrt(energy) dimension.

Nobody know this model of vacuum is right or not, but the SU(2) symmetry never goes wrong---that is matter interesting for us.

We may write the interaction matrix as

$$\begin{bmatrix} R^{(1)} \\ R^{(2)} \end{bmatrix} = (1+S)\begin{bmatrix} p^{(1)} \\ p^{(2)} \end{bmatrix} = (1+G\begin{bmatrix} g_3 & g_1 - ig_2 \\ g_1 + ig_2 & -g_3 \end{bmatrix})\begin{bmatrix} \varphi^{(1)} \\ \varphi^{(2)} \end{bmatrix}. \tag{325}$$

Substituting it into the energy density of the system, we get

$$E = \begin{bmatrix} R^{(1)} \\ R^{(2)} \end{bmatrix}^{+} \begin{bmatrix} R^{(1)} \\ R^{(2)} \end{bmatrix} = \left|1 + G\begin{bmatrix} g_3 & g_1 - ig_2 \\ g_1 + ig_2 & -g_3 \end{bmatrix}\begin{bmatrix} \varphi^{(1)} \\ \varphi^{(2)} \end{bmatrix}\right|^2. \tag{326}$$

It was found that the scale field doublet can be reduced into singlet in a minimal energy state as

$$\begin{bmatrix} \varphi^{(1)} \\ \varphi^{(2)} \end{bmatrix} = \frac{1}{\sqrt{2}}\begin{bmatrix} 0 \\ \upsilon \end{bmatrix}. \tag{327}$$

Then the system is

$$E = \left|1 + \frac{G}{\sqrt{2}}\begin{bmatrix} g_3 & g_1 - ig_2 \\ g_1 + ig_2 & -g_3 \end{bmatrix}\begin{bmatrix} 0 \\ \upsilon \end{bmatrix}\right|^2 = \frac{G^2\upsilon^2}{8}[g_1^2 + g_2^2 + g_3^2]. \tag{328}$$

We compare these terms with a typical mass term of a boson, we find the Highs mass

$$M = \frac{G\upsilon}{2}. \tag{329}$$

As we know, the $\sigma_2$ represents imagine contribution

$$\sigma_2 = \begin{bmatrix} 0 & -i \\ i & 0 \end{bmatrix} \to demands \to \varphi = \varphi^{(Re)} + i\varphi^{(Im)}. \tag{330}$$

To cope with this failure, we propose a modification that consists of two ideas: (1) the scale fields are conventionally real numbers; (2) smearing off the term $g_2\sigma_2$ and compensating with a real matrix $g_4\sigma_4$ which play the equivalent function of the dismissed term $g_2\sigma_2$, we call the $\sigma_4$ as the prosthesis matrix.

$$\begin{bmatrix} R^{(1)} \\ R^{(2)} \end{bmatrix} = (1 + \frac{g}{r} \begin{bmatrix} g_3 + g_4 & g_1 \\ g_1 & -g_3 + g_4 \end{bmatrix}) \begin{bmatrix} \varphi^{(1)} \\ \varphi^{(2)} \end{bmatrix}, \quad (331)$$

where the $g_4\sigma_4$ represents the prosthesis duo to cutting off the $\sigma_2$, in a hope that the prosthesis works well with an expected precision. Analyzing experimental data, after installed a prosthesis, ones find that the fields must mix in such a way as

$$\begin{aligned} A &= g_3 \sin\theta_w + g_4 \cos\theta_w \\ Z &= g_3 \cos\theta_w - g_4 \sin\theta_w \end{aligned}, \quad (332)$$

Or

$$\begin{aligned} g_3 &= A\sin\theta_w + Z\cos\theta_w \\ g_4 &= A\cos\theta_w - Z\sin\theta_w \end{aligned}, \quad (333)$$

where $\theta_w$ is called the Weinberg angle or weak mixing angle We may write the weak interaction matrix as

$$\begin{aligned} \begin{bmatrix} R^{(1)} \\ R^{(2)} \end{bmatrix} &= (1+S) \begin{bmatrix} \varphi^{(1)} \\ \varphi^{(2)} \end{bmatrix} \\ &= (1+G \begin{bmatrix} (A\sin\theta_w + Z\cos\theta_w) + (A\cos\theta_w - Z\sin\theta_w) & g_1 \\ g_1 & -(A\sin\theta_w + Z\cos\theta_w) + (A\cos\theta_w - Z\sin\theta_w) \end{bmatrix}) \begin{bmatrix} \varphi^{(1)} \\ \varphi^{(2)} \end{bmatrix} \end{aligned}. \quad (334)$$

After these modifications, we also get the same Higgs mass that is the product of the SU(2) symmetry.

## 17. Strong interaction

In the preceding chapter, let $\alpha_1\sigma_1$ to work while $\alpha_2\sigma_2$ and $\alpha_3\sigma_3$ to idle, then regarding $\alpha_2\sigma_2$ and $\alpha_3\sigma_3$ as additive effects for the collision, we have studied the electron-hadron interaction as follows

$$\begin{bmatrix} R^{(1)} \\ R^{(2)} \end{bmatrix} = (1+S) \begin{bmatrix} p^{(1)} \\ p^{(2)} \end{bmatrix} = (1+\alpha_1\sigma_1 + 0 + 0) \begin{bmatrix} p^{(1)} \\ p^{(2)} \end{bmatrix}. \quad (335)$$

$$\begin{aligned} \alpha_1 &= \frac{1}{4\pi\varepsilon_0} \frac{q^{(1)}q^{(2)}}{rm^{(2)}c^2} = \frac{1}{m^{(2)}c^2} V(r) \\ A_\mu &= \frac{1}{4\pi\varepsilon_0} \frac{q^{(2)}u_\mu^{(2)}}{c^2 r} = \frac{\alpha_1 p_\mu^{(2)}}{q^{(1)}} \end{aligned}. \quad (336)$$

We derived out the Rosenbluth scattering formula

$$\frac{d\sigma}{d\Omega} = \left(\frac{d\sigma}{d\Omega}\right)_{Mott} [1 - \frac{q^2}{2m^{(2)2}c^2} \sin^2(\theta/2)]. \quad (337)$$

The modified Rosenbluth scattering formula is given by

$$\frac{d\sigma}{d\Omega} = \left(\frac{d\sigma}{d\Omega}\right)_{Mott} [F_1(\theta, E) - \frac{q^2}{2m^{(2)}c^2} F_2(\theta, E) \sin^2(\theta/2)] \ . \tag{338}$$

Where $F_1$ and $F_2$ are the form factors, as we know, the form factors work very well in experiments of high energy particle physics. The first form factor originates from electric interaction; the second form factor originates from magnetic interaction.

In our derivation, the hadron is regarded as a point-particle without internal structure. But for two form factors $F_1$ and $F_2$, early physicists might had read as there is charge distribution in the hadron and there is magnetic moment with the hadron, finally they believed in that the hadron consists of several partons or quarks (u,d,s,…). Now we are going to investigate the quark model by thinking the quarks as the prosthesis of the hadron.

The probing particle is named as the particle 1, the hadron is named as the particle 2, they satisfy

$$\begin{bmatrix} R^{(1)} \\ R^{(2)} \end{bmatrix} = (1 + \alpha_1\sigma_1 + 0 + \alpha_3\sigma_3) \begin{bmatrix} p^{(1)} \\ p^{(2)} \end{bmatrix} + (\alpha_4\sigma_4) \begin{bmatrix} p^{(1)} \\ p^{(2)} \end{bmatrix} \ . \tag{339}$$

Cutting off imaginary $\alpha_2\sigma_2$, we have to compensate it with real $\alpha_4\sigma_4$ --- hypercharge interaction, and introducing the three quarks (u,d,s) to replace the hadron --- prosthesis, by coupling $\alpha_n = Gg_n$, we have real interaction matrix as

$$\begin{bmatrix} R^{(1)} \\ R^{(2)} \end{bmatrix} = (1 + G\begin{bmatrix} g_3 + g_4 & g_1 \\ g_1 & -g_3 + g_4 \end{bmatrix}) \begin{bmatrix} p^{(1)} \\ p^{(2)} \end{bmatrix}$$

$$\Updownarrow \tag{340}$$

$$\begin{bmatrix} R^{(1)} \\ R^{(u)} \\ R^{(d)} \\ R^{(s)} \end{bmatrix} = (1 + \begin{bmatrix} S_{11} & S_{12} & S_{13} & S_{14} \\ S_{21} & S_{22} & S_{23} & S_{24} \\ S_{31} & S_{32} & S_{33} & S_{34} \\ S_{41} & S_{42} & S_{43} & S_{44} \end{bmatrix}) \begin{bmatrix} p^{(1)} \\ p^{(u)} \\ p^{(d)} \\ p^{(s)} \end{bmatrix}$$

Where the momenta are all in real number form (with respect to the first three components), $g_4$ is the hypercharge --- the most important property of the hadron for the construction of quark concept, the equally important property of the hadron is its electric charge --- $g_1$, they are all quantized. What is about $g_3$? We have mentioned in the preceding chapter that $g_3$ represents the Pauli exclusive interaction, analyzing hadron spectrum as shown in Fig.5, we find that the $g_3$ is connected with the $I_3$ quantity after quantization --- the third component of isospin.

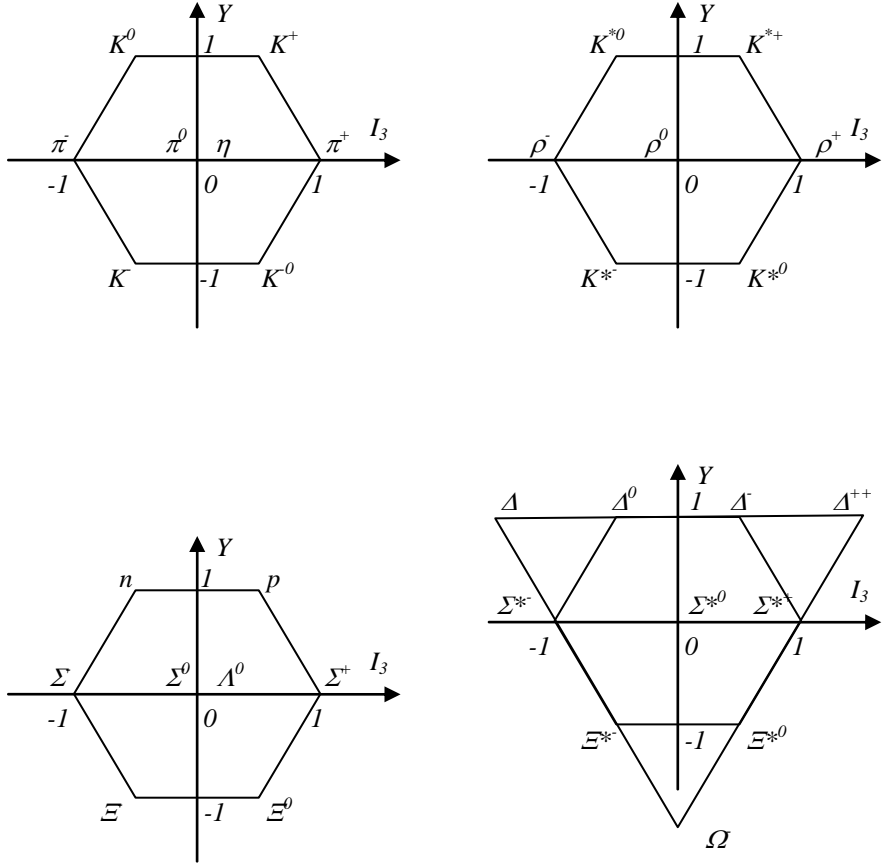

Fig.5    hadron spectrum

We can immediately write down three laws: (1)charge conservation law about $g_1$ ;(2) the hypercharge conservation law about $g_4$; (3) the $I_3$ conservation law about $g_3$, which will be further discussed in chapter 6.

Making prosthesis is a kind of art, concerning on the physics and concerning on aesthetics.

(1) The prosthesis with SU(3) group symmetry

Analyzing hadron spectrum as shown in Fig.5, Gell-Mann realized that the three quarks u, d and s are with SU(3) group symmetry. In the interaction matrix, its right-bottom 3X3 sub-matrix are with SU(3) group symmetry.

$$S = \begin{bmatrix} S_{11} & S_{12} & S_{13} & S_{14} \\ S_{21} & S_{22} & S_{23} & S_{24} \\ S_{31} & S_{32} & S_{33} & S_{34} \\ S_{41} & S_{42} & S_{43} & S_{44} \end{bmatrix} . \quad (341)$$

$$S^{(uds)} = \begin{bmatrix} S_{22} & S_{23} & S_{24} \\ S_{32} & S_{33} & S_{34} \\ S_{42} & S_{43} & S_{44} \end{bmatrix} = \sum_{a=1}^{8} \frac{1}{2} \alpha_a \lambda_a$$

Where $\alpha_a$ is eight real small coefficients of the expansion, $\lambda_a$ are the eight independent Gell-Mann matrices called as the eight generators of SU(3) group, the Gell-mann matrices are

$$\lambda_1 = \begin{bmatrix} 0 & 1 & 0 \\ 1 & 0 & 0 \\ 0 & 0 & 0 \end{bmatrix} \quad \lambda_2 = \begin{bmatrix} 0 & -i & 0 \\ i & 0 & 0 \\ 0 & 0 & 0 \end{bmatrix} \quad \lambda_3 = \begin{bmatrix} 1 & 0 & 0 \\ 0 & -1 & 0 \\ 0 & 0 & 0 \end{bmatrix}, \quad (342)$$

$$\lambda_4 = \begin{bmatrix} 0 & 0 & 1 \\ 0 & 0 & 0 \\ 1 & 0 & 0 \end{bmatrix} \quad \lambda_5 = \begin{bmatrix} 0 & 0 & -i \\ 0 & 0 & 0 \\ i & 0 & 0 \end{bmatrix} \quad \lambda_6 = \begin{bmatrix} 1 & 0 & 0 \\ 0 & 0 & 1 \\ 0 & 1 & 0 \end{bmatrix}, \quad (343)$$

$$\lambda_7 = \begin{bmatrix} 0 & 0 & 0 \\ 0 & 0 & -i \\ 0 & i & 0 \end{bmatrix} \quad \lambda_8 = \frac{1}{\sqrt{3}} \begin{bmatrix} 1 & 0 & 0 \\ 0 & 1 & 0 \\ 0 & 0 & -2 \end{bmatrix}. \quad (344)$$

But, we need to cut off $\lambda_2$, $\lambda_5$, $\lambda_7$, because they contain imaginary entries, and compensate them with three real matrices $\lambda_9$, $\lambda_{10}$, $\lambda_{11}$, so that we obtain

$$S^{(uds)} = \begin{bmatrix} S_{22} & S_{23} & S_{24} \\ S_{32} & S_{33} & S_{34} \\ S_{42} & S_{43} & S_{44} \end{bmatrix}$$

$$= (\alpha_1 \lambda_1 + 0 + \alpha_3 \lambda_3 + \alpha_4 \lambda_4 + 0 + \alpha_6 \lambda_6 + 0 + \alpha_8 \lambda_8) + (\alpha_9 \lambda_9 + \alpha_{10} \lambda_{10} + \alpha_{11} \lambda_{11}) \quad (345)$$

Thus the quarks u,d,s are not completely with SU(3) group symmetry, indeed early physicists took many time to study the breaking of SU(3) group symmetry for the quarks.

How to determine the real prosthesis matrices? Not only we need to analyze the experimental data, but also we have to consider the asymptotic freedom of quark and the confinement of quark.

(2) The asymptotic freedom of quark

The three quarks are regarded as three prosthesis, we indeed design them as free particles with a hope that they satisfy

$$S^{(uds)} \approx \begin{bmatrix} S_{22} & 0 & 0 \\ 0 & S_{33} & 0 \\ 0 & 0 & S_{44} \end{bmatrix} \quad (346)$$

At least the smeared entries should be small enough in comparison with the diagonal entries, this strategy will greatly simplify dynamical calculation over quark problems.

(3) The confinement of quark

It is easy to understand that the three prosthesis have integrity: we can not separate one free quark from real body --- the hadron. So that the quarks must confine in the hadron considered. You know, prosthesis leave its master body, they also lose themselves, like fish in water. In prosthesis design, the fundamental conservation laws must be hold.

$$\begin{array}{l} \text{Charge conservation law (quantized } \alpha_1 \sigma_1) \\ \text{Hypercharge conservation law (quantized } \alpha_4 \sigma_4) \\ I_3 \text{ conservation law (quantized } \alpha_3 \sigma_3) \end{array} \quad (347)$$

These conservation laws confine quarks in the hadron, it is impossible to find the so-called fractal charge in experiment.

## 18. Prosthesis in time axis

In chapter 1, we have derived out the relativistic matter wave from the Newton's second law as follows

$$\phi = \exp\left(\frac{i}{\hbar}\int_{x_0}^{x}(mu_\mu + qA_\mu)dx_\mu\right) = \exp\left(\frac{i}{\hbar}\int_{x_0}^{x} R_\mu dx_\mu\right) \tag{348}$$

where the integral takes from the initial point $x_0$ to the final point $x$ by an arbitrary mathematical path in the velocity field

Consider a particle beam incident on a potential barrier $V(x)$ with the initial energy $E_0$ and initial momentum $p_0$ in one dimensional axis $x$, as shown in Fig.6. Particles come in at the origin $x=-L/2$ at time $t=0$, through the barrier, reach at the observer at $x=L/2$ at time $t=T$.

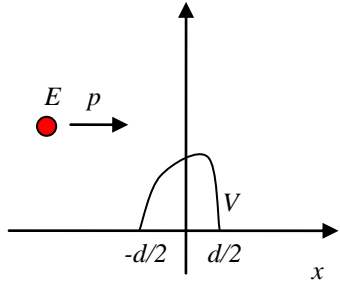

Fig.6 A particle beam incident on a potential barrier.

Since the total mechanical energy keeps on a constant $E_0$, the matter wave is given by

$$\phi = \exp\left(\frac{i}{\hbar}\int_{x_0}^{x} pdx - \frac{i}{\hbar}E_0 t\right) = \exp\left(\frac{i}{\hbar}\int_{x_0}^{x}\sqrt{2m(E_0 - mc^2 - V)}dx - \frac{i}{\hbar}E_0 t\right) \tag{349}$$
$$|\phi| = 1$$

There may be a controversy over it: at one hand, $|\phi|=1$ means we find a particle at everywhere in the beam with the same probability; at another hand, the particle moves with a varying velocity in the barrier range, it results in a broken particle continuity: the current $j=\rho v$ is not constant for the one dimensional motion. This problem needs to develop the prosthesis theory to cope with.

Actually, there is another explanation for the matter wave which is equivalently given by

$$\phi = \exp\left(\frac{i}{\hbar}\int_{x_0}^{x}(mu_\mu + qA_\mu)dx_\mu\right) = \exp\left(\frac{i}{\hbar}p_0 x - \frac{i}{\hbar}(E_0 + V)t\right) \tag{350}$$
$$|\phi| = 1$$

In this formula, the energy is regarded as an open quantity in the time axis, the particle moves with the constant velocity $p_0/m$, with the same probability $|\phi|=1$ at everywhere, the current $j=\rho v$ keeps on a constant at everywhere. In this picture, we must admit that the energy $E_0+V$ is not the conventional mechanical energy, because the particle involves in new characteristic quantity in the time axis, the term $Vt$ is called as the prosthesis of the particle in the time axis.

At the first, we will prove that the above two explanations are equivalent to each other. Consider $V(x)$ as a perturbation for the system, that is

$$\begin{aligned}
\phi &= \exp\left(\frac{i}{\hbar}\int_{x_0}^{x}\sqrt{2m(E_0-mc^2-V)}dx - \frac{i}{\hbar}E_0 t\right) \\
&= \exp\left(\frac{i}{\hbar}\int_{x_0}^{x}\sqrt{p_0^2-2mV}dx - \frac{i}{\hbar}E_0 t\right) \\
&= \exp\left(\frac{i}{\hbar}\int_{x_0}^{x}(p_0 - \frac{mV}{p_0})dx - \frac{i}{\hbar}E_0 t\right) \\
&= \exp\left(\frac{i}{\hbar}p_0 x + \int_{x_0}^{x}(-\frac{mV}{p_0})\frac{dx}{dt}dt - \frac{i}{\hbar}E_0 t\right)
\end{aligned} \tag{351}$$

We invoke $dx/dt=p_0/m$, we have

$$\phi = \exp\left(\frac{i}{\hbar}p_0 x + \int_{x_0}^{x}(-V)dt - \frac{i}{\hbar}E_0 t\right) = \exp\left(\frac{i}{\hbar}p_0 x - \frac{i}{\hbar}(E_0+V)t\right) \tag{352}$$

The proof finished.

Next, all troubles come from the prosthesis term, if we impose the prosthesis into the momentum space like the first explanation, then the prosthesis lets the momentum of the particle vary over the range of the barrier, and the beam current continuity is broken; if we leave the prosthesis in the time axis like the second explanation, then the prosthesis lets the particle uniformly move, but the energy sum in the time axis opens; this is just the spirit of the Heisenberg uncertainty principle.

The third, for visualization of the matter wave, we can use the following transformation

$$\int_{t_0}^{t}V(x)dt = \int_{x_0}^{x}V(x)\frac{dt}{dx}dx = \int_{x_0}^{x}\frac{mV(x)}{p}dx \simeq \int_{x_0}^{x}\frac{mV(x)}{p_0}dx \tag{353}$$

Thus the visualized relativistic matter wave is given by

$$\phi = \exp\left(\frac{i}{\hbar}p_0 x - \int_{x_0}^{x}\frac{mV(x)}{p_0}dx - \frac{i}{\hbar}E_0 t\right) \tag{354}$$

From that on, we can see what the prosthesis term works out in the real spatial ground.

Finally, we conclude that the term

$$\int_{t_0}^{t}V(x)dt \tag{355}$$

lives in the time axis, we lack enough knowledge or experience on its behavior in the time axis. Usually, we transform this term into the spatial ground, like

$$\int_{x_0}^{x}\sqrt{2m(E-mc^2-V)}dx \tag{356}$$

what we saw is the prosthesis of this term.